\documentclass[sn-mathphys-num]{sn-jnl}

\usepackage{graphicx}%
\usepackage{multirow}%
\usepackage{amsmath,amssymb,amsfonts}%
\usepackage{amsthm}%
\usepackage{mathrsfs}%
\usepackage[title]{appendix}%
\usepackage{xcolor}%
\usepackage{textcomp}%
\usepackage{manyfoot}%
\usepackage{booktabs}%
\usepackage{algorithm}%
\usepackage{algorithmicx}%
\usepackage{algpseudocode}%
\usepackage{listings}%

\usepackage[true]{anonymous-acm}

\usepackage{latexsym, bbm}
\usepackage{amstext}
\usepackage{multicol}
\usepackage{tabularx}
\usepackage{hyperref}
\usepackage{url}
\usepackage{color}
\usepackage{array}
\usepackage{footnote}
\usepackage{subcaption}
\usepackage{abstract}

\begin{document}

\title[Don't Trust a Single Gerrymandering Metric]{Don't Trust a Single Gerrymandering Metric }

\author[1]{\fnm{Thomas} \sur{Ratliff}}\email{ratliff\_thomas@wheatoncollege.edu}
\equalcont{Authors listed alphabetically, per convention.}

\author[2]{\fnm{Stephanie} \sur{Somersille}}\email{ssomersille@gmail.com}
\equalcont{Authors listed alphabetically, per convention.}

\author*[3]{\fnm{Ellen} \sur{Veomett}}\email{eveomett@usfca.edu}
\equalcont{Authors listed alphabetically, per convention.}

\affil*[1]{\orgdiv{Department of Mathematics and Computer Science}, \orgname{Wheaton College}}

\affil[2]{\orgdiv{Somersille Math \& Consulting Services}}

\affil[3]{\orgdiv{Department of Computer Science}, \orgname{University of San Francisco}}

\begin{titlepage}
    \maketitle 
    \vspace{2cm}

\bmhead{Acknowledgments}  
This work is supported by the National Science Foundation under Grant No. DMS-1928930, while Ellen Veomett and Stephanie Somersille were in residence at the Mathematical Sciences Research Institute in Berkeley, California, during the semester/year of Fall/2023.

\bmhead{Data Availability}
All data used to conduct the empirical analyses were obtained from the Metric Geometry Gerrymandering Group's GitHub organization at \url{https://github.com/mggg-states}.

\bmhead{Code Availability}
We have made all Python code used to run the empirical analyses publicly available at the following GitHub repository: \url{https://github.com/stem-redistricting/ShortBursts/tree/main}.
    
    \vfill 
\end{titlepage}

\newpage
\begin{abstract}
\vspace{0.5cm}
In recent years, in an effort to promote fairness in the election process, a wide variety of techniques and metrics have been proposed to determine whether a map is a partisan gerrymander.  The most accessible measures, requiring easily obtained data, are metrics such as the Mean-Median Difference, Efficiency Gap, Declination, and GEO metric.  But for most of these metrics, researchers have struggled to describe, given no additional information, how a value of that metric on a single map indicates the presence or absence of gerrymandering.

Our main result is that each of these metrics is gameable when used as a single, isolated quantity to detect gerrymandering (or the lack thereof). That is, for each of the four metrics, we can find district plans for a given state with an extremely large number of Democratic-won (or Republican-won) districts while the metric value of that plan falls within a reasonable, predetermined bound. We do this by using a hill-climbing method to generate district plans that are constrained by the bounds on the metric but also maximize or nearly maximize the number of districts won by a party.  

In addition, extreme values of the Mean-Median Difference do not necessarily correspond to maps with an extreme number of districts won.  Thus, the Mean-Median Difference metric is particularly misleading, as it cannot distinguish more extreme maps from less extreme maps.  The other metrics are more nuanced, but \textit{when assessed on an ensemble},
none perform substantially differently from simply measuring number of districts won by a fixed party. 

One clear consequence of these results is that they demonstrate the folly of specifying a priori bounds on a metric that a redistricting commission must meet  in order to avoid gerrymandering. 

\end{abstract}


\vspace{0.5cm}

\section{Introduction}\label{sec:intro}

Many legislative bodies elect their members through single member districts where jurisdictions are divided up into districts and voters elect one representative for their district. Perhaps the most well-known representatives elected in this way in the United States are the members of the House of Representatives, but districts are also used to elect many state senators, city council members, and others. These representatives are elected with the expectation (hope?) that they will represent their communities' interests.

The United States has a long history of partisan gerrymandering where one party draws the maps to gain a durable, lasting advantage. Examples range from the original use of the term gerrymander in Massachusetts in 1812 to particularly egregious instances in the last twenty years  which have led to multiple lawsuits that have reached the Supreme Court \cite{GillWhitford, RuchoCommonCause, LamoneBenisek}. While the Supreme Court ruled in \cite{RuchoCommonCause} that partisan gerrymandering is non-justiciable by the federal courts\footnote{A matter is non-justiciable by federal courts if it is a matter upon which the federal court is not capable or not allowed to give a ruling.}, it remains justiciable in state courts where some states prohibit partisan gerrymandering. Furthermore, racial demographics and partisan preference are often intertwined and used as stand-ins for each other \cite{RacePolitics}.  Hence, the prevention of partisan gerrymandering can be a way to improve access and opportunity for historically disadvantaged racial groups.

A naturally appealing idea is to develop a metric to quantify gerrymandering: If a map has a metric value within some pre-determined bounds, then the map is reasonable, and if the metric value falls outside these bounds, then the map is gerrymandered. The existence of such a metric is very attractive since it would give citizens a means to evaluate any map produced by the party in control of the redistricting process. In addition, the metric could provide guidance for independent redistricting commissions in states that have such bodies, such as Arizona, California, Colorado, and Michigan.

The purpose of this paper is to examine four such metrics that have been proposed in the last 10 years: the Mean-Median Difference, Efficiency Gap, Declination, and GEO metric. The first three metrics are ubiquitous in the realm of the study of gerrymandering.  For example, they can all be found in the GerryChain library \cite{GerryDetails}, a Python library developed to implement a Markov Chain Monte Carlo process constructing an ensemble of potential redistricting maps that could be used in comparison to a proposed redistricting map.  They can all be found in Dave's Redistricting App \cite{DRA} and the PlanScore site \cite{PlanScore}, which are sites that allow the public to evaluate potential redistricting maps that they create or evaluate pre-existing maps.  And these metrics can all be found in the Princeton Gerrymandering Project's report cards \cite{PGP}, which are publicly viewable assessments of redistricting maps, done by the Princeton Gerrymandering Project team. For all of these resources (\cite{GerryDetails, DRA, PlanScore, PGP}), the metrics are touted as indicating more extreme gerrymanders when the metric values are more extreme (farther from 0).  The Geography and Election Outcome (GEO) metric is a more recently proposed metric that uses both map data as well as partisan data  \cite{geo_paper}.

Our first main result is that each of these metrics is gameable when used as a single, isolated quantity to detect gerrymandering. That is, for each of the four metrics, we can find district plans for a given state with an extremely large number of Democratic-won (or Republican-won) districts while the metric value of that plan falls within a reasonable, predetermined bound. Indeed, we can find a single map with an extremely large number of Democratic-won districts on which all four of the metrics fall within the reasonable range.

Our second main result is that the Mean-Median Difference cannot distinguish between more extreme maps and less extreme maps at all. We describe the results more precisely in the following sections, but the main takeaway is that it is unrealistic to expect that a problem with as many dimensions as partisan gerrymandering could be reduced to a single quantity between \( -1 \) and \( 1 \).  These metrics may have a role to play as part of a more computationally intensive ensemble analysis, but we should not place any weight on the value of a single metric without further context. And the Mean-Median Difference should not be used at all, not even as part of a more computationally intensive ensemble analysis.

We perform three different studies which allow us to conclude these two main results, described in Sections \ref{subsec:Gameability}, \ref{subsec:MetricValsExtremeMaps}, and \ref{subsec:MetricsAndEnsembles}.

\subsection{Gameability Study: Finding Extreme Maps that aren't Out of Bounds}\label{subsec:Gameability}

We define a map as ``extreme'' using the most straightforward of terms: it results in an extreme number of districts won by a given party, while still maintaining the basic redistricting criteria of population balance, contiguity and some measure of compactness.  Thus, in order to determine whether a metric is ``gameable,'' we must determine if extreme maps (maps with a high number of districts won by a particular party) are achievable while keeping the metric within acceptable bounds.  In the case of the Efficiency Gap, the creators were careful not to prescribe the ``acceptable bounds''. However, they used a value of $\pm 8\%$ for the bound between ``presumptive validity and invalidity" in their paper introducing the metric \cite{PartisanGerrymanderingEfficiencyGap}.   For consistency, we use the corresponding values here for all of our metrics: $\pm 16\%$ for metrics whose range is $\left[-1, 1\right]$ (the GEO Metric Ratio\footnote{A value obtained using the two GEO metric scores, which we define later.} and Declination), and $\pm 8\%$ for metrics whose range is  $\left[-0.5, 0.5\right]$ (the Efficiency Gap and Mean-Median Difference).  While we needed to fix a consistent range of metric values in order to complete our Gameability Study, in Section \ref{sec:conclusions} we explain why this choice does not impact our ultimate conclusions.

In order to create these extreme maps, we utilize a hill-climbing method dubbed the ``Short Burst'' method, developed by Cannon et al \cite{cannonShortBursts}.  The Short Burst method uses small runs of a Markov Chain Monte Carlo process to find extreme maps.  Cannon et al showed that this Short Burst method finds maps with extreme numbers of majority-minority districts more successfully than other typically used methods, such as biased Markov Chains \cite{cannonShortBursts}.  While Cannon et al applied the Short Burst method to find districting plans with large numbers of majority-minority districts, we use it to find districting plans with largest number of majority-Democratic (or majority-Republican)
districts.  In section \ref{sec:shortbursts} we describe this method.

In Section \ref{subsec:results_metrics_restricted}, we discuss our results on the kinds of extreme maps that are allowed by each of the four metrics we consider.  As we will see, each of these metrics is gameable for at least some state/map.

\subsection{Extreme Metric Values versus Extreme Maps}\label{subsec:MetricValsExtremeMaps}

While our Gameability Study focuses on extreme maps that have metric values within an acceptable range, one might also ask the even more basic question: do more extreme maps correspond to more extreme metric scores?  This question is related to Stephanopoulos and McGee's description of the \emph{efficiency principle}.  McGee introduced the efficiency principle in \cite{MeasuringEfficiency}, and based on Stephanopoulos and McGhee's explanation of the efficiency principle in \cite{MeasureMetricDebate}, Veomett \cite{2018arXiv180105301V} fleshed out the definition of the eficiency principle to have two parts: 

\begin{quote}
    The efficiency principle states that a measure of partisan gerrymandering must indicate both
    \begin{description}
        \item[EP1]  a greater advantage for (against) a party when the seat share for that party increases (decreases) without any corresponding increase (decrease) in its vote share
        \item[EP2]  a greater advantage against (for) a party when the vote share increases (decreases) without any corresponding increase (decrease) in the seat share.
    \end{description}
\end{quote}
In our exploration of extreme maps (and in every Markov Chain Monte Carlo process that creates an ensemble of maps), the same partisan data is used on every map in order to compare the maps to each other.  Thus, the question of whether more extreme maps correspond to more extreme metric scores is similar to asking whether a metric satisfies a more flexible version of EP1; one in which the most extreme maps are correlated to extreme metric values.

In order to address this correlation question, we again use the Short Burst method.  Without any restrictions (other than the required population balance, contiguity and loose compactness constraints) we create short bursts seeking maps with a large number of Democratic-won districts (and then repeat the process searching for maps with a large number of Republican-won districts).  From the extreme maps that are generated, we separate them into maps with the maximum number of districts won $M$.  We also consider maps with $M-1, M-2, M-3, $ and $M-4$ districts won.  For each of those numbers, we look at the range of values for each metric on those maps to see how that range changes (or does not change) as the number of districts won changes.  

In Section \ref{subsec:results_no_restriction}, we discuss the range of metric values on maps with the maximum or near-maximum number of Democratic-won (or Republican-won) districts.  We consider the same metrics that we considered for the Gameability Study of Section \ref{subsec:Gameability}.
We shall see that extreme values of the Mean-Median Difference do \emph{not} correspond to maps with an extreme number of districts won (although extreme values of the other metrics do correspond to maps with an extreme number of districts won).

\subsection{Metrics and Ensembles}\label{subsec:MetricsAndEnsembles}

Our third and final study is based in ensemble analysis.  Ensemble analysis is a well-studied, widely agreed-upon technique in the modern study of gerrymandering.  Groups from the Metric Geometry Gerrymandering Group \cite{mgggWebpage} to the Quantifying Gerrymandering group \cite{quantifyingGerrymanderingWebpage} to individual expert witness testimony \cite{SS_PSC_MT} use ensemble analysis as a tool to detect gerrymandering.

Ensemble analysis is performed by creating a large number (an ensemble) of potential redistricting maps, which are expected to be a reasonable sample from the space of all redistricting maps.  The proposed map is compared to the maps in the ensemble in some way, such as the number of districts won by a particular party, the order statistics of the vote shares in the districts, or the value of some other metric on those maps.  If the proposed map looks like an outlier among the sample of maps, this is seen as indication that it was potentially drawn with partisan intent.\footnote{This is the typical application of ensemble analysis, which has substantial support/usage in academic articles \cite{DukeNC, DukeWisconsin, RecomMGGG, ColoradoInContext} and expert witness reports \cite{EW_ClarkevWisconsin}.}

Using ensemble analysis on ``number of districts won'' by a particular party is a straightforward comparison when it comes to gerrymandering: we are determining whether the proposed map has an extreme number of districts won, compared to the other maps in the ensemble.  The implications are easy to interpret.  However, if we use ensemble analysis on another metric, 
it is less clear how to interpret the results.  Does an outlier Efficiency Gap value imply the map has an extreme number of districts won by some party?  What about an outlier Mean-Median Difference value?  Our results of this study are in Section \ref{subsec:ensembles}.  In short, we see that the answer is ``no'' for the Mean-Median Difference (that is, outlier values of the Mean-Median Difference do not correspond to maps with an extreme number of districts won) and ``yes'' for the other three metrics we consider.

\section{Metrics and Detecting Gerrymandering}
\label{sec:metrics_studied}

Each of the four metrics we study assumes that there are two political parties vying for seats.  
For completeness, we define these four metrics here, but first we highlight why we ground our understanding of the existence of gerrymandering in ``districts won.''

\subsection{``Districts Won'' as the Foundation of Gerrymandering}


Gerrymandering is commonly understood as the manipulation of district boundaries to produce an extreme number of districts won (or predicted to be won) by a particular party.
For example, the Brennan Center for Justice, in their ``Gerrymandering Explained''  report \cite{BrennanCenterGerrymanderingExplained}, describe gerrymanders in terms of the number of additional seats a party won due to gerrymandering.  Intuition and common understanding aside, Political Scientists and Social Scientists also describe gerrymandering in terms of seats won: ``Partisan gerrymanderers use their knowledge of voter preferences and their ability to draw favorable redistricting plans to maximize their party’s seat share \cite{626161}.''    Mathematicians and Computer Scientists agree with this definition: ``[T]he express intent of a partisan gerrymander is to secure for their party as many
seats as possible under the constraints of voter geography and the other rules of redistricting" \cite{ImplementingPartisanSymmetry}.  While there may be other factors that appear in a gerrymandered map, such as durability of the map over time or incumbency protection, we hold that the baseline goal of a gerrymander is to push the number of districts won towards extreme values for the favored party.

\subsection{Mean-Median Difference}

The Mean-Median Difference is a partisan symmetry metric \cite{ImplementingPartisanSymmetry} which was highlighted by MacDonald and Best in a 2015 article \cite{doi:10.1089/elj.2015.0358}.  This metric is calculated from partisan data only.  Specifically, suppose a state has $n$ districts, whose estimated vote share for party $A$ is $V_i$ in district $i$ (such an estimation can be made using prior election data or any other partisan index estimating support for the two parties).  The Mean-Median Difference is the median of the $V_i$s, minus the mean:
\begin{equation*}
    \text{median}\left\{V_i: i=1, 2, \dots, n\right\} - \text{mean}\left\{V_i: i=1, 2, \dots, n\right\}
\end{equation*}
A large positive Mean-Median Difference value indicates a larger median than mean, and the proponents of the Mean-Median Difference suggest that it indicates a map that is beneficial for party $A$.   

The Mean-Median Difference is called a partisan symmetry metric because, under the assumption of ``uniform partisan swing'' (that when voters swing towards some party, they do so uniformly across all districts), the Mean-Median Difference measures an asymmetry in the way that the map treats the two parties. Specifically, it measures how far party $A$ can fall below a 50\% vote share, while still receiving a 50\% seat share \cite{ImplementingPartisanSymmetry}\footnote{To give a simple example, if party $A$ has a median vote share among the districts of $50\%$ and a mean vote share of $40\%$ (or a Mean-Median Difference of $0.1$) then party $A$ wins half the seats with only $40\% $ of the total vote share.}. Thus, the Mean-Median Difference is always between -0.5 and 0.5.  

\subsection{Efficiency Gap}

The Efficiency Gap is also a metric which is calculated from partisan data only.  It was proposed by  Stephanopoulos and McGhee \cite{PartisanGerrymanderingEfficiencyGap} and was used to argue the existence of gerrymandering in Wisconsin in the Supreme Court case Gill v. Whitford \cite{GillWhitford}.

The Efficiency Gap utilizes the idea of a ``wasted vote\footnote{``Wasted votes'' is terminology from  Stephanopoulos and McGhee's paper, and not of our creation.}.''  For a political party whose candidate lost in a district, all votes for that candidate are considered ``wasted votes'' for that party.  For a political party whose candidate who won in a district, that party's number of wasted votes from that district corresponds to the number of votes their candidate received \emph{beyond 50\%} of the votes in that district.  That is, a winning candidate's party's wasted votes is the number of votes that candidate received, minus half of the turnout of that district. 

To calculate the Efficiency Gap (EG), one sums party $B$'s wasted votes across all districts, subtracts the sum of party $A$'s wasted votes across all districts, and divides by the total votes:
\begin{equation*}
    EG = \frac{\text{party } B \text{'s wasted votes } - \text{ party } A \text{'s wasted votes}}{\text{Total votes}}
\end{equation*}

A large positive EG value indicates more wasted votes for party $B$, which is intended to indicate a map that is beneficial for party $A$.  

Note that exactly half of the votes in each district are wasted by one party or another.  Thus, $-0.5 \leq EG \leq 0.5$.

\subsection{Declination}

The Declination is a metric introduced by Warrington, which also uses only partisan data \cite{WarringtonDeclinationELJ}.  This metric is intended to find ``packing'' and ``cracking,'' which are how maps are frequenty described to be gerrymandered.  Packing and cracking occur when an opposing party's voters are ``packed" into a small number of districts that they win with an overwhelming majority.  The remaining opposing party's voters are ``cracked" or split among many districts, so that they lose by relatively slim margins in those districts.  It is by this packing and cracking that a mapmaker can attempt to win as many districts as possible for her preferred party.

To calculate the Declination, we order the vote shares $V_i$ in all districts $i=1, 2, \dots, n$ from smallest vote share for party $B$ to largest vote share for party $B$:

\begin{equation*}
V_1 \leq V_2 \leq \cdots \leq V_k < \frac{1}{2} < V_{k+1} \leq \cdots \leq V_n
\end{equation*}
Note that here we have party $B$ losing $k$ districts and winning $n-k$ districts.  

We then define two sets $\mathcal{A}$ and $\mathcal{B}$ as follows:
\begin{align*}
\mathcal{A} &= \left\{\left(\frac{i}{n}-\frac{1}{2n}, V_i,\right): i = 1, 2, \dots, k\right\} \\
\mathcal{B} &= \left\{\left(\frac{j}{n}-\frac{1}{2n}, V_j\right): j = k+1, k+2, \dots, n\right\}
\end{align*}

Let $F$ be the center of mass of points in $\mathcal{A}$, $G = \left(\frac{k}{n},\frac{1}{2}\right)$, and $H$ the center of mass of points in $\mathcal{B}$.  We define angle $\theta_P$ to be the angle that the line through $F$ and $G$ makes with the line $y = \frac{1}{2}$, and $\theta_Q$ to be the angle that the line through $G$ and $H$ makes with the line $y = \frac{1}{2}$.  These can be visualized in Figure \ref{fig:declination}, which is from Warrington's article \cite{WarringtonDeclinationELJ}:

\begin{figure}[h]
    \centering
    \includegraphics[width = 4in]{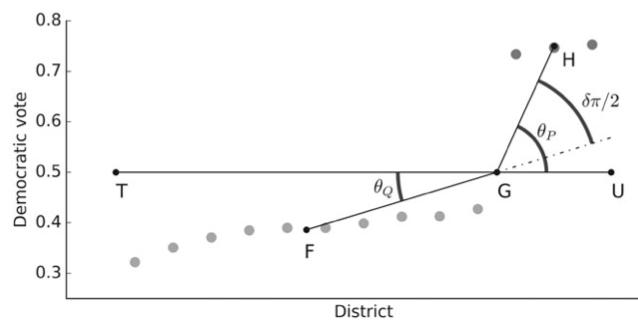}
    \caption{Image from \cite{WarringtonDeclinationELJ}, using data from North Carolina's 2014 Congressional election}
    \label{fig:declination}
\end{figure}

Then the declination $\delta$ is calculated as:
\begin{equation*}
\delta = \frac{2}{\pi}\left(\theta_P-\theta_Q\right)
\end{equation*}
Note that the scaling by $\frac{2}{\pi}$ guarantees that $-1 \leq \delta \leq 1$.  A large positive $\delta$ indicates that if the $V_i$'s correspond to party $B$'s vote shares, then party $B$'s ``average'' losing district\footnote{represented by point $F$} has a vote share near 50\%, and party $B$'s ``average'' winning district\footnote{represented by point $H$} has a very high vote share (which translates into party $B$ being ``packed and cracked'').  Thus, a large positive $\delta$ is intended to indicate a map that is beneficial for party $A$.

\subsection{Geography and Election Outcome (GEO) Metric}\label{subsec:geo_definition}

The GEO metric is the only metric we are aware of that can be computed by hand, and which also requires not just partisan data but also map data.  It was proposed by Campisi, Ratliff, Somersille, and Veomett \cite{geo_paper}.  The GEO metric gives a score to each party which is a non-negative integer.  

The calculation of party $X$'s GEO score involves swaps of vote shares between neighboring districts in a way that keeps any current wins for party $X$, but allows some lost districts to become competitive (have a 50/50 vote share split), without changing the vote shares so much as to be statistically unreasonable.  The value of party $X$'s GEO score is a \emph{count}: it is the number of additional districts that can be made competitive for party $X$ through this vote share swapping process (see \cite{geo_paper} for more details).  Thus, if party $X$'s GEO score is very large relative to the total number of districts, that party could potentially have many additional competitive districts with reasonable changes to the map (indicating the map is not favorable to party $X$).  Conversely, if party $X$'s GEO score is close to 0, there is very little that can be done to improve party $X$'s chance of being competitive in additional districts (indicating the map is optimized for party $X$).

We note here that, for the Gameability Study and Extreme Metric Values versus Extreme Maps Study (described in Sections \ref{subsec:Gameability} and \ref{subsec:MetricValsExtremeMaps}), we needed each metric to have a bounded range of values centered at 0 in order to compare ``apples to apples'' with respect to the \emph{values} of the metrics.   Thus, for those two studies, we used a ``GEO ratio'', which is the difference of the two parties' GEO scores, divided by the total number of districts in the state:
\begin{equation*}
    \text{GEO ratio} = \frac{\text{GEO(B) - GEO(A)}}{\text{number of districts}}
\end{equation*}
Since the GEO score for each party is an integer between 0 and the total number of districts, this GEO ratio is guaranteed to be between -1 and 1.  The creators of the GEO metric do not recommend using the GEO ratio itself as a metric, and promote that the two GEO scores (along with the number of districts) be considered themselves when assessing a map \cite{geo_paper}.  Nevertheless, we included this GEO ratio so that we could include a variant of the GEO metric that gave a single score which was on a comparable scale to the other metrics.  From the discussion above, we can see that a large positive GEO ratio indicates a map beneficial to party $A$.

\section{Hill Climb Implementation: Short Bursts}
\label{sec:shortbursts}

 There are many methods available to use in conjunction with a Markov Chain process to find plans that maximize a given metric. For example, one could accept a subsequent plan only when its metric value is greater than the last plan's value. However there are many drawbacks to this method. It can be slow and computationally costly. Also, if the current position is a local maximum, one move in the chain to a higher valued plan may not be possible. Another commonly used method is a biased random walk. In a biased random walk, when moving to the next plan in the chain, a bias is imposed so that plans with higher values of the desired attribute are more likely to be chosen.

Cannon and her collaborators developed the Short Burst method as an alternative to a biased random walk \cite{cannonShortBursts}.  Specifically, it starts with a seed map and runs a Markov Chain Monte Carlo process for  10 steps (for example), to create 10 new redistricting maps.  Among those maps, the map with the most extreme metric value is selected to be the seed of the next ``short burst" of size 10; if two maps have the same extreme metric value, then the most recently visited map with that extreme metric value is retained as the next seed.  This process is repeated thousands of times.\footnote{ Note that the Short Burst method can be thought of as a non-deterministic version of beam search on the metagraph of redistricting maps (edges are between two maps that can be obtained from each other through a single ReCom move in GerryChain), using depth-first search instead of breadth-first, and in which a random selection of maps near to the current map are explored.}

In \cite{cannonShortBursts},
they then used the Short Burst method in a study of Louisiana’s state House of Representatives plans. Their goal was, in the context of a Voting Rights Act study, to attempt to find the maximum number of majority-minority districts a single districting plan could sustain.  They found that, in the context of maximizing majority-minority districts, Short Bursts were more successful in finding extreme maps than biased random walks.

 As stated in Section \ref{sec:intro}, we use the Short Burst method in order to create our ``extreme'' maps. i.e. the maps with the most Democratic-majority (Republican-majority) districts.  Like Cannon et al, we use the freely available GerryChain library to implement the Markov Chain Monte Carlo (MCMC) process \cite{GerryChain}.  Based on the results in \cite{cannonShortBursts}, one can see that bursts of length 10 were able to find the most majority-minority districts, so we also used bursts of length 10 for our studies.  In their study comparing short bursts to biased random walks (Section 5.2.3 of \cite{cannonShortBursts}), the metric values drastically slowed in increase after about 5,000 steps.  For this reason, and also because computing constraints, we limited our runs to 5,000 steps, completing 10 total runs (5,000 steps each) for each map.  Given the fact that metric values continued to increase  only slightly after 5,000 steps in the experiments from \cite{cannonShortBursts}, we expect that the number of districts won that we find in our experiments will be quite close to, but perhaps not equal to, the maximum possible number of districts won for that party and map.  We used and heavily edited the code that Cannon's team shared \cite{SBGitHub}; our code is also available on GitHub at \textanon{\url{https://github.com/stem-redistricting/ShortBursts} \cite{SB_our_repo}}{(anonymized for review)}.  

\section{Methods}

\subsection{States Chosen}

Our choice of which states to evaluate was based on two main factors.  Firstly, we needed to use states for which there was reliable data on which we could run GerryChain's ReCom MCMC process.  The Metric Geometry Gerrymandering group \cite{MGGGstates} has an excellent GitHub page with this data for some states, and we chose our states from among those that they had available.  Our studies are not intended to focus on any particular map, but rather to focus on the metrics themselves.  Thus, we are unconcerned that these states' maps and partisan data are 6-10 years old.

Secondly, in order to assess how well metrics work in general, we wanted to choose states with various types of political geographies on which to test the metrics.  Some states are fairly homogeneous, with most parts of the state having the same partisan breakdown.  Other states have pockets that more heavily lean Republican and other pockets that more heavily lean Democratic; such states may have an overall clear partisan lean (Democratic or Republican) or they may be more ``purple.'' We chose at least one state from most of these combinations, as outlined in Table \ref{tab:state_types}.

\begin{table}
\begin{tabular}{|c|c|c|c|c|c|}\hline
State & leans Dem & leans Rep & no lean & Homogenous & Non-homogeneous \\ \hline\hline
Massachusetts & X& & &X & \\ \hline
Oklahoma & &X & & X& \\ \hline
Pennsylvania & & &X & & X\\ \hline
Michigan & & & X & & X\\ \hline
Texas & & X& & & X\\ \hline
Oregon & X& & & & X\\ \hline
\end{tabular}
\caption{Types of States in Analysis \label{tab:state_types}}
\end{table}

\begin{figure}[h]
\includegraphics[width = 2in]{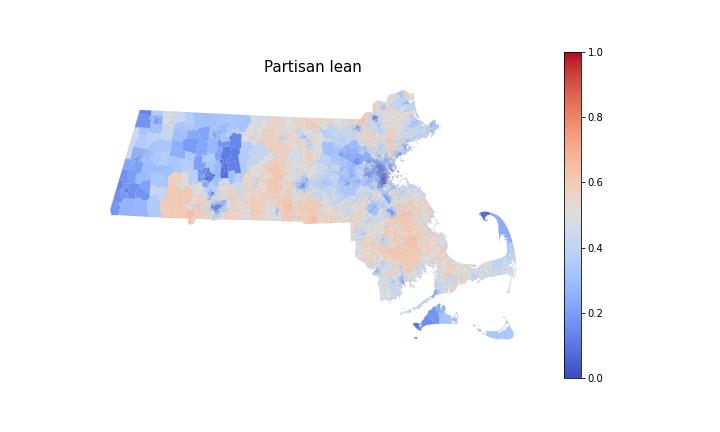} 
\includegraphics[width = 2in]{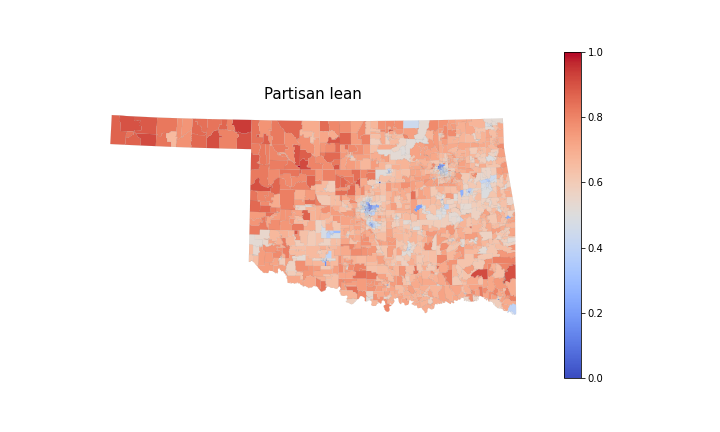}
\includegraphics[width = 2in]{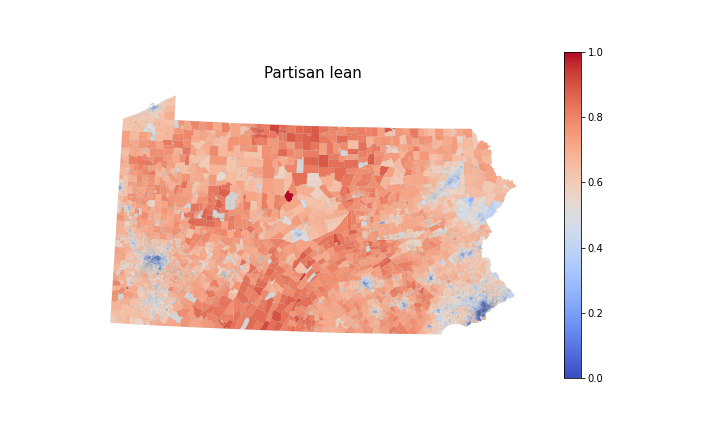} 
\includegraphics[width = 2in]{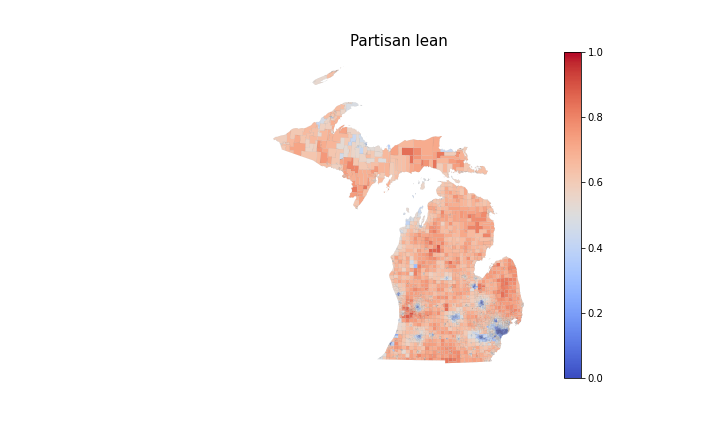}
\includegraphics[width = 2in]{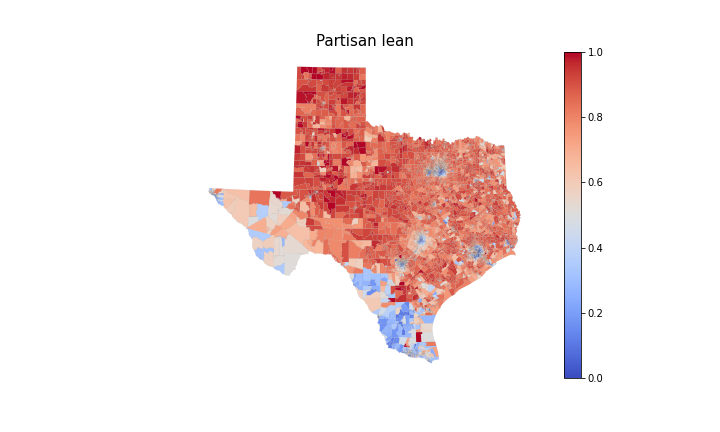} 
\includegraphics[width = 2in]{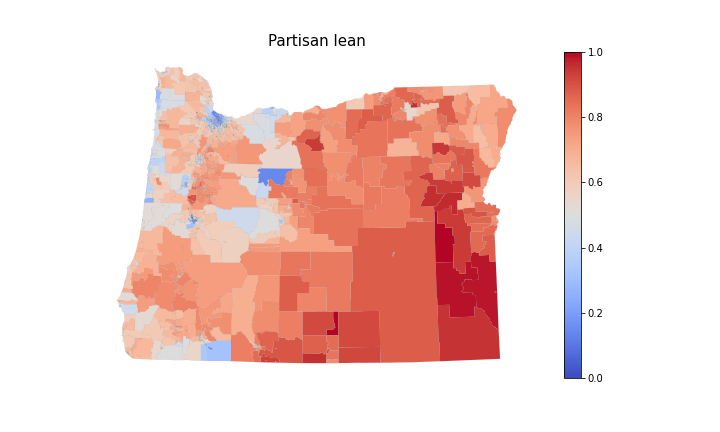}
\caption{Chloropleths of the states in our analysis, using the election results from Table \ref{tab:burst_table}.  They are, in clockwise order from the top left, Massachusetts, Oklahoma, Michigan, Oregon, Texas, and Pennsylvania.}
\label{image:choropleths}
\end{figure}

The reader may agree or disagree with these categorizations, but we believe that it is clear that these states form a varied array of political geographies, and thus are reasonable states on which to conduct a metric assessment.  We include in Figure \ref{image:choropleths} a choropleth of each state (using the election data that we used for our analysis) for the reader's reference.  

\subsection{Short Burst Runs}\label{sec:sb_runs_methods}

For each state analyzed, we did two kinds of Short Burst runs.  The first kind of run for our Gameability Study (described in Section \ref{subsec:Gameability}) had a metric $m$ restricted to be within a fixed set of bounds.  Those runs \emph{only} produced maps with metric values $m$(map) satisfying
\begin{equation}\label{eq:metric_bounds}
0.16\text{inf}(m) \leq m\left(\text{map}\right) \leq 0.16\text{sup}(m)\footnote{Here, $\text{inf}(m)$ ($\text{sup}(m)$) is the infimum (supremum) of all possible values of the metric $m$}
\end{equation}
and retained the map with the most districts won by a particular political party.  Those runs were searching for the largest number of wins that a particular party could have in a state, with the fixed partisan data, \emph{and} with the metric in consideration $m$ considering the map to be reasonably ``fair.'' As stated in Section \ref{subsec:Gameability}, the reason that we chose the bounds for $m$ to be 16\% of the infimal/supremal values of $m$ is solely the fact that those are the only bounds that have been proposed as potential bounds for a gerrymandering metric; they were proposed as bounds for the Efficiency Gap in \cite{PartisanGerrymanderingEfficiencyGap}\footnote{We note that the infimal/supremal value of the Efficiency Gap is -1/2 and 1/2 respectively; the bounds stated in \cite{PartisanGerrymanderingEfficiencyGap} were $\pm 8\%$}.  In order to have an ``apples to apples'' comparison between the metrics, we chose to keep the bounds within the same percentage of the infimal/supremal values for each metric.

The second kind of run for our Extreme Metric Values versus Extreme Maps Study (described in Section \ref{subsec:MetricValsExtremeMaps}) had no restrictions on the metrics, and simply retained the map with the most districts won by a particular political party in each burst.  These runs were simply searching for the largest number of wins that a political party could have in a state, with the fixed partisan data.  

We ran each of these two kinds of short bursts on three maps for each state: the lower house districting map, the upper house districting map, and the congressional districting map.  For congressional maps, we allowed up to 5\% deviation from ideal population in each district.  For the lower and upper house maps, we allowed up to 12\% deviation from ideal population.   As noted in \cite{RecomMGGG}, ``Excessively tight requirements for population balance can spike the rejection rate of the Markov chain and impede its efficiency or even disconnect the search space entirely.'' 
 In \cite{RecomMGGG}, they use a deviation of 5\% for their congressional map of Arkansas, which is why we chose 5\% for our congressional maps here.  State legislatures have significantly looser restrictions on population deviation; Brown v Thomson \cite{BrownvThomson} states that state legislative districts can deviate by up to 10\% in population, and many states do have population deviations of that magnitude.  
 We found that the Markov Chain process for many state legislative maps quickly became stuck when allowing a deviation up to 10\%, but loosening the restriction to 12\% enabled the creation of many additional sample maps.

 As described in \cite{RecomMGGG}, the ReCom process in GerryChain naturally samples plans with more compact districts with higher probability.  Outside of this built-in preference for compact maps, we did not impose additional compactness constraints.

We chose a statewide election data as a proxy for party preference for each state.  Statewide data is important to use here (rather than, say, district election results) because it eliminates confounding variables such as incumbency or uncontested races.  Since our goal was to evaluate the metrics, rather than the maps themselves, the choice of which statewide election was used is unimportant; our choices can be found in Table \ref{tab:burst_table}.

For each state, each party, each districting map, and each metric bound (including no metric bound) we ran 10 short burst samples, each of size 5,000, and each having burst length of size 10.  See Section \ref{sec:shortbursts} for a justification of these parameters.  This produced potentially 45,000 different maps\footnote{Since the last maximal map of each burst is the first map in the next burst, but the seed at the beginning of each sample is the same.}
for all of the state, party, map, and metric combinations. 

There are two exceptions where we could not run bursts of the length described above.  The first exception is when the MCMC process got stuck and could not continue producing new maps; this only happened with Massachusetts' lower legislative body.  This likely happened because the process reached a map from which it could not produce another new map satisfying the population, contiguity,  and metric constraints\footnote{The Massachusetts state house has a large number of districts: 160.  This makes satisfying all of the constraints more difficult than when a state has a smaller number of districts.}.  In an effort to keep the population deviation lower (no more than 12\% for lower and upper legislative houses), but include as many results as possible, we did keep the maps created with metric constraints that we were able to produce for the MA lower house.

The second exception is for the states where we could not, in an unbiased GerryChain run of size 5,000, find a single map with metric value within ``reasonable`` bounds.  Specifically, the Declination value is never defined on Massachusetts' congressional districting map, since the unbiased run of size 5,000 did not find a single map where a single Republican seat was won.  (The details for finding seed maps are below in Section \ref{sec:seeds}).  And the unbiased runs of size 5,000 for both the Massachusetts and Oklahoma congressional maps did not produce a map with Efficiency Gap value between -0.08 and 0.08.

Table \ref{tab:burst_table} includes the details of which runs were completed for each state/map.  For the Massachusetts's lower house runs with metrics restricted, we list the number of runs of size 5,000 that we were able to produce. e.g. for MA, Lower House,  Efficiency Gap Restricted runs we were did five runs maximizing democratic seats (5D) and six runs maximizing republican seats (6R).

\begin{table}[h]
\begin{tabular}{|p{2.3cm}||p{1.7cm}|p{1.5cm}|p{1.6cm}|p{1.6cm}|p{1.6cm}|p{1.5cm}|}\hline
State & MA & OK & PA & MI & TX & OR \\ \hline
Election Used & 2018  \newline US Senate & 2018\newline Governor & 2016 \newline US Senate & 2016 \newline Presidential &  2014\newline US Senate & 2018\newline Governor \\ \hline
No metric \newline restriction &C \newline L \newline U &C \newline L\newline U & C \newline L \newline U& C \newline L \newline U& C \newline L \newline U& C \newline L \newline U\\ \hline
Efficiency Gap Restricted & $\quad$ \newline L: 5D, 6R \newline U& $\quad$ \newline L \newline U & C \newline L \newline U& C \newline L \newline U& C\newline L \newline U & C\newline L \newline U\\ \hline
Declination Restricted & $\quad$ \newline L: 3D, 2R \newline U&C \newline L \newline U & C \newline L \newline U& C \newline L \newline U& C\newline L \newline U & C\newline L \newline U\\ \hline
GEO ratio \newline Restricted &C \newline L: 10D, 4R \newline U&C \newline L  \newline U & C \newline L \newline U& C \newline L \newline U& C \newline L \newline U& C\newline L \newline U\\ \hline
Mean-Median Restricted & C \newline L: 5D, 6R \newline U&C \newline L \newline U & C \newline L \newline U& C \newline  L \newline U& C \newline L \newline U& C\newline  \newline U\\ \hline
\end{tabular}
\caption{Types of Short Bursts for redistricting maps, each run for both the Democratic and Republican party. C is congressional map, L is lower state legislative body map, U is upper state legislative body map.  Unless otherwise noted, 10 total runs of size 5,000 were run for each party.  \label{tab:burst_table}}
\end{table}

Finally, we remind the reader that, as stated in Section \ref{subsec:geo_definition}, we used a GEO metric ratio in these Short Burst runs.  Again, this was in an effort to compare ``apples to apples'' when limiting the metric values to be within certain ranges.

\subsection{Construction of Seeds}\label{sec:seeds}

For each state, we needed a seed map that we could use for all of our downstream analyses.  Many of our analyses required that a metric value $m$ be within bounds from Equation \eqref{eq:metric_bounds} for each map in the burst.  This meant that the seed value had to also satisfy those bounds.  Thus, for each state and each map for that state, we needed to create a seed map that had all of those metric bounds satisfied, for each metric.  (This way, each short burst run started from the same seed.)

In order to create those seeds, we simply ran a GerryChain ReCom chain with no bias and no restrictions (aside from contiguity constraints, loose compactness constraints, and the population deviation of 5\% for congressional maps and 12\% for lower and upper state house maps) until a map satisfying all restrictions was found.  If, for a particular metric $m$, no map was found in a chain of length 5,000 satisfying Equation \eqref{eq:metric_bounds}, then we did not run a short burst chain restricting that metric on that map at all.  This was the case only for the congressional maps when $m = \text{Efficiency Gap}$ for the for MA and OK, and for $m = \text{Declination}$ for MA (as noted in section \ref{sec:sb_runs_methods}).

\section{Results and Discussion}\label{sec:results_discussion}

\subsection{Sample output: Texas congressional}\label{subsec:results_sample}
We will first look in detail at the output from our experiments for the Texas congressional map, as an example. The results of the Gameability Study (described in Section \ref{subsec:Gameability}) are shown in Figure \ref{fig:results_TX_metric_compare}. This demonstrates that it is possible to create districting plans where the Democrats win 13 of the 36 possible seats and the Efficiency Gap, Declination, and GEO ratio stay within the bounds of Equation~{\eqref{eq:metric_bounds}. Similarly, we can create a districting plan where the Democrats win 16 seats and the Mean-Median Difference stays within in the acceptable range. Note that these may be four distinct plans\footnote{However, for some of our most extreme results, we find a single map with all metrics within reasonable bounds.  See Section \ref{subsec:results_metrics_restricted} for details.}. The maximum number of districts won by Republicans where we stay within the bounds of Equation~{\eqref{eq:metric_bounds} are 31 for the Declination and GEO ratio, 30 for the Efficiency Gap, and 34 for the Mean-Median Difference. In both of these instances the Mean-Median Difference allows more extreme maps than the other metrics while staying within bounds. 

\begin{figure}[h]
    \centering
    \includegraphics[width=0.45\linewidth]{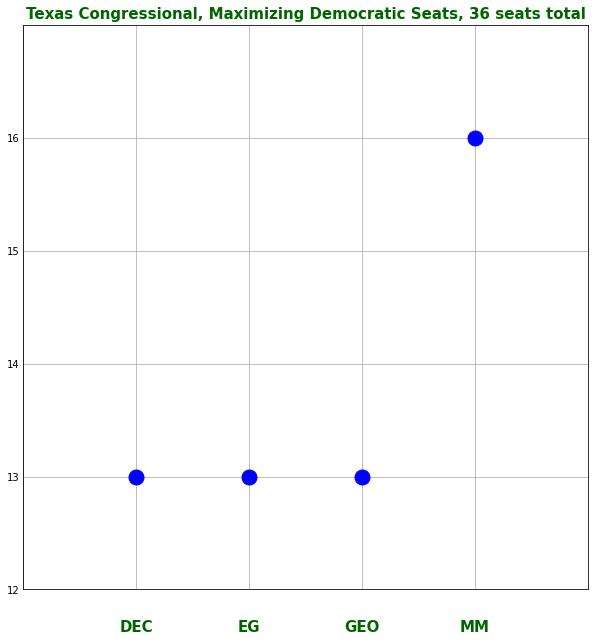}
    \includegraphics[width=0.45\linewidth]{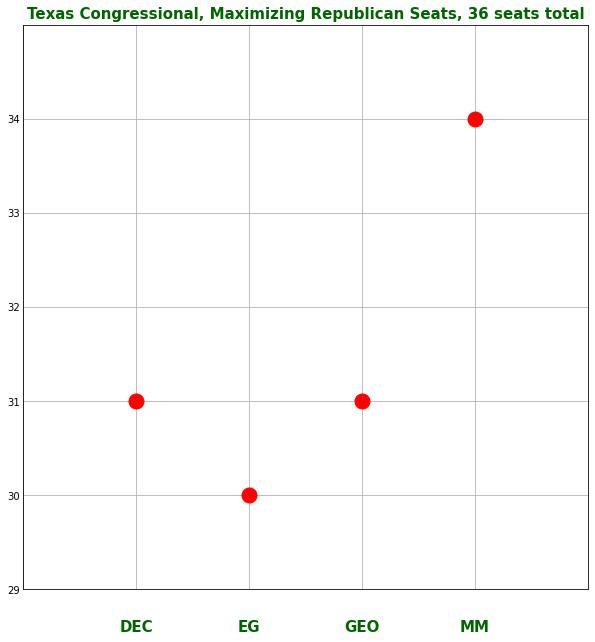}
    \caption{Results of the Gameability Study for TX congressional map: Maximum number of seats won by each party while staying within acceptable bounds (\( \pm0.08 \) for Efficiency Gap and Mean-Median Difference, \( \pm0.16 \) for Declination and GEO ratio)}

    \label{fig:results_TX_metric_compare}
\end{figure}

Recall that for the Extreme Metric Values versus Extreme Maps Study (described in Section \ref{subsec:MetricValsExtremeMaps}), we are maximizing the number of seats won by a party without any restrictions based on the metrics. The range of the metric values when maximizing the number of seats won by Democrats is shown in Figure~\ref{fig:results_TX_none_declination}. This shows that the seed for this run began with the Democrats winning 7 of the possible 36 seats and that it is possible to push this to 16 seats won by the Democrats. Notice that when the Democrats win 7, 14, 15, or 16 seats, the Declination is entirely outside of the acceptable range, but for the other values 8--13 there are plans where the Declination is in the acceptable range. The Efficiency Gap and GEO ratio follow a similar pattern to the Declination, with the exception that some plans with 7 Democratic seats fall in the acceptable range. 

However, notice that for the Mean-Median Difference, \textit{every} plan in the ensemble falls within the acceptable range! In addition, there is significant overlap in the values that the Mean-Median Difference takes, regardless of the number of districts won. This means that there are plans with a Mean-Median Difference of approximately \( -0.003 \) and the Democrats win anywhere between 8 and 16 seats. As we shall see, this is typical behavior for the Mean-Median difference.

\begin{figure}[h]
    \centering
    \includegraphics[width=0.45\linewidth]{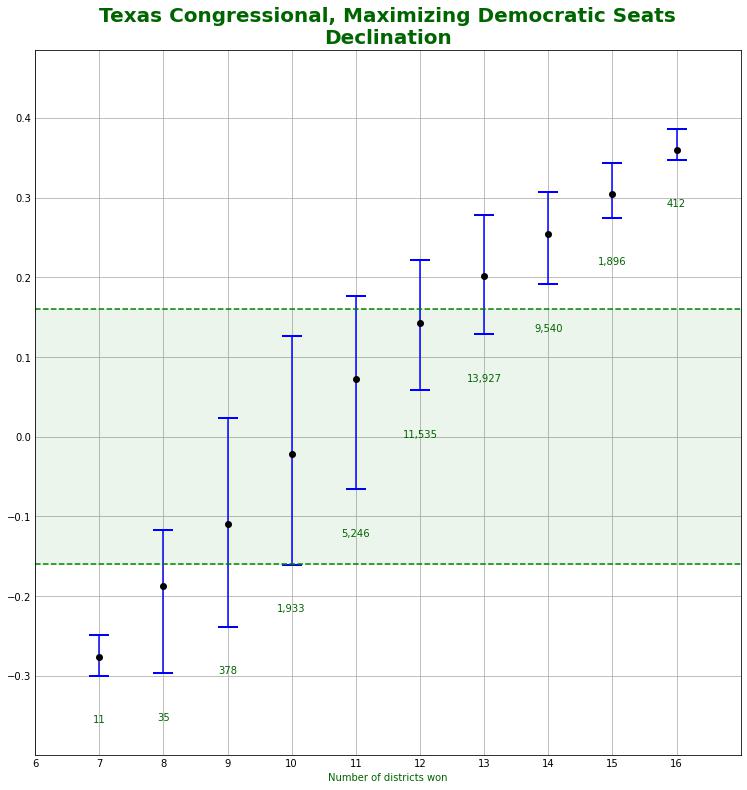}
    \includegraphics[width=0.45\linewidth]{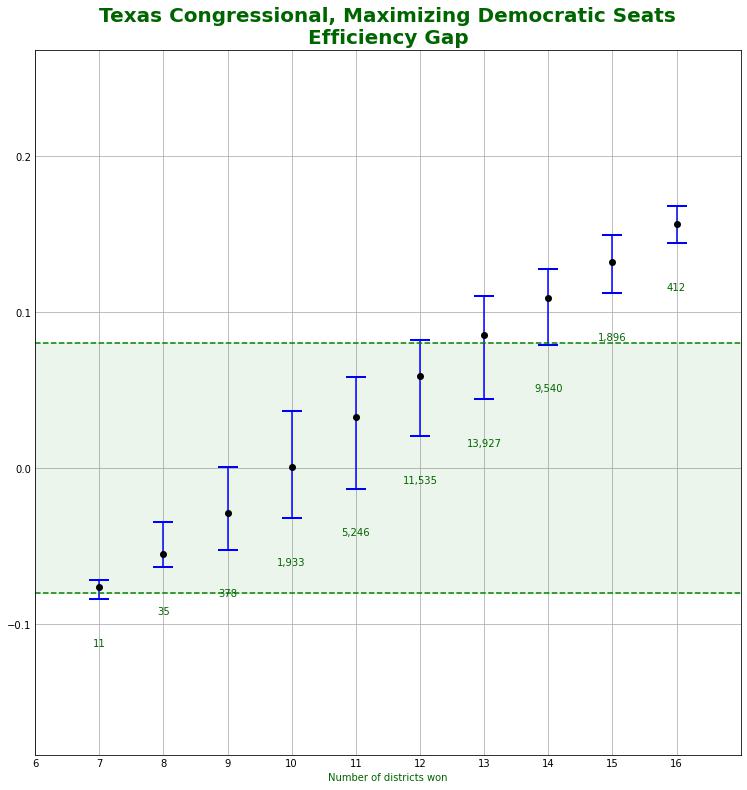}
    \includegraphics[width=0.45\linewidth]{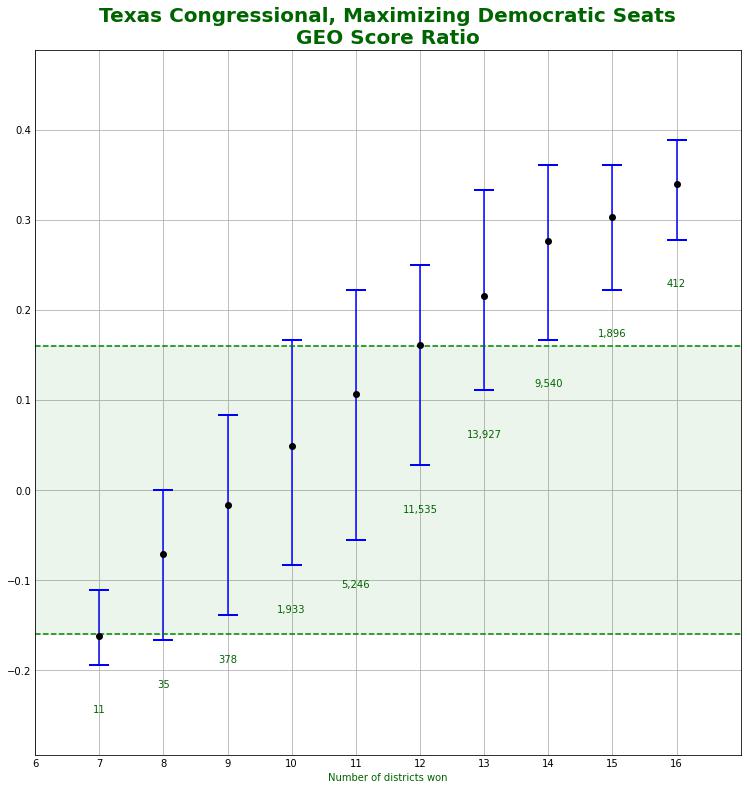}
    \includegraphics[width=0.45\linewidth]{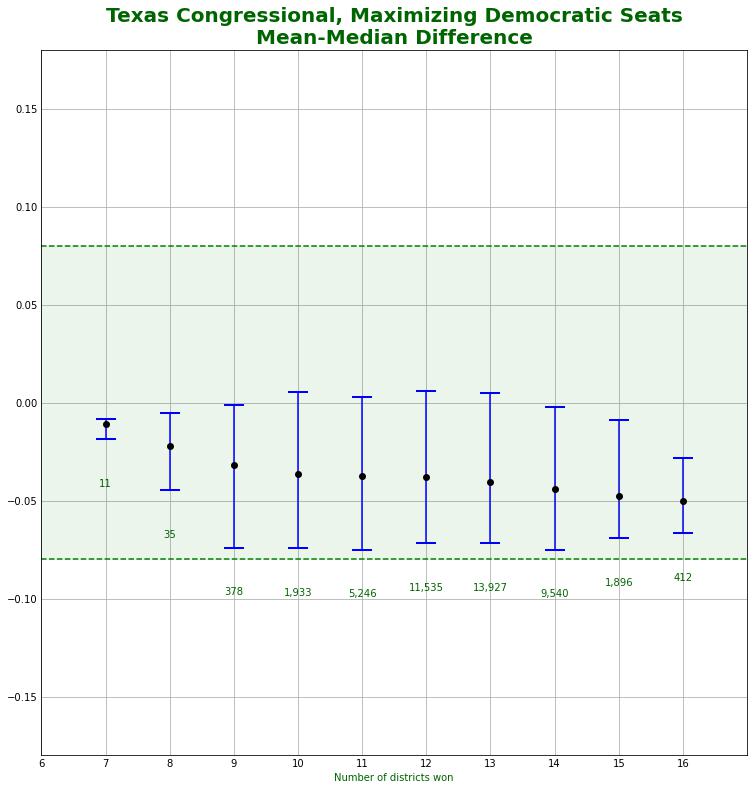}
    \caption{Results of Extreme Metric Values versus Extreme Maps Study for TX congressional map: Range of scores for each metric when maximizing Democratic seats won. The green band shows the acceptable range. The value under each whisker indicates the number of plans with that number of districts won. }

    \label{fig:results_TX_none_declination}
\end{figure}

We have similar plots when maximizing the number of Republican seats won and for the other 17 maps given in Appendix \ref{appendix:figures}.

\subsection{Results: Gameability Study}\label{subsec:results_metrics_restricted}

The results of the Gameability Study described in Section \ref{subsec:Gameability} can be found in Table \ref{table:maxvalues}.  For each party and each metric, Table \ref{table:maxvalues} lists the \emph{maximum number of seats won} by that party (among all maps constructed with the Short Bursts described in Section \ref{sec:sb_runs_methods}}) when restricting the metric value as in Equation \eqref{eq:metric_bounds}.  This is compared to the maximum number of seats won in the short bursts where no metric is restricted (as described in Section \ref{subsec:MetricValsExtremeMaps}).

\begin{table}[h]
\caption{Results from Gameability Study}
\label{table:maxvalues}
\begin{tabular}{|l |l ||l |l |l |l |l ||l |l |l |l |l |l |l |}
\hline
& Num   &  Dem Max & & & & &  Rep Max & & & & \\ \hline
& Seats &  None    & Dec       & EG        & GEO       & MM        &  None    & Dec       & EG        & GEO       & MM        \\ \hline
MA cong  & 9   &   9   &     &     & 9   & 9   &   1   &     &     & 1   & 1   \\ \hline
MA lower & 160 &   148 & 142 & 136 & 134 & 148 &   41  & 38  & 41  & 42  & 41  \\ \hline
MA upper & 40  &   40  & 40  & 35  & 36  & 40  &  10  & 8   & 10  & 10  & 10  \\ \hline \hline
MI cong  & 13  &   10  & 8   & 8   & 8   & 10  &   12  & 8   & 8   & 10  & 12  \\ \hline
MI lower & 110 &   59  & 60  & 60  & 60  & 59  &   78  & 62  & 65  & 69  & 78  \\ \hline
MI upper & 38  &   24  & 22  & 22  & 21  & 24  &   29  & 22  & 22  & 25  & 29  \\ \hline \hline
OK cong  & 5   &   2   & 1   &     & 1   & 2   &   5   & 4   &     & 5   & 5   \\ \hline
OK lower & 101 &   41  & 40  & 40  & 39  & 41  &   83  & 73  & 71  & 76  & 83  \\ \hline
OK upper & 48  &   21  & 20  & 21  & 20  & 21  &   43  & 37  & 35  & 39  & 43  \\ \hline \hline
OR cong  & 5   &   4   & 3   & 3   & 4   & 4   &   4   & 2   & 2   & 3   & 4   \\ \hline
OR lower & 60  &   40  & 37  & 38  & 36  & 40  &   39  & 32  & 33  & 36  & 39  \\ \hline
OR upper & 30  &   22  & 19  & 19  & 19  & 22  &   22  & 16  & 16  & 18  & 22  \\ \hline \hline
PA cong  & 18  &   11  & 10  & 9   & 9   & 10  &   15  & 11  & 11  & 12  & 15  \\ \hline
PA lower & 203 &   98  & 98  & 98  & 98  & 98  &   145 & 120 & 123 & 130 & 145 \\ \hline
PA upper & 50  &   27  & 27  & 27  & 27  & 27  &   41  & 30  & 31  & 33  & 41  \\ \hline \hline
TX cong  & 36  &   16  & 13  & 13  & 13  & 16  &   34  & 31  & 30  & 31  & 34  \\ \hline
TX lower & 150 &   58  & 58  & 57  & 59  & 58  &   124 & 109 & 123 & 113 & 124 \\ \hline
TX upper & 31  &   14  & 11  & 12  & 11  & 14  &   29  & 27  & 26  & 27  & 29  \\ \hline
\end{tabular}
\end{table}

 One clear takeaway message is that, at least using the bounds from Equation \eqref{eq:metric_bounds} that were suggested by \cite{PartisanGerrymanderingEfficiencyGap}, every metric can be gamed. 
 This is assuming that the definition of ``gaming the metric'' implies that there is a legislative body where the maximum number of seats won by a party (indicated in the `None' column) can be obtained by a districting plan that remains within the acceptable bounds from Equation \eqref{eq:metric_bounds}.  
 
 For example, if we consider the Oklahoma lower chamber, the maximum number of Democratic districts found with no metric restrictions is 41, and we can find a districting plan where the Mean-Median Difference stays within the acceptable bounds  and Democrats win 41 seats.  In fact, the Pennsylvania lower and upper chambers are examples where the maximum Democratic outcome is obtainable by \textit{every} metric. This suggests that, even if one were to utilize more than a single metric, this would not limit an extreme number of districts won for all states and maps.  Further, upon inspecting by hand the results of our bursts with no metric restricted (discussed further in Section \ref{subsec:results_no_restriction}), we found some maps attaining the maximum number of Democratic-won districts and \emph{all metrics} were within the acceptable bounds from Equation \eqref{eq:metric_bounds}. We found such maps for both of the Pennsylvania lower and upper chambers.

We also had three cases (Massachusetts lower chamber, Republican, Michigan lower chamber Democratic, Texas lower Democratic) where the Short Burst runs on a constrained metric obtained \textit{one more} seat than the Short Burst runs with no metric restrictions. (See Table \ref{table:maxvalues}) While somewhat unexpected, this was not entirely surprising.  Given that our Short Bursts were 5,000 steps in length, as explained in Section
\ref{sec:shortbursts}, we felt that 5,000 steps was sufficient to get near to, but perhaps not reach, the maximum possible number of districts won for each party. Therefore, it is possible that the unrestrained run would have found an additional seat under different parameters, starting from a different seed, etc\footnote{It is known that due to the size of the sample space of permissible maps that none of these methods are sampling the entire space. However, much can be said without mixing time results; for example, see \cite{chikina2017assessing}.}.

The observations in this section indicate that we should \textit{never} rely upon a single measurement to fully evaluate a districting plan. The metric may give a value that is considered acceptable, yet the plan gives the most extreme partisan outcome possible.  Indeed, keeping the metric within ``reasonable bounds'' may even skew the space of maps towards more extreme numbers of districts won, as we shall see in examples like Figure \ref{fig:results_TXcong_ensemble} in Section \ref{subsec:ensembles}.  This strongly indicates that one should use an ensemble analysis or other tool to determine if a specific plan is an outlier. We discuss this further in Section \ref{subsec:ensembles}. 

While each metric had at least some legislative body for which it considered extreme maps acceptable, for the most part, restricting the metric value limited the number of seats won to a number less than the maximum number of seats won when no metric was restricted.  The one significant exception is the Mean-Median Difference, which allowed the number of seats won to reach the maximum value in all but one map: the PA congressional map. 

We also note that the number of seats won by the Republican party is generally restricted farther from the max value (in the ``None'' column) by each metric than the number of seats won by the Democratic party is restricted from its max value.  We suspect that this has to do with the typical nature of political geography in the United States: in general, the rural areas tend to be majority Republican (but not an overwhelming majority), while urban areas tend to be majority Democratic (with an overwhelming majority).  Of course, these are generalizations and each region has its nuances.  But it is noteworthy that Massachusetts is the one state where metrics allowed the number of Republican won districts to reach the maximum seen when no metric was restricted (Mean-Median Difference aside).  Massachusetts is well-known to have weaker Republican support in rural areas than other states \cite{MA_DemDomination}; this can be seen visually in the choropleth map in Figure \ref{image:choropleths}.

We conclude that, while each metric had \emph{some} state/legislative body on which it could be gamed, the Mean-Median Difference seems to be gameable on nearly \emph{every} map.  

We now address the following question: why not just restrict the metrics further?  Perhaps the actual bounds for the metric are the real issue, rather than the metrics themselves; why not change the bounds?  As we shall see in the next section (Section \ref{subsec:results_no_restriction}), this would indeed limit the gameability of all of the proposed metrics . . . except for the Mean-Median difference.

\subsection{Results: Extreme Metric Values versus Extreme Maps}\label{subsec:results_no_restriction}

The results of the extreme maps study discussed in Section \ref{subsec:MetricValsExtremeMaps} can be seen in the images in Appendix \ref{appendix:figures}; most are similar to the sample maps in Figure \ref{fig:results_TX_none_declination}.  It is quite clear from those results that, for each metric except the Mean-Median Difference, more extreme maps are associated with more extreme values.  Thus, if we were to further restrict the bounds of each metric, this would eliminate some extreme maps.

However we see severe issues with taking this approach.  It is worth noting that restricting a metric to be within certain bounds likely raises legal issues, which we leave to the legal scholars to debate.  Setting the legal issues aside, the results we see themselves don't allow for a clear way to introduce metric bounds.  Firstly, these metrics in this restricted study are all symmetric: one party's score is the negative of the other party's score.  Should we have restrictions that are also symmetric?  That is, if we say the Declination cannot be more than 0.05, must we also say it cannot be less than -0.05, thereby treating the parties symmetrically?  One need only look at the results from Michigan's state house in Figure \ref{fig:results_MIlower_short_bursts_all} to see that this could easily be seen as beneficial to the Democratic party.  The fact that these metric values are not truly centered at 0 (which will be seen in the ensemble results in Section \ref{subsec:ensembles}, such as in Figures \ref{fig:results_TXcong_ensemble}, \ref{fig:results_PAlower_ensemble} and \ref{fig:results_PAcong_ensemble}) are what gives rise to this first issue.

Secondly, different states themselves have different ranges of metric values.  This can be seen by, for instance, comparing the Efficiency Gap values between Figures \ref{fig:results_TXcong_ensemble} and \ref{fig:results_PAlower_ensemble}.  For Texas's congressional map, the Efficiency Gap takes on both positive and negative values, with a maximum near 0.075 and minimum near -0.1.  However Pennsylvania's house map (and Pennsylvania's congressional map in Figure \ref{fig:results_PAcong_ensemble}) have \emph{no maps} with positive values!  Thus: how would one find a range of values for a metric which is reasonable for a particular state/map?  One would need to create an ensemble of potential redistricting maps in order to find that appropriate range, since each state's political geography is so different.  This completely defeats the purpose of an easy-to-calculate metric.

The Mean-Median Difference quite notably does not seem to have much relationship between extreme maps and extreme values of the metric.  Note that the authors of \cite{ImplementingPartisanSymmetry} gave evidence of some of the flaws of the Mean-Median Difference in an analysis of Utah, Texas, and North Carolina.  Our study (which considers 18 maps from states with a wide array of political geographies) gives significantly more evidence that the Mean-Median Difference does not detect extreme maps for nearly any state, any legislative body.  Thus, restricting the Mean-Median Difference to be in a smaller interval around 0 will do nearly nothing to avoid the most extreme of maps, as can be seen from Figure \ref{fig:results_TX_none_declination} and the many whisker plots in Appendix \ref{appendix:figures}.

\subsection{Ensembles: no metric much better than districts won}\label{subsec:ensembles}

From the results discussed in Section \ref{subsec:results_no_restriction}, we concluded that metrics whose value for Democrats is the negative of the value for Republicans (such as the Mean-Median Difference, Efficiency Gap and Declination) are uninterpretable on their own.  Each state's political geography determines a range of ``typical'' values for that metric.  Those values are not symmetric about 0, and they differ from state to state.  Thus, one cannot fix a single ``reasonable range'' of that metric's values that can be applied universally.  One must construct an ensemble of potential redistricting maps in order to determine such a reasonable range (again, this defeats the purpose of a single metric value which is easy to calculate).

Given this conclusion, we were curious as to how each of these metrics behaved within an ensemble.  For each of our 18 state/legislative body maps, we created a neutral ensemble of 20,000 maps using the GerryChain library implementing a ReCom Markov Chain.  We chose 20,000 steps because it could be completed in under two hours for each map, and the authors of \cite{RecomMGGG} stated that ``tens of thousands of recombination steps give stable results on practical-scale problems whether we
work with the roughly 9000 precincts of Pennsylvania or the roughly 100,000 census blocks in our Virginia
experiment.''  For partisan data, we used the same statewide election results for each state that was listed in Table \ref{tab:burst_table}.  We looked at the histograms of each metric (including number of Democratic seats won), and where the metric value of the enacted map lay within that histogram.  Sample histograms for Texas's congressional map are in Figure \ref{fig:results_TXcong_ensemble}.  Note that, to consistently have that each metric's larger values correspond to a map that is better for Democrats (and smaller values correspond to a map that is better for Republicans), the Democratic GEO score is negated in its histogram\footnote{Note that we use the GEO scores here (rather than the GEO ratio which we used previously) because the creators of the GEO metric encouraged usage of the scores directly (not the GEO ratio).  Since we no longer need to have the metric produce a number which is is in a bounded interval centered at 0 (in order for the apples-to-apples comparison with the other metrics) we directly use the Democratic GEO score and Republican GEO score here.}. 

\begin{figure}[h]
    \centering
    \includegraphics[width=0.3\linewidth]{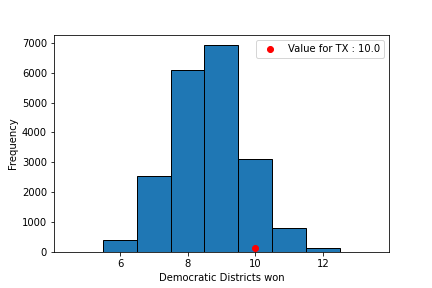}
    \includegraphics[width=0.3\linewidth]{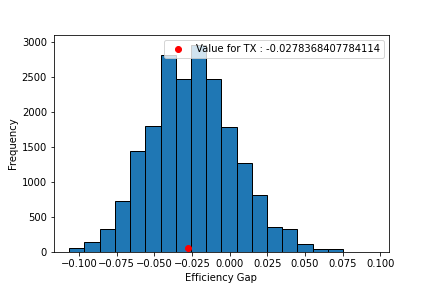}
    \includegraphics[width=0.3\linewidth]{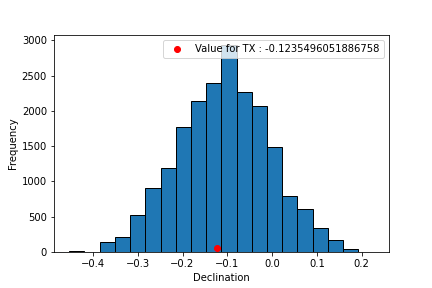}
    \includegraphics[width=0.3\linewidth]{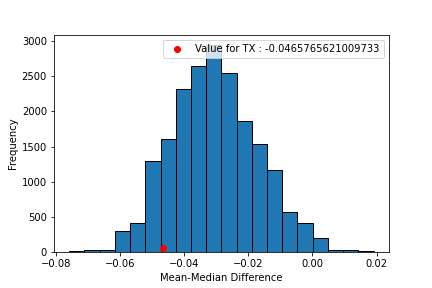}
    \includegraphics[width=0.3\linewidth]{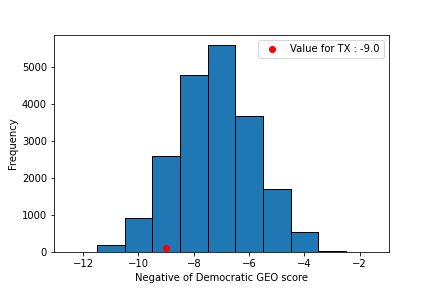}
    \includegraphics[width=0.3\linewidth]{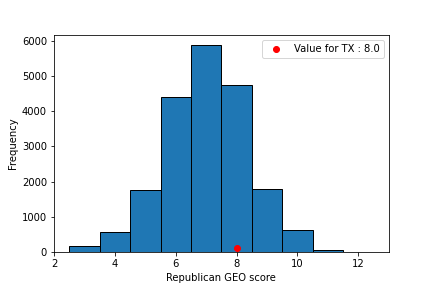}
    \caption{Neutral ensembles for each metric on TX congressional districts; the red dot indicates the value on the enacted map.  Each metric gives information similar to ``Democratic Districts won,'' specifically: the map is not an outlier compared to other maps in the ensemble.}
    \label{fig:results_TXcong_ensemble}

\end{figure}

Overall, the GEO metric, Declination, and Efficiency Gap all had values within their respective histograms that mimicked the histogram for number of seats won by Democrats.  That is, if the number of Democratic won seats for the enacted map looked like (or did not look like) an outlier compared to the number of Democratic won seats for all maps created by the ensemble, the same was true for the GEO metric, Declination, and Efficiency Gap.  The Mean-Median Difference value appeared to be an outlier on three maps, where all other metrics did not seem to be outliers:  PA lower, MI lower, and TX upper.  The results for the PA lower house are in Figure \ref{fig:results_PAlower_ensemble}.  (Images for the MI lower and TX upper houses are in Appendix \ref{appendix:figures}.)

\begin{figure}[h]
    \centering
    \includegraphics[width=0.3\linewidth]{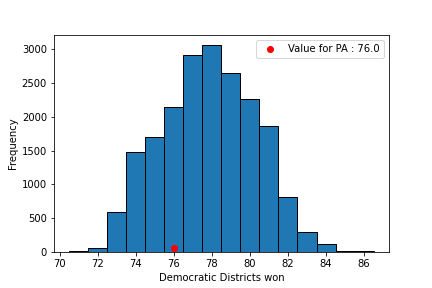}
    \includegraphics[width=0.3\linewidth]{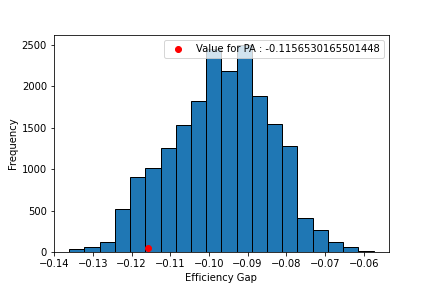}
    \includegraphics[width=0.3\linewidth]{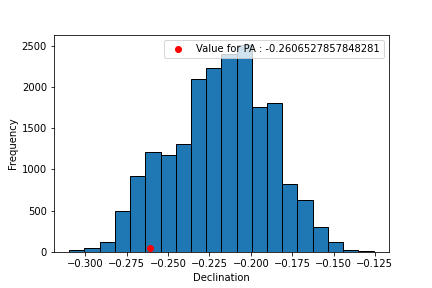}
    \includegraphics[width=0.3\linewidth]{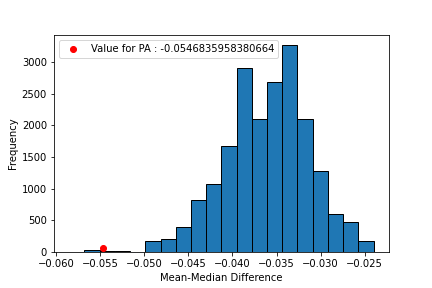}
    \includegraphics[width=0.3\linewidth]{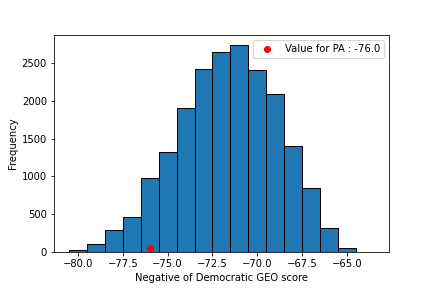}
    \includegraphics[width=0.3\linewidth]{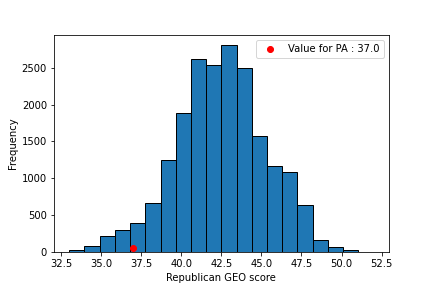}
    \caption{Neutral ensembles for each metric on PA lower state house districts; the red dot indicates the value on the enacted map.  Each metric gives information similar to ``Democratic Districts won,'' (that the map is not an outlier) except for Mean-Median, which indicates the map is an outlier.}
    \label{fig:results_PAlower_ensemble}

\end{figure}

Finally, while the PA congressional map did seem to be an outlier for the other metrics (and the Pennsylvania Supreme court \cite{LeagueWomenPA}), the Mean-Median Difference value was not an outlier within its ensemble.  Those results are in Figure \ref{fig:results_PAcong_ensemble}.  

\begin{figure}[h]
    \centering
    \includegraphics[width=0.3\linewidth]{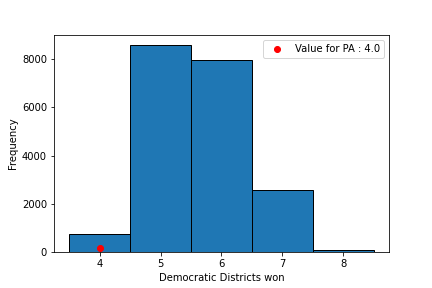}
    \includegraphics[width=0.3\linewidth]{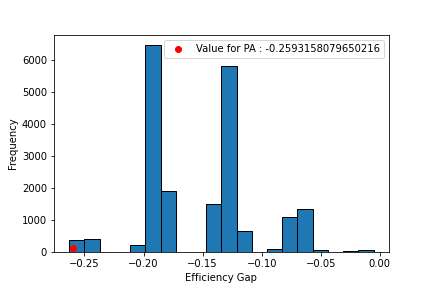}
    \includegraphics[width=0.3\linewidth]{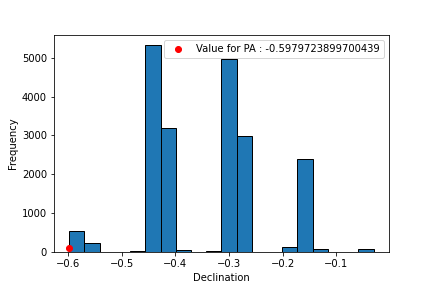}
    \includegraphics[width=0.3\linewidth]{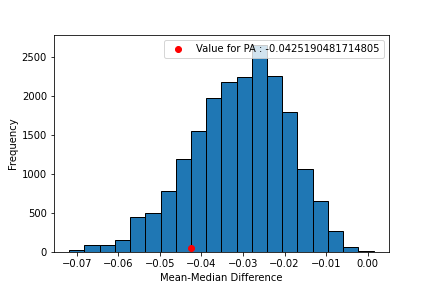}
    \includegraphics[width=0.3\linewidth]{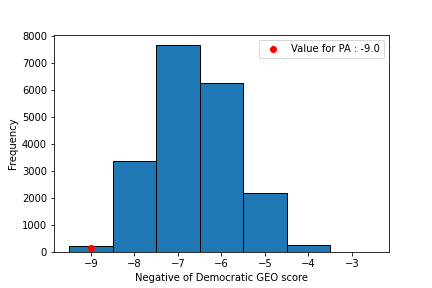}
    \includegraphics[width=0.3\linewidth]{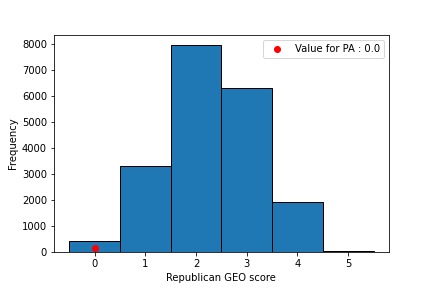}
    \caption{Neutral Ensembles for each metric on PA congressional districts; the red dot indicates the value on the enacted map.  Each metric gives information similar to ``Democratic Districts won,'' (that the map is an outlier) except for Mean-Median, which indicates the map is \emph{not} an outlier. }
    \label{fig:results_PAcong_ensemble}

\end{figure}

Our results from the ensemble analyses that we performed confirm:

\begin{enumerate}
    \item The Efficiency Gap, Declination, and GEO metric are no better than simply the number of districts won, when looking at an ensemble.
    \item The Mean-Median Difference is not a helpful metric, and is potentially misleading, even when considered in the context of an ensemble of maps.
\end{enumerate}

\subsection{When using a single metric}

There may be situations in which the creation of an ensemble is difficult or undesired.  The GerryChain library has made ensemble creation much easier \cite{GerryChain}; difficulty in creating an ensemble is primarily in the data collection, cleaning, and preparation process.  When such data is difficult to obtain, or time-consuming to prepare, one might want to be able to make a quick initial analysis of a map.

In such a situation, we propose that the best metric to use is the GEO metric.   We've shown several issues with the Mean-Median Difference here.  The Efficiency Gap prefers maps with a proportionality \cite{2018arXiv180105301V} that has no basis in law  \cite{DavisBandemer} and may be impossible to achieve \cite{MA_DemDomination}.  Both the Efficiency Gap and the Declination idealize a symmetry that doesn't exist within the realities of American political geography; furthermore, insisting on symmetry in gerrymandering metrics also raises many ``problems and paradoxes'' \cite{ImplementingPartisanSymmetry}.  The GEO metric does not insist on symmetry, is interpretable as a count of a potential number of additional competitive districts for a party, and gives additional information on particular districts that may be suspect \cite{geo_paper}.  As stated above, the GEO metric can be gamed (like all the other metrics) and so should not be used as the sole indicator of gerrymandering.  But it can be a useful tool when ensembles are not available or desirable.

\section{Ethical Considerations}

 In this study, we are finding maps that have extreme numbers of districts won but are still ``in bounds" with respect to given metrics. Therefore, it should be emphasized that this is not a guide for gerrymanderers seeking to find extreme maps. A would-be gerrymanderer hardly needs such tutelage. Just as ethical hacking attempts to prevent hacking, using the methods described here to determine which metrics are more vulnerable can help policy makers avoid using metrics that are misleading when drawing or evaluating maps. Furthermore, once a number of districts won by one party is identified as extreme, we can be more wary when a district plan approaches that number.

We reiterate, our goal is to support courts and policy makers to accurately detect partisan gerrymanders.  As discussed in Section \ref{sec:intro}, there are many metrics that have been proposed to detect gerrymandering, and little research to help courts know whether these metrics are gameable.  Preventing partisan gerrymanders enfranchises voters, but it is important know exactly what guidelines will accurately detect and prevent gerrymanders.  We would like to guard against ineffective legislation such as the Clean Missouri act, the creators of which did not understand that two of its guidelines intended to enfranchise voters (requiring small Efficiency Gap and competitive districts) could not be simultaneously achieved \cite{Competitiveness}.  We would similarly like to prevent redistricting commissions from adopting required bounds for the Mean-Median Difference (as they were encouraged to do by Handley \cite{CommissionHandley}), with the expectation that this would guard against extreme gerrymanders.  Our results here clearly show that bounding the Mean-Median Difference would do no such thing.

\section{Conclusions}\label{sec:conclusions}

As described in Section \ref{sec:intro}, the goal of a metric to detect gerrymandering is to give an easy-to-understand value that reasonably accurately determines whether or not a map is gerrymandered.  A metric can be ``gamed'' if it can stay within ``reasonable values,'' and yet an extreme number of districts is won by some party.  We have shown here that, if we consider ``reasonable values'' to be the values described by the creators of the Efficiency Gap (and the apples-to-apples conversion to the other metrics in consideration) each of the metrics we consider (the Efficiency Gap, GEO metric, Declination, and Mean-Median Difference) can be gamed.  Further, we have shown that \emph{all} metrics can be gamed on a single map, suggesting that utilizing more than a single metric does not fully address this issue.

Of course, one could simply claim that the bounds of ``reasonable values'' are not tight enough, and that with tighter bounds the metrics cannot be gamed.  There are two issues with this approach.  The first issue applies to all the metrics we consider.  This first issue is that the values achieved by these metrics are not symmetric in the parties. That is, the values of the Mean-Median Difference, Efficiency Gap, and Declination can take on for a single state, given that state's political geography, are not centered at 0.  (This can be seen in the histograms from our neutral ensembles, such as in Figures \ref{fig:results_TXcong_ensemble}, \ref{fig:results_PAlower_ensemble}, and \ref{fig:results_PAcong_ensemble}).  Additionally, the histograms for the Democratic GEO score is not centered around the same values as the histograms for the Republican GEO score.  (This can be seen again in the histograms in Figures \ref{fig:results_TXcong_ensemble}, \ref{fig:results_PAlower_ensemble}, and \ref{fig:results_PAcong_ensemble}).  Thus, either the metric values allowed for each party must be the same (that is, we restrict the Mean-Median Difference, Efficiency Gap, and the Declination to be in an interval centered at 0, and restrict the values for the Democratic and Republican GEO metric to be in the same interval) or the values allowed for each party must be different.  In the first case, we run into the issue that this may simply be impossible; for example, in Figure \ref{fig:results_PAlower_ensemble}, the Mean-Median Difference, Efficiency Gap, and Declination almost never take on positive values!  And in the second case, because the values allowed for each party would be different, the public and the courts may perceive that one party is being treated unfairly (not to mention that the bounds chosen will likely need to be different for each state/map, since we can see in Figures \ref{fig:results_TXcong_ensemble}, \ref{fig:results_PAlower_ensemble}, and \ref{fig:results_PAcong_ensemble} that ``typical values'' for each of these metrics varies quite a bit by state/map).

The second issue with the approach of requiring tighter bounds is a larger issue, and one which is focused on the Mean-Median Difference.  The issue is that, as can be seen from the results of Section \ref{subsec:results_no_restriction}, extreme values of the Mean-Median Difference metric do not correlate to extreme maps.  That is, when we look at maps with any number of districts won by a party, the Mean-Median Difference takes on similar ranges of values, regardless of the number of districts won.  The Mean-Median Difference simply cannot detect when a map has an extreme number of districts won.  We note that \cite{DeFordVeomett2025Bounds} provides ranges for the values that the Mean-Median Difference (and the Partisan Bias) can take on for various Seats-Votes pairs. The authors there use proofs and empirical studies to show that the values that these metrics are able to take on ``do not generally become more extreme as the number of seats won becomes more extreme,'' thus addressing why the Mean-Median Difference behaves so poorly with regards to detecting an extreme number of districts won.

 Given the issues just described, we looked to see if, within an ensemble of potential redistricting maps, each of these metrics could detect gerrymandered maps (detected when the proposed map is an outlier in the ensemble) and not inaccurately flag maps that have not been considered to be gerrymandered.  The Efficiency Gap, GEO metric, and Declination were able to accurately detect gerrymanders with these ensembles, but here again the Mean-Median Difference fails.  Even within an ensemble of potential maps, the Mean-Median Difference fails to notice that Pennsylvania's 2011 congressional map is a gerrymander.  And it flagged three maps that have not been claimed to be gerrymandered (Pennsylvania State House, Michigan State House, and Texas Senate).  We conclude that not only is the Mean-Median Difference value inaccurate at detecting gerrymandered maps, when considered on an ensemble of maps, the Mean-Median Difference cannot accurately detect when a map does or does not have an extreme number of districts won by some party.

 No metric is a \emph{better} indicator of a map being an outlier in an ensemble analysis than the number of districts won by a fixed party.  We note that, in cases where there are a small number of districts, determining whether a map has an extreme number of districts won from the histogram of ``number of districts won'' may be less visually clear or appealing; this is simply because there are fewer bars on the bar graph.  While non-integer metrics may give the impression of a larger distribution (and thus a more visually appealing histogram), those metrics potentially introduce meaningless precision.


 We hope that the above findings serve as a caution for the public, the courts, and those giving expert testimony.  In particular, we strongly caution against groups such as the Michigan Citizens' Redistricting Commission fixing a range of values for the Mean-Median Difference in an effort to curb gerrymandering.  We have provided sufficient convincing evidence that restricting the value of the Mean-Median Difference will not avoid even the most extreme maps.

We understand the desire to have a simple metric whereby partisan gerrymandering can be detected with a single value.  However, our results show that metrics are not the simple solution that many researchers and practitioners hoped they would be. The metrics we study here (and we suspect all metrics) can be gamed.


\bmhead{Acknowledgments}    
\textanon{This material is based upon work supported by the National Science Foundation under Grant No. DMS-1928930 and by the Alfred P. Sloan Foundation under grant G-2021-16778, while Ellen Veomett and Stephanie Somersille were in residence at the Simons Laufer Mathematical Sciences Institute (formerly MSRI) in Berkeley, California, during the Fall 2023 semester.}{Anonymized for review.}

\bmhead{Conflict of Interests}
On behalf of all authors, the corresponding author states that there is no conflict of interest.

\bibliography{SB_analyze_metrics}

\begin{appendices}
\section{Additional Figures}\label{appendix:figures}

\begin{figure}[h]
    \centering
    \includegraphics[width=0.3\linewidth]{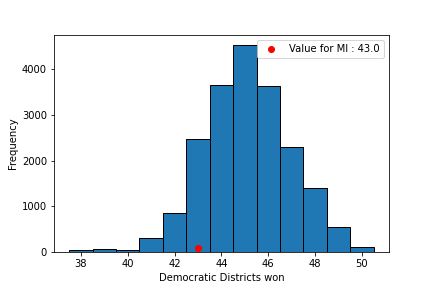}
    \includegraphics[width=0.3\linewidth]{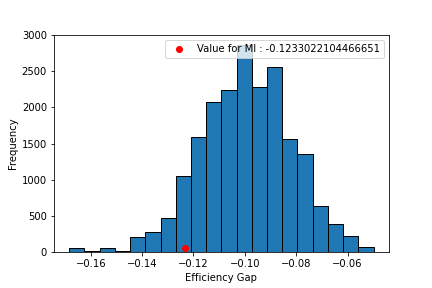}
    \includegraphics[width=0.3\linewidth]{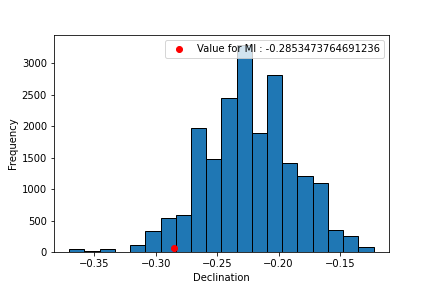}
    \includegraphics[width=0.3\linewidth]{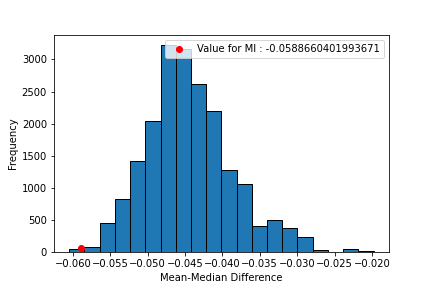}
    \includegraphics[width=0.3\linewidth]{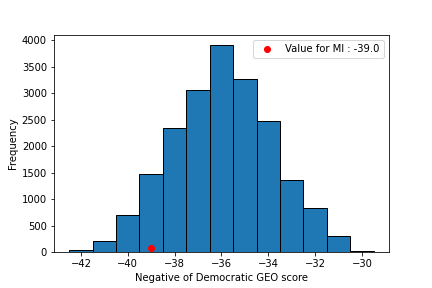}
    \includegraphics[width=0.3\linewidth]{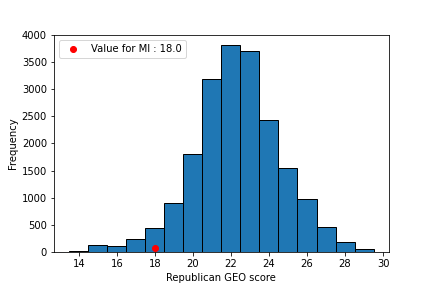}
    \caption{Neutral ensembles for each metric on MI lower state house districts; the red dot indicates the value on the enacted map.  Each metric gives information similar to ``Democratic Districts won,'' (that the map is not an outlier) except for Mean-Median, which indicates the map is an outlier.}
    \label{fig:results_MIlower_ensemble}

\end{figure}

\begin{figure}[h]
    \centering
    \includegraphics[width=0.3\linewidth]{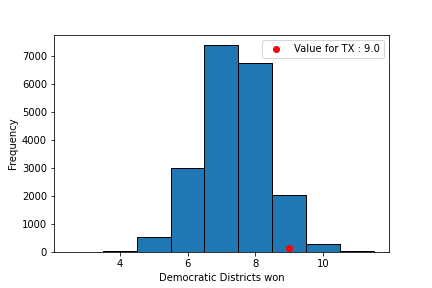}
    \includegraphics[width=0.3\linewidth]{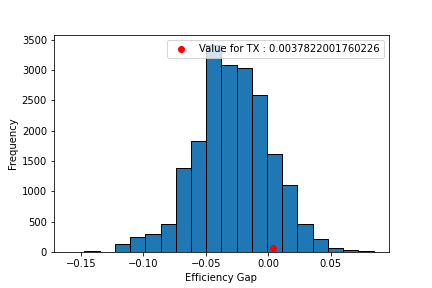}
    \includegraphics[width=0.3\linewidth]{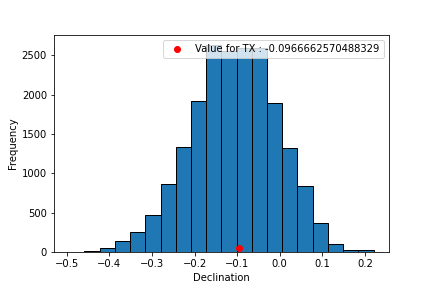}
    \includegraphics[width=0.3\linewidth]{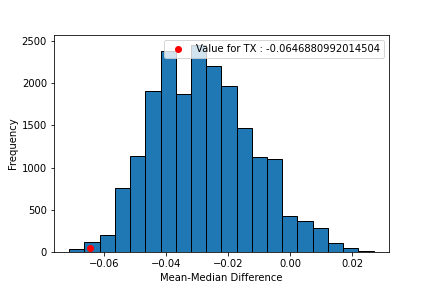}
    \includegraphics[width=0.3\linewidth]{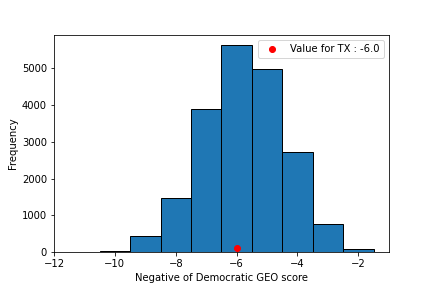}
    \includegraphics[width=0.3\linewidth]{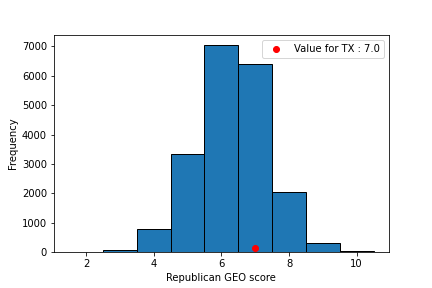}
    \caption{Neutral ensembles for each metric on TX upper state house districts; the red dot indicates the value on the enacted map.  Each metric gives information similar to ``Democratic Districts won,'' (that the map is not an outlier)  except for Mean-Median, which indicates the map is an outlier.}
    \label{fig:results_TXupper_ensemble}

\end{figure}

\begin{figure}[h]
    \centering
    \includegraphics[width=0.29\linewidth]{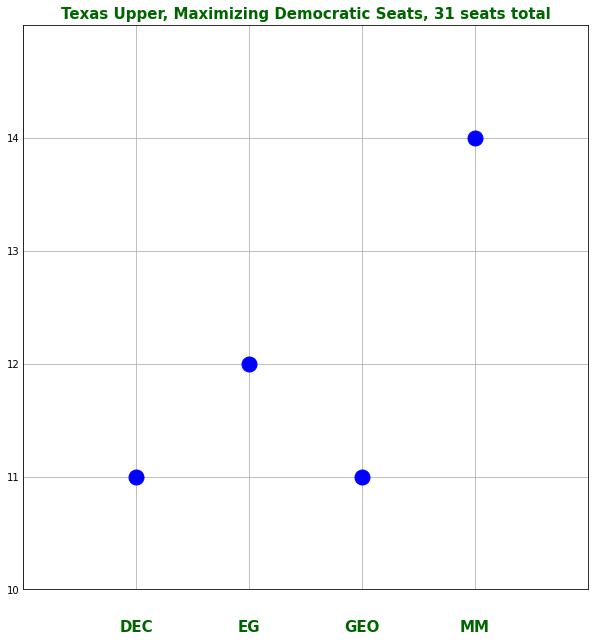}
    \includegraphics[width=0.29\linewidth]{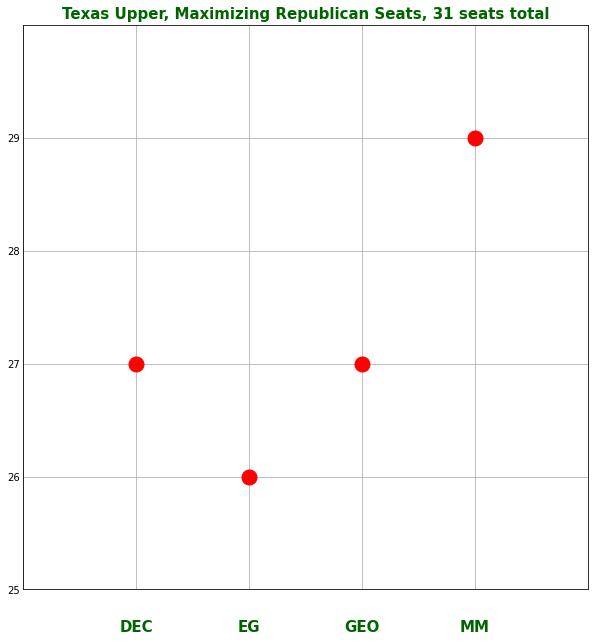}
    \newline
    \includegraphics[width=0.29\linewidth]{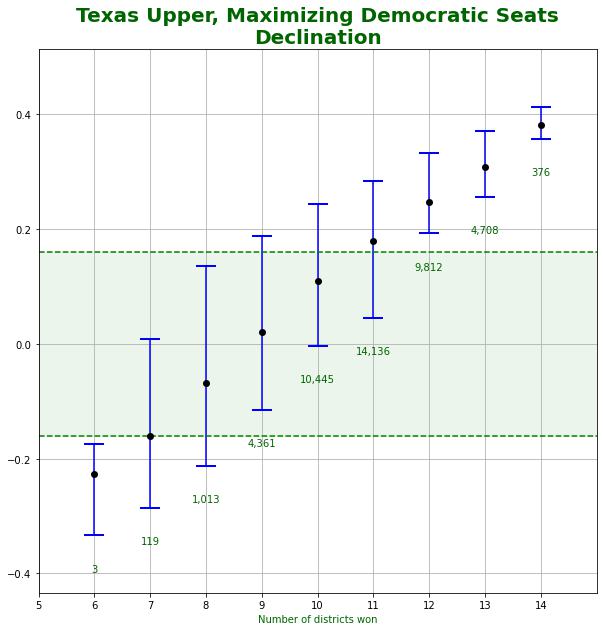}
    \includegraphics[width=0.3\linewidth]{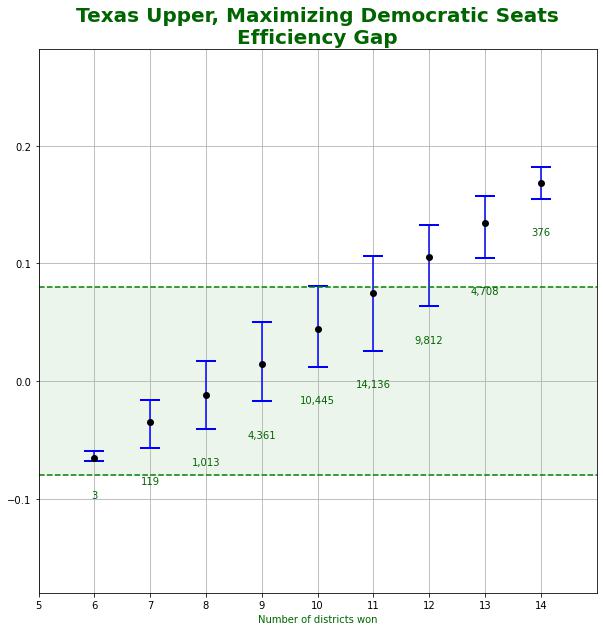}
    \includegraphics[width=0.29\linewidth]{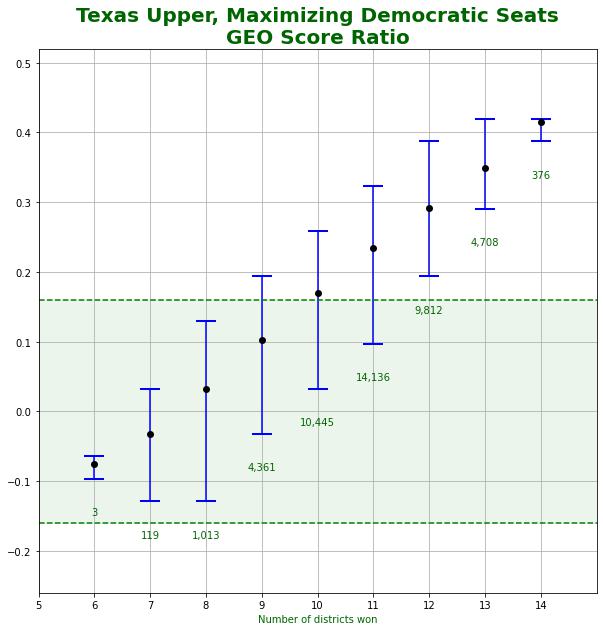}
    \includegraphics[width=0.29\linewidth]{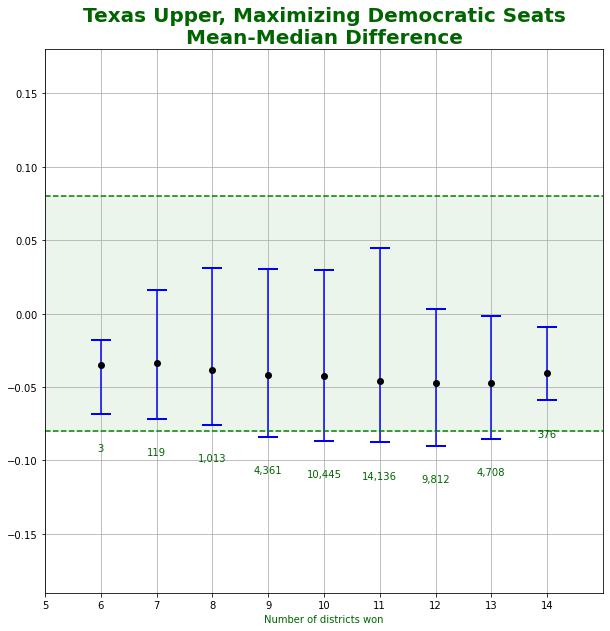}
    \includegraphics[width=0.29\linewidth]{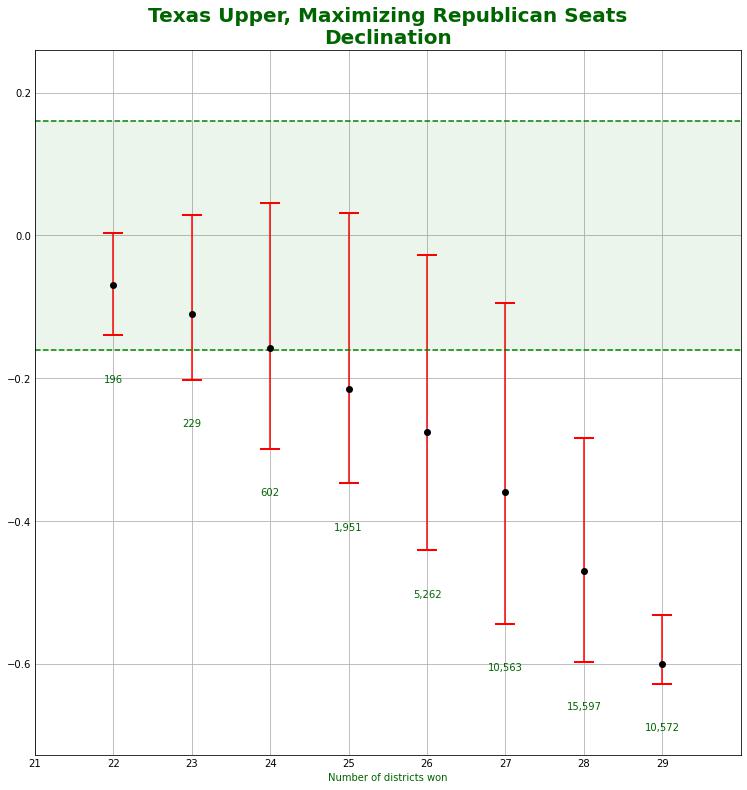}
    \includegraphics[width=0.29\linewidth]{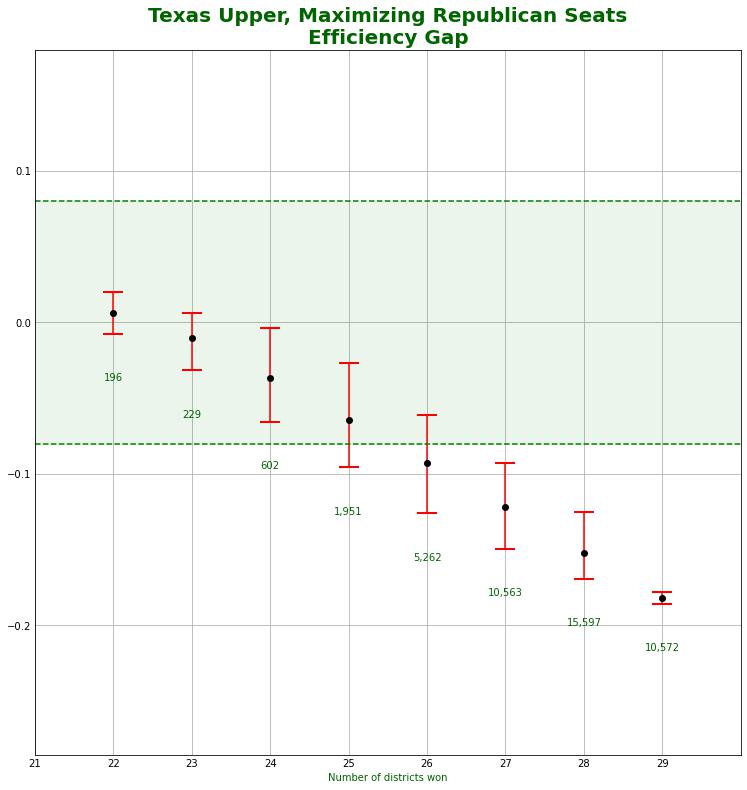}
    \includegraphics[width=0.29\linewidth]{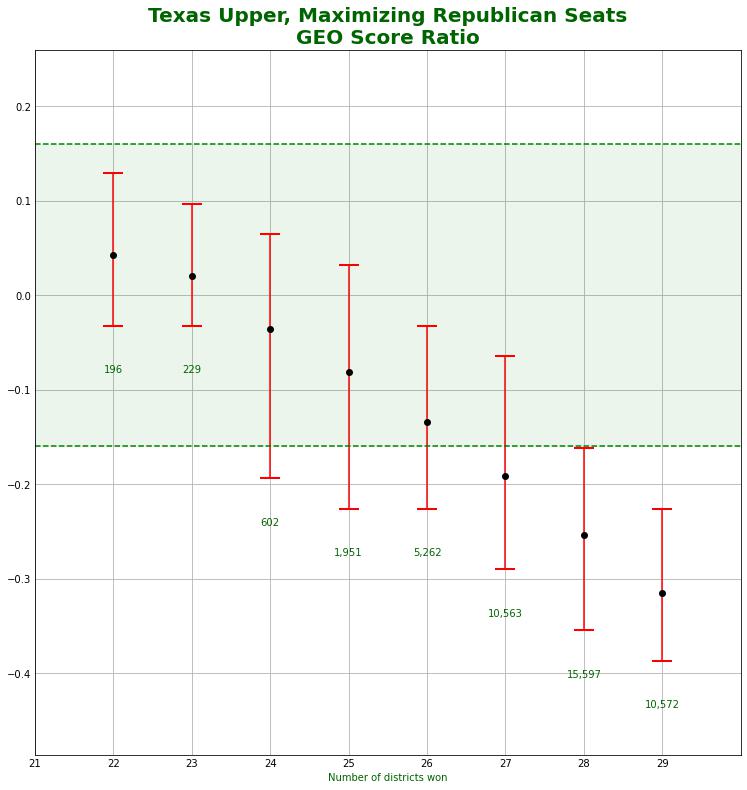}
    \includegraphics[width=0.29\linewidth]{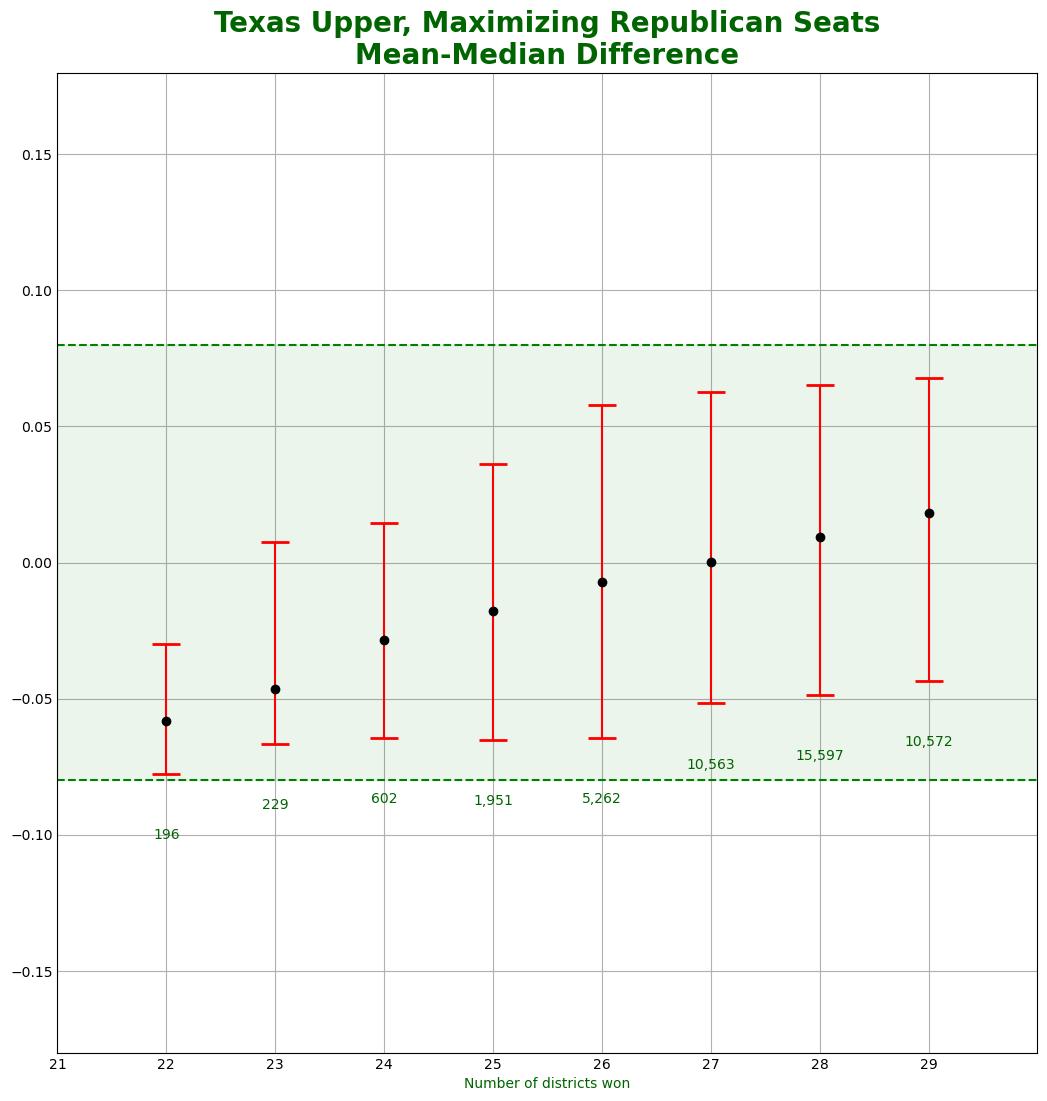}    
    \caption{Results of Short Burst runs for Texas upper house districts.}

\end{figure}

\begin{figure}[h]
    \centering
    \includegraphics[width=0.29\linewidth]{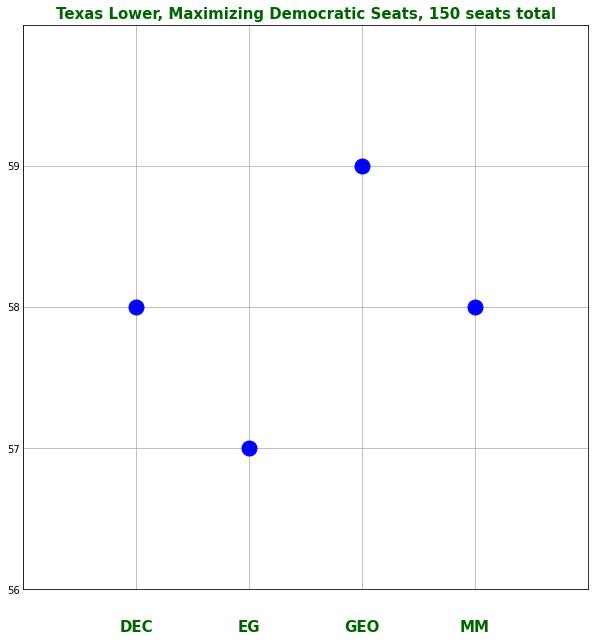}
    \includegraphics[width=0.29\linewidth]{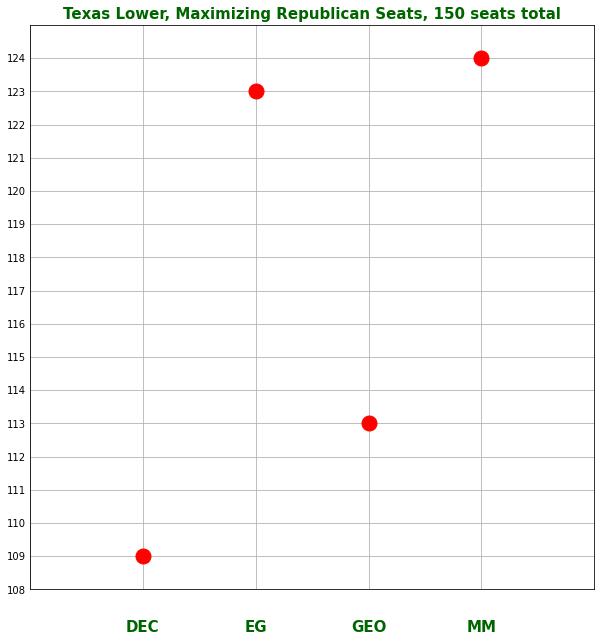}
    \newline
    \includegraphics[width=0.29\linewidth]{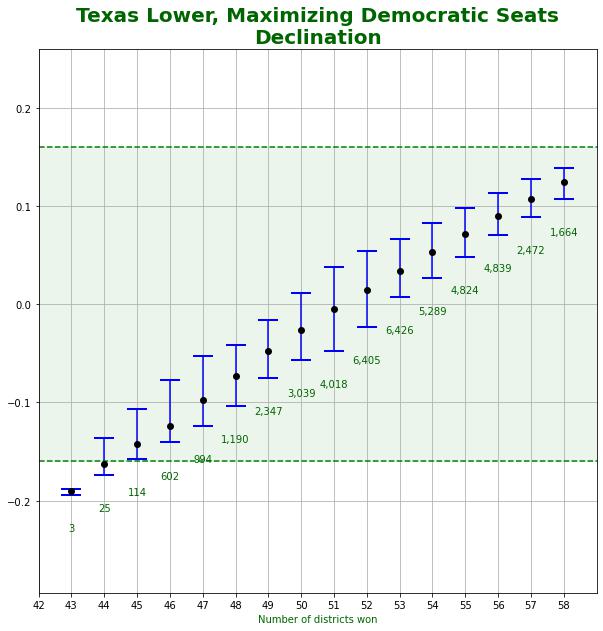}
    \includegraphics[width=0.29\linewidth]{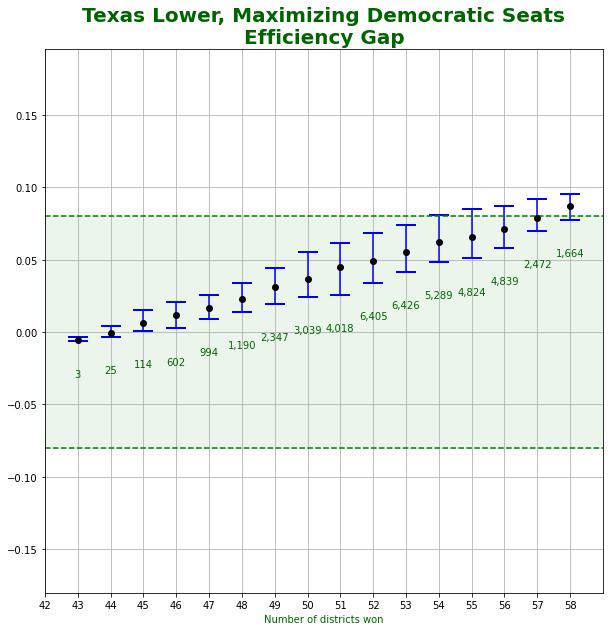}
    \includegraphics[width=0.29\linewidth]{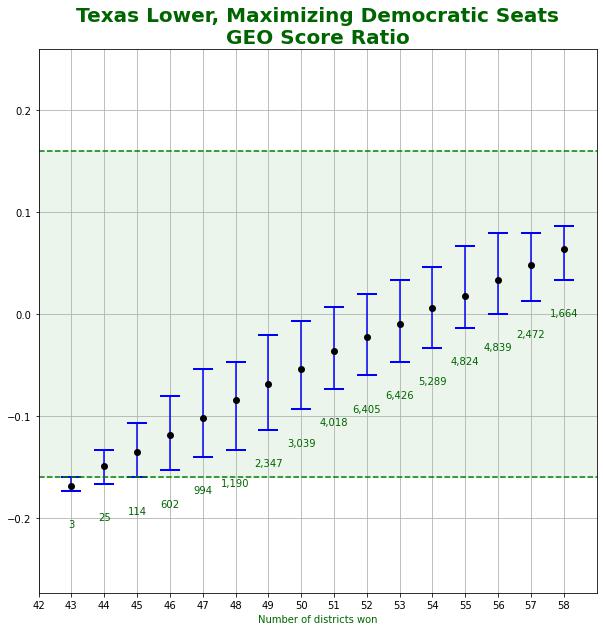}
    \includegraphics[width=0.29\linewidth]{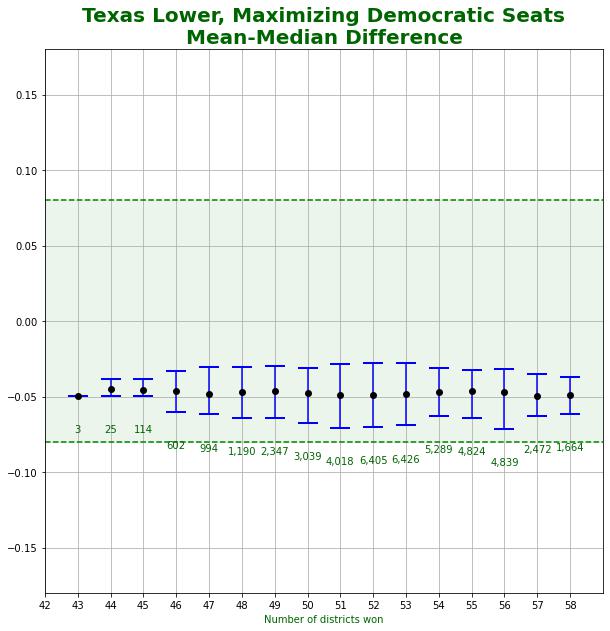}
    \includegraphics[width=0.29\linewidth]{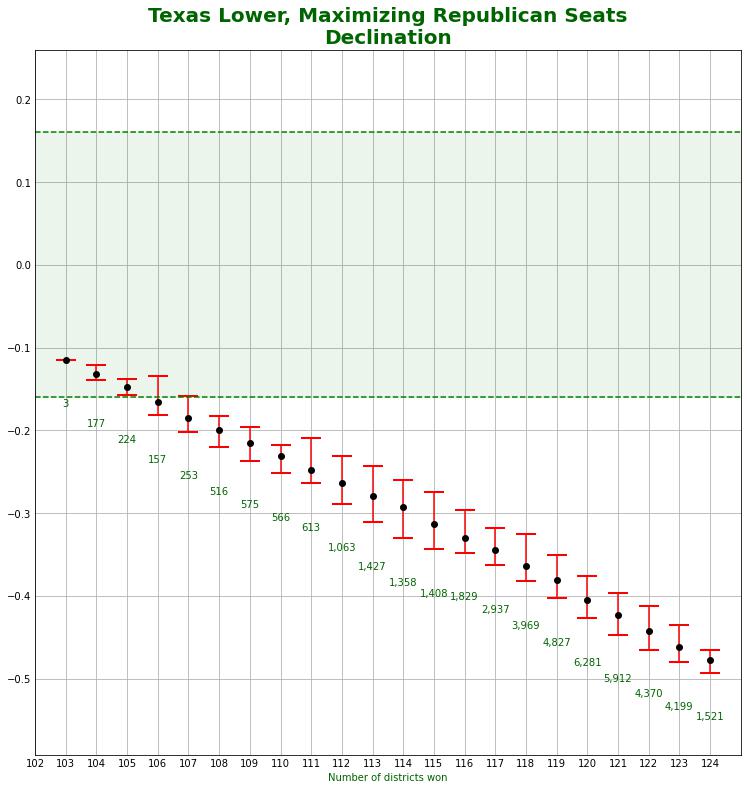}
    \includegraphics[width=0.29\linewidth]{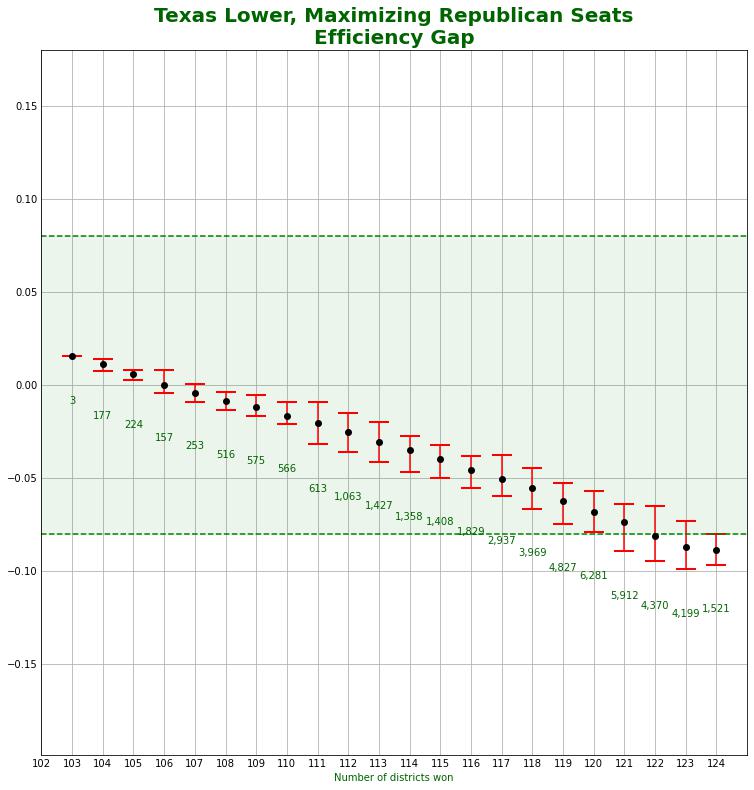}
    \includegraphics[width=0.29\linewidth]{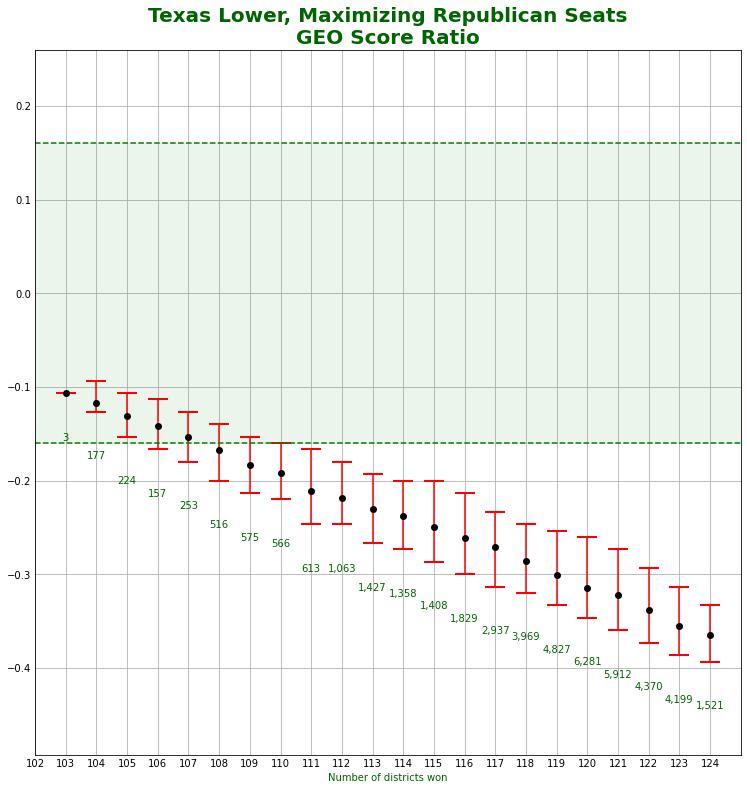}
    \includegraphics[width=0.29\linewidth]{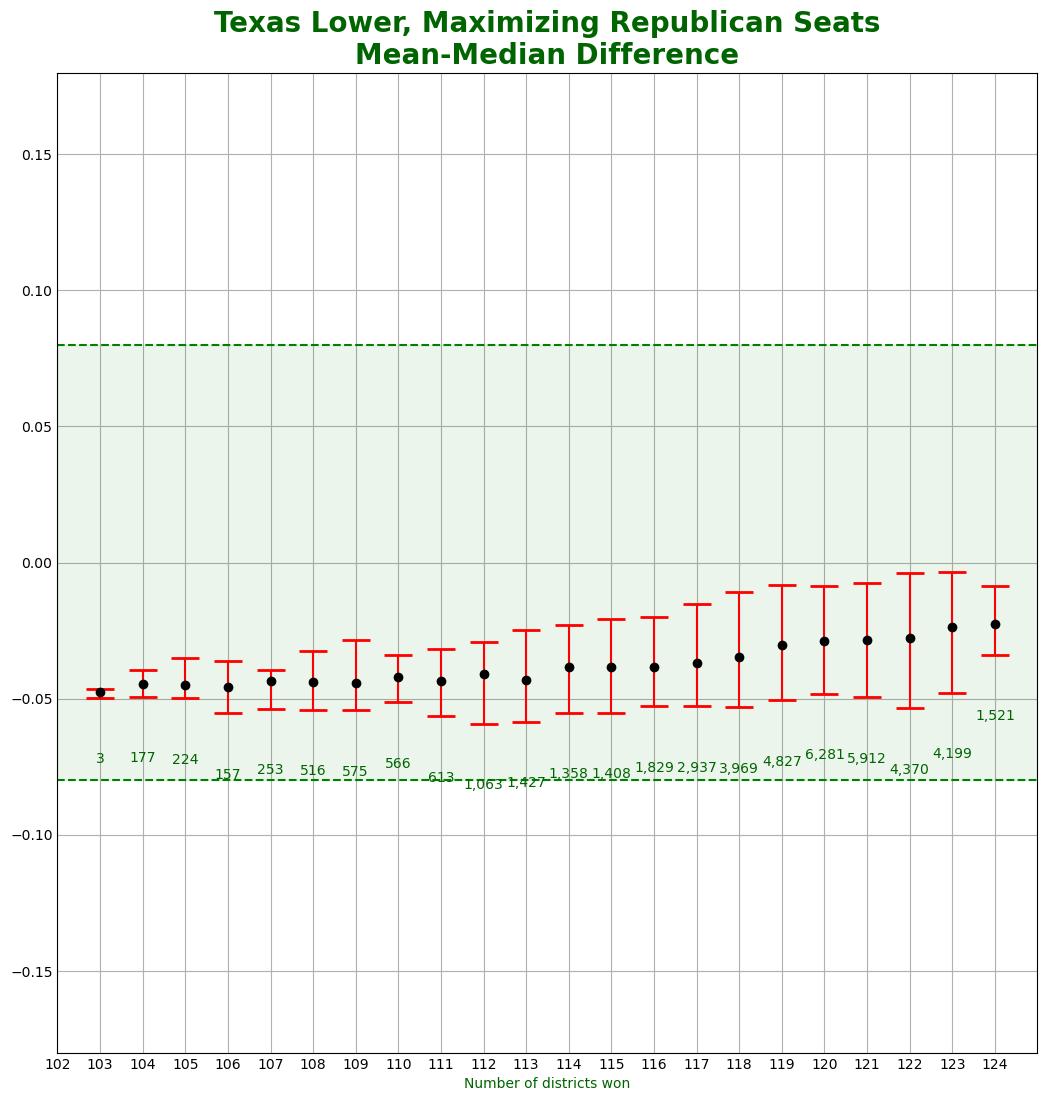}
    \caption{Results of Short Burst runs for Texas lower house districts.}
    \label{fig:results_TXlower_short_bursts_all}
\end{figure}

\begin{figure}[h]
    \centering
    \includegraphics[width=0.29\linewidth]{Images/TXcong_dists36_SEN14Dopt_5.0__5000_sbl10_score0_biasFALSE_.jpg}
    \includegraphics[width=0.29\linewidth]{Images/TXcong_dists36_SEN14Ropt_5.0__5000_sbl10_score0_biasFALSE_.jpg}
    \newline
    \includegraphics[width=0.29\linewidth]{Images/TXcong_dists36_SEN14Dopt_5.0__5000_sbl10_score0_None_biasFalse_D_Declination.jpg}
    \includegraphics[width=0.29\linewidth]{Images/TXcong_dists36_SEN14Dopt_5.0__5000_sbl10_score0_None_biasFalse_D_EfficiencyGapwithwastedvotes.jpg}
    \includegraphics[width=0.29\linewidth]{Images/TXcong_dists36_SEN14Dopt_5.0__5000_sbl10_score0_None_biasFalse_D_GEOscoreratio.jpg}
    \includegraphics[width=0.29\linewidth]{Images/TXcong_dists36_SEN14Dopt_5.0__5000_sbl10_score0_None_biasFalse_D_Mean-Median.jpg}
    \includegraphics[width=0.29\linewidth]{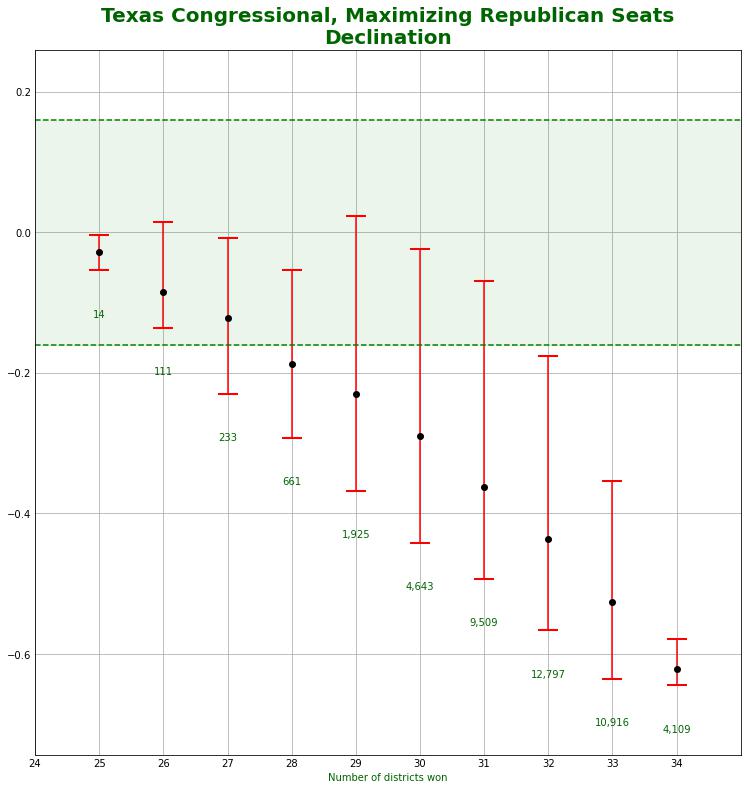}
    \includegraphics[width=0.29\linewidth]{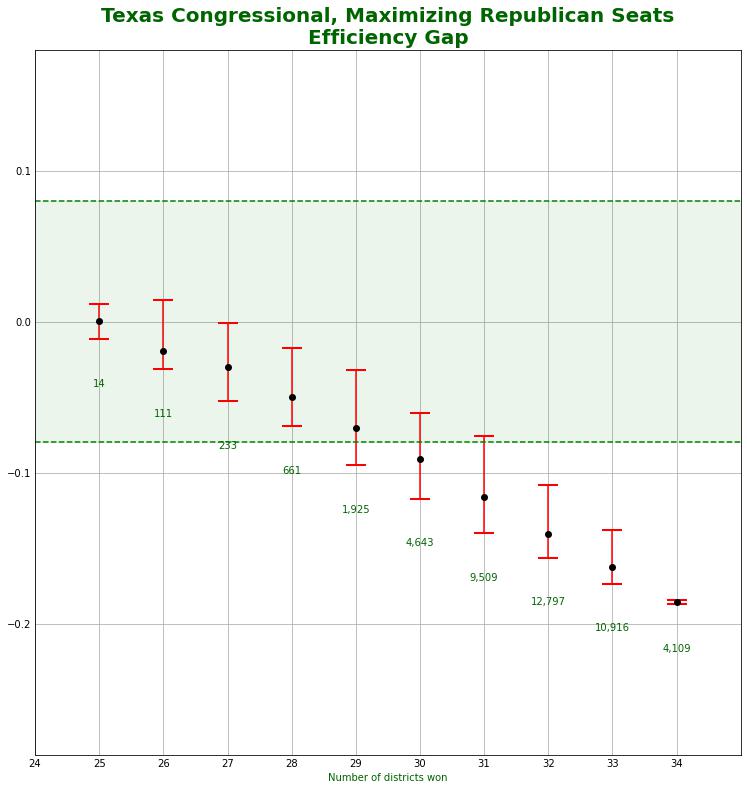}
    \includegraphics[width=0.29\linewidth]{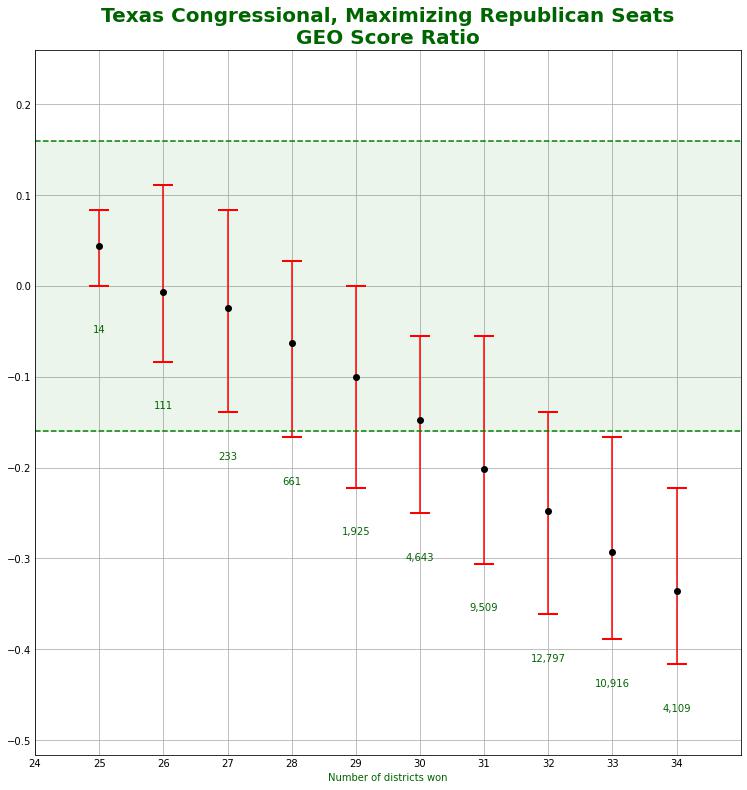}
    \includegraphics[width=0.29\linewidth]{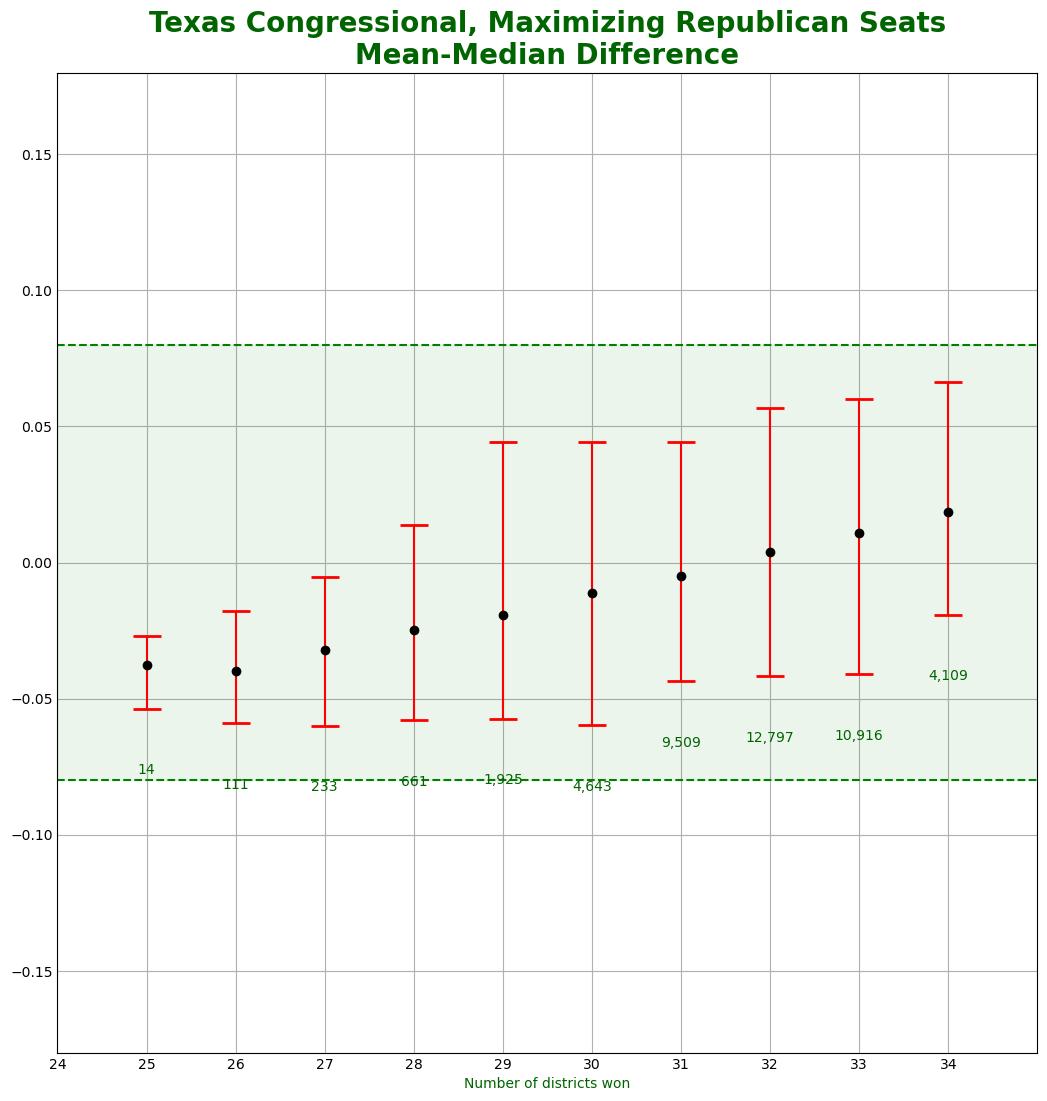}
    \caption{Results of Short Burst runs for Texas Congressional districts.}
    \label{fig:results_TXcong_short_bursts_all}
\end{figure}

\begin{figure}[h]
    \centering
    \includegraphics[width=0.29\linewidth]{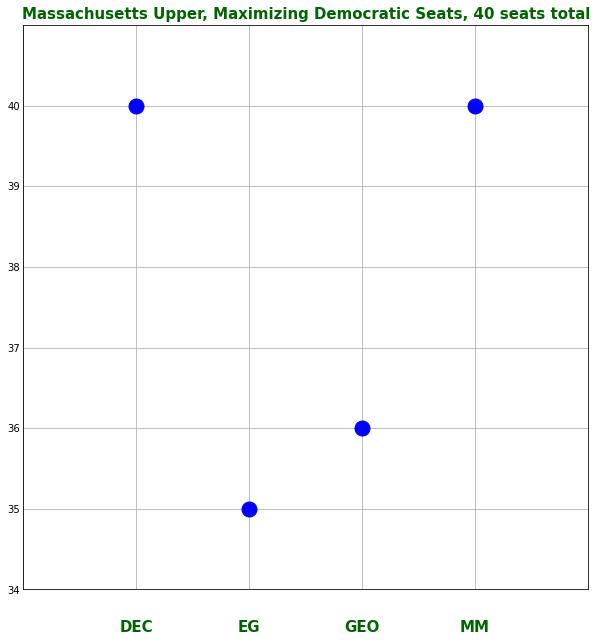}
    \includegraphics[width=0.29\linewidth]{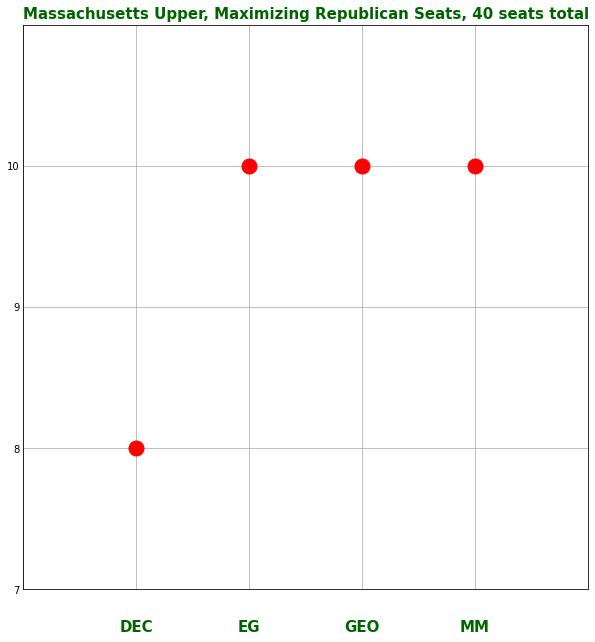}
    \newline
    \includegraphics[width=0.29\linewidth]{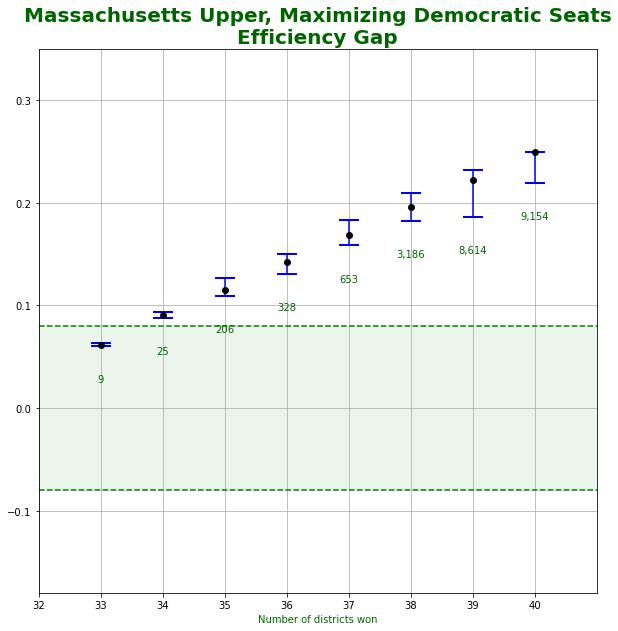}
    \includegraphics[width=0.29\linewidth]{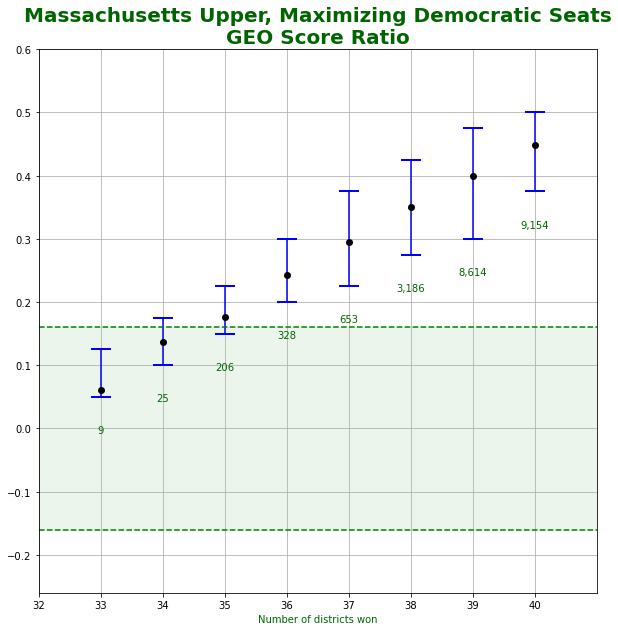}
    \includegraphics[width=0.29\linewidth]{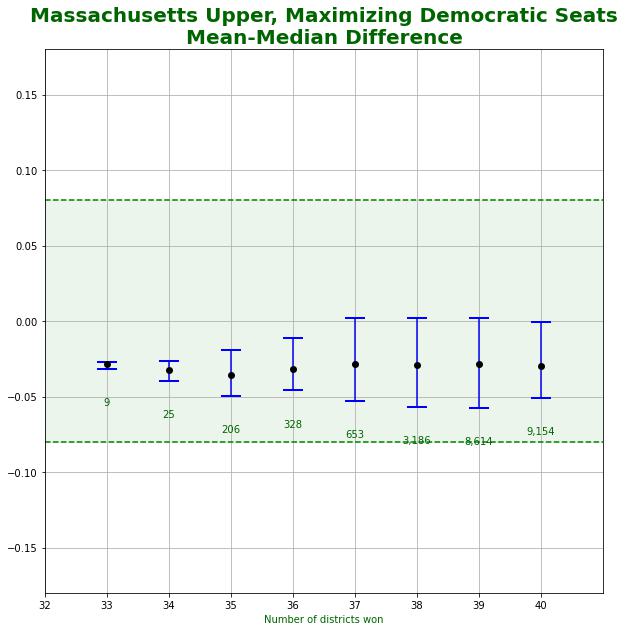}
    \includegraphics[width=0.29\linewidth]{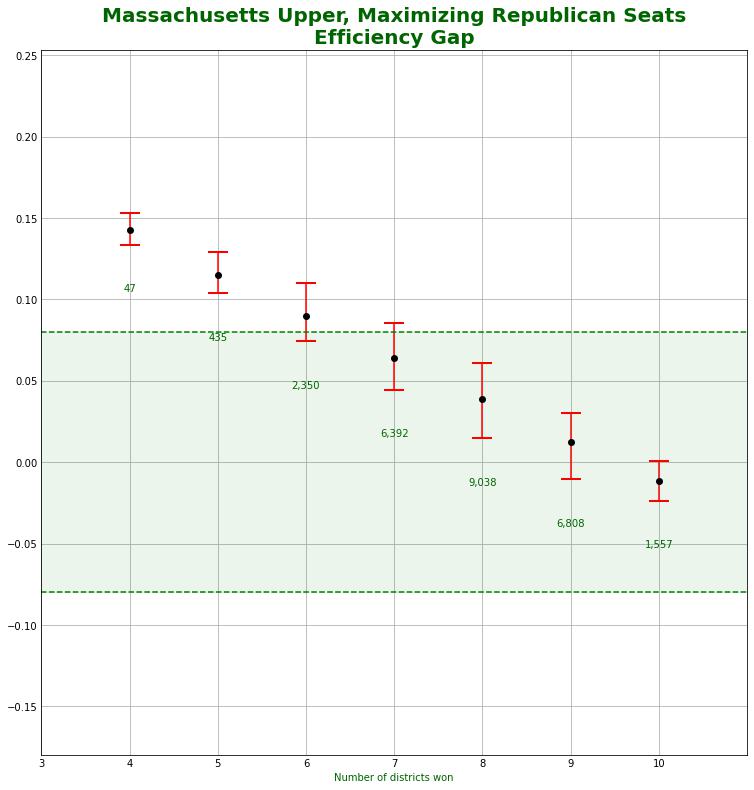}
    \includegraphics[width=0.29\linewidth]{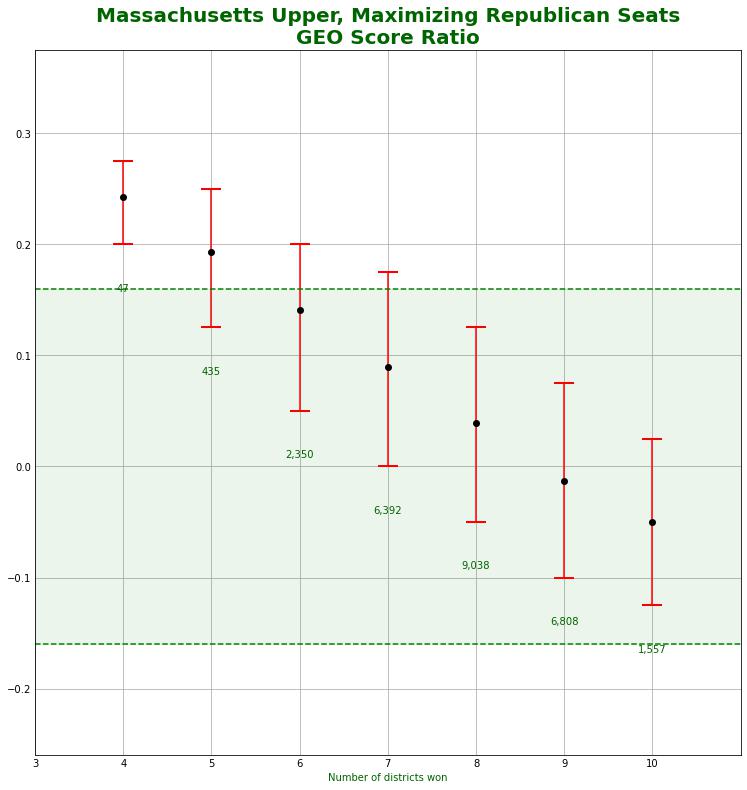}
    \includegraphics[width=0.29\linewidth]{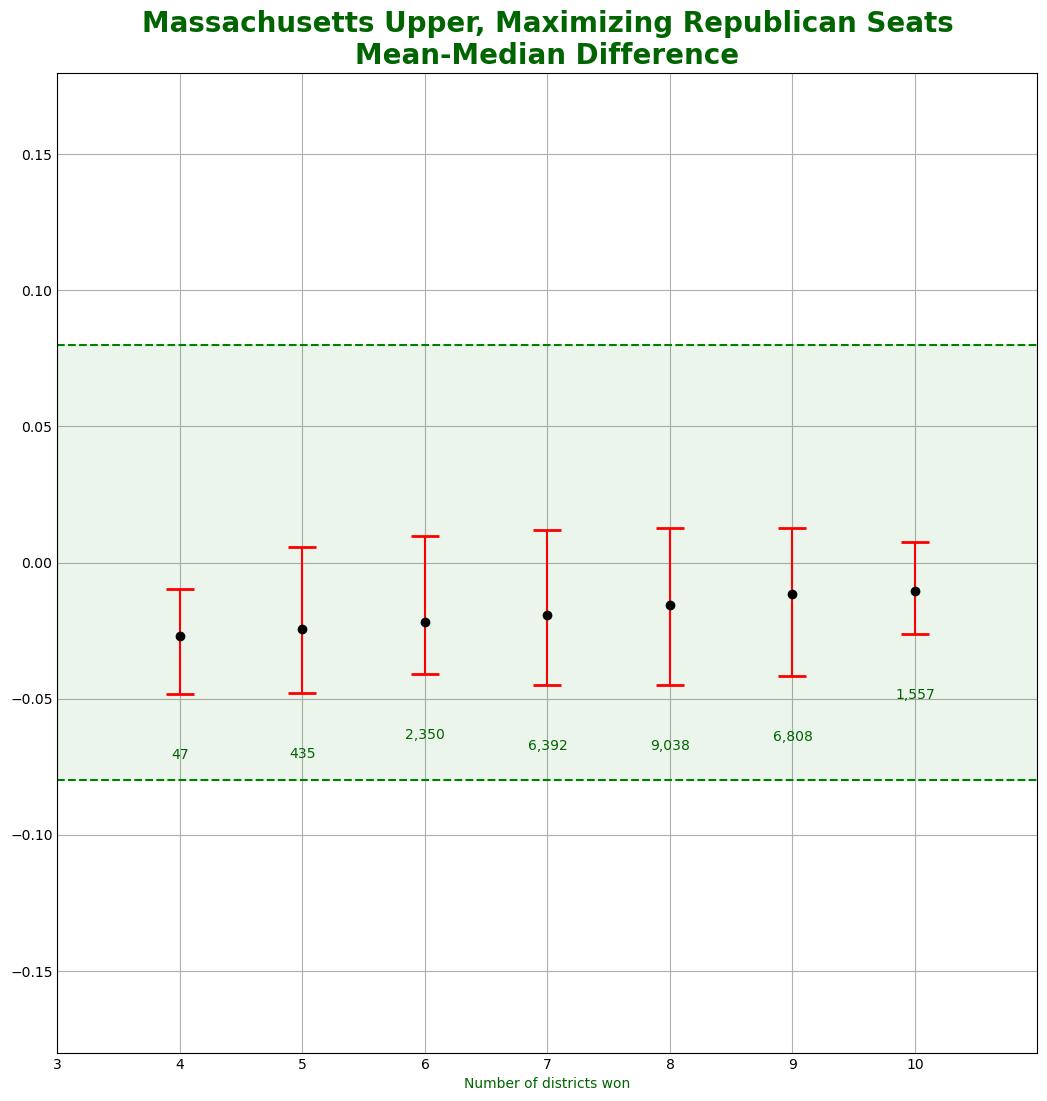}
    \caption{Results of Short Burst runs for Massachusetts upper house districts.}
    
\end{figure}

\begin{figure}[h]
    \centering
    \includegraphics[width=0.29\linewidth]{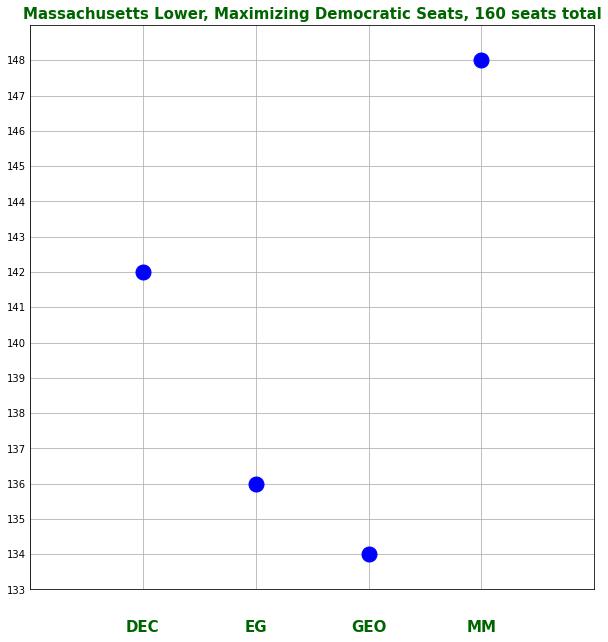}
    \includegraphics[width=0.29\linewidth]{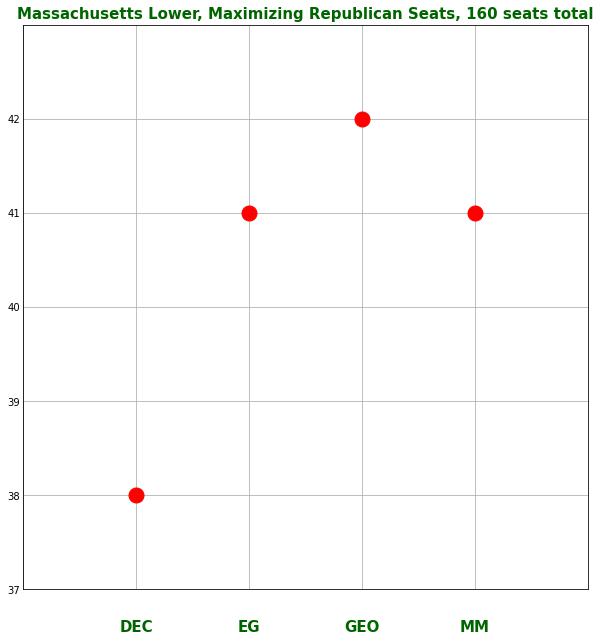}
    \newline
    \includegraphics[width=0.29\linewidth]{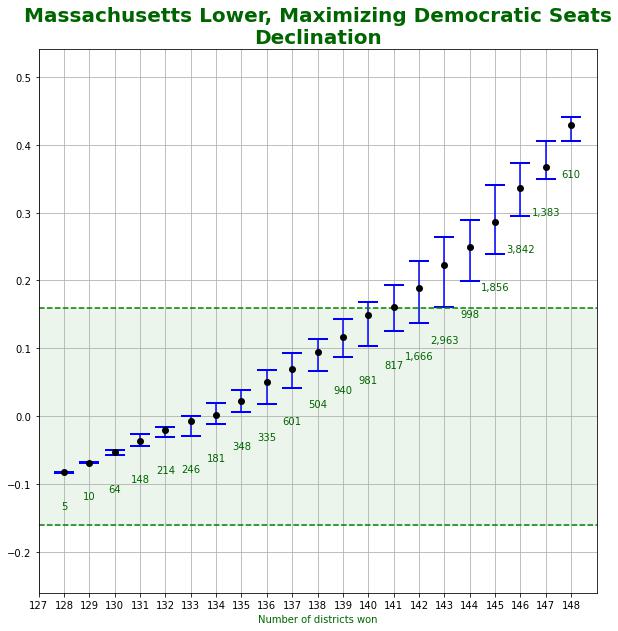}
    \includegraphics[width=0.29\linewidth]{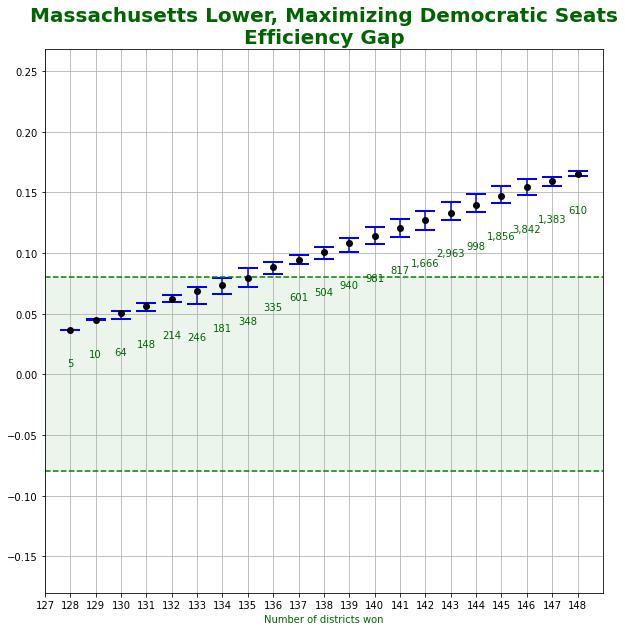}
    \includegraphics[width=0.29\linewidth]{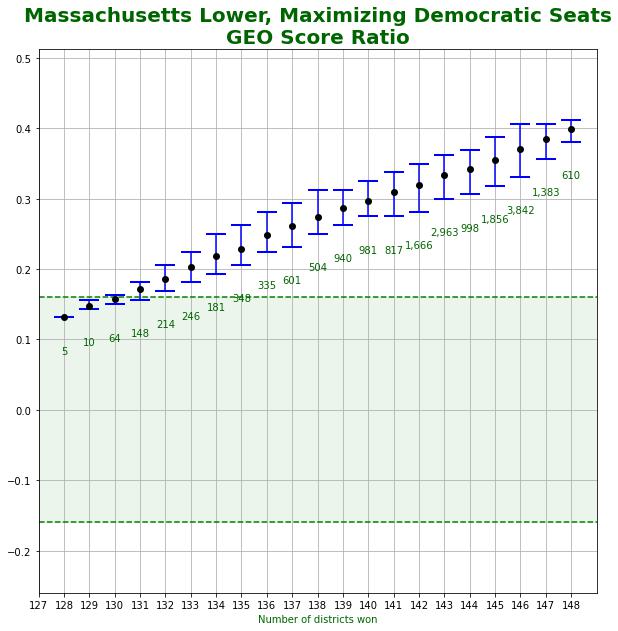}
    \includegraphics[width=0.29\linewidth]{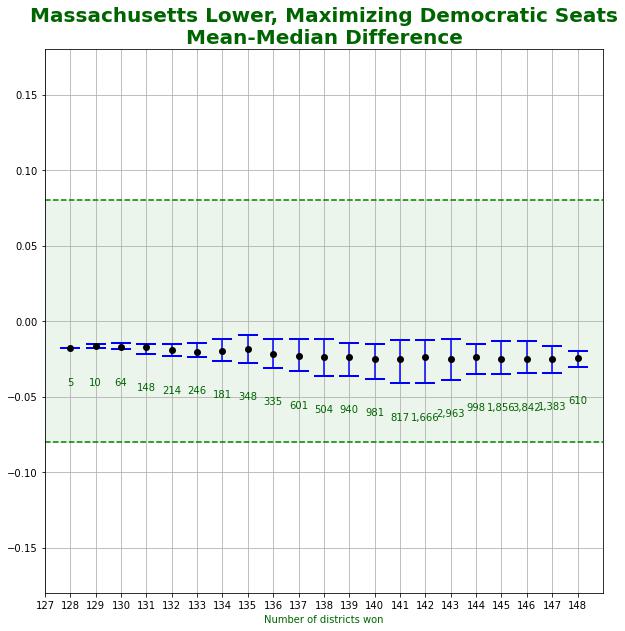}
    \includegraphics[width=0.29\linewidth]{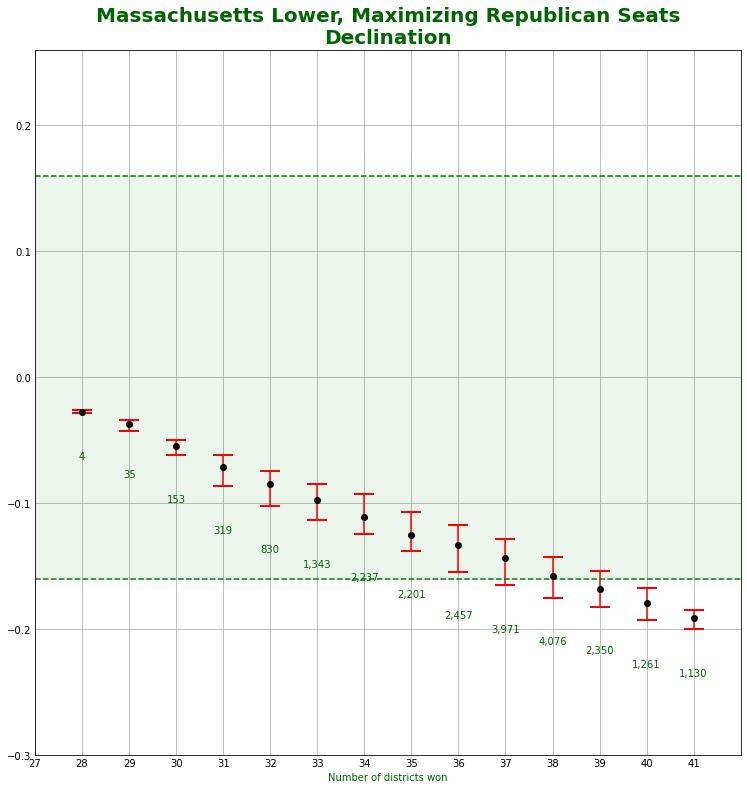}
    \includegraphics[width=0.29\linewidth]{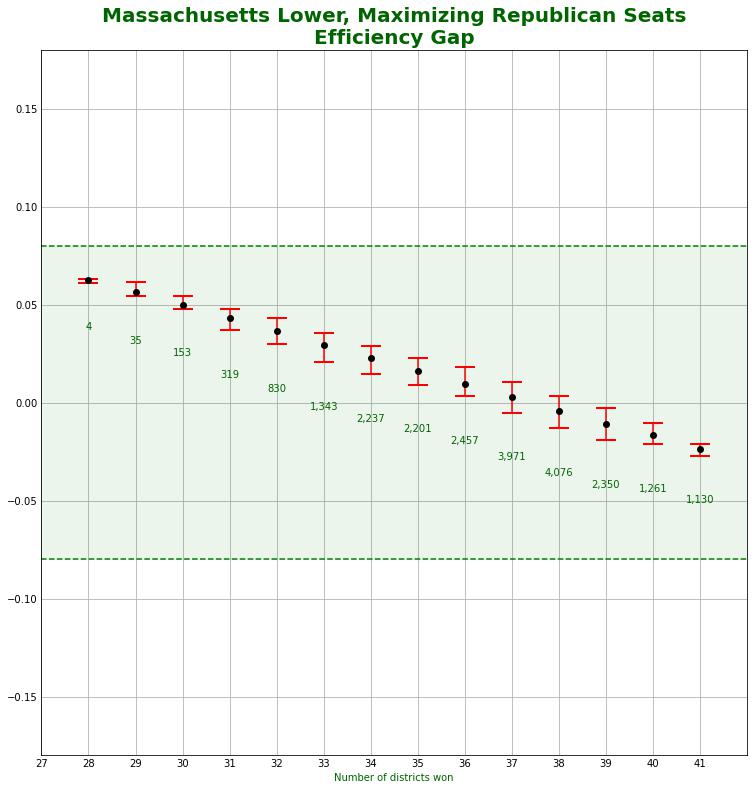}
    \includegraphics[width=0.29\linewidth]{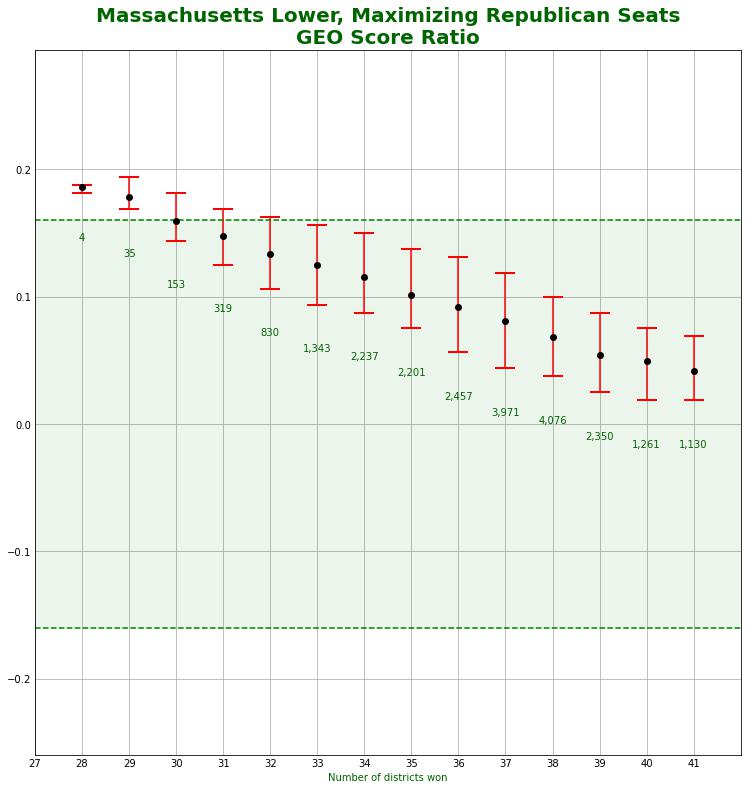}
    \includegraphics[width=0.29\linewidth]{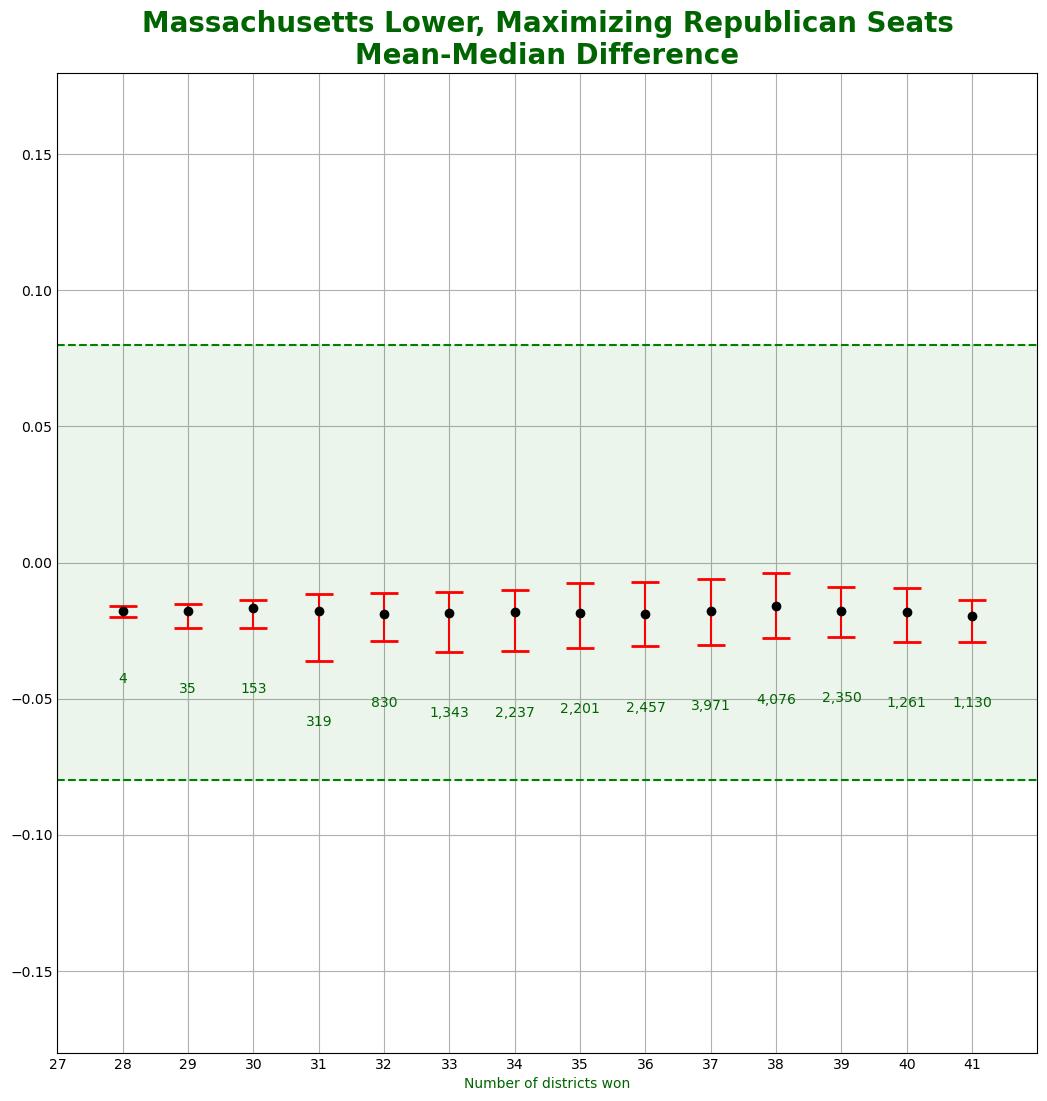}
    \caption{Results of Short Burst runs for Massachusetts lower house districts.}

    \label{fig:results_MAlower_short_bursts_all}
\end{figure}
  
\begin{figure}[h]
    \centering
    \includegraphics[width=0.29\linewidth]{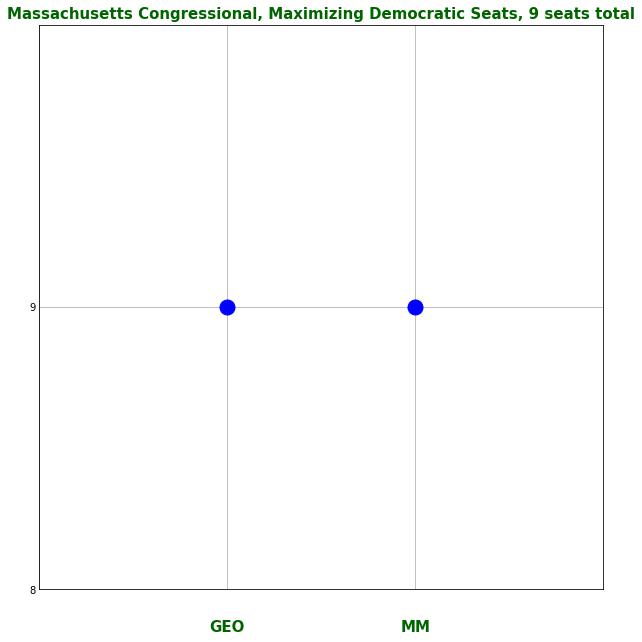}
    \includegraphics[width=0.29\linewidth]{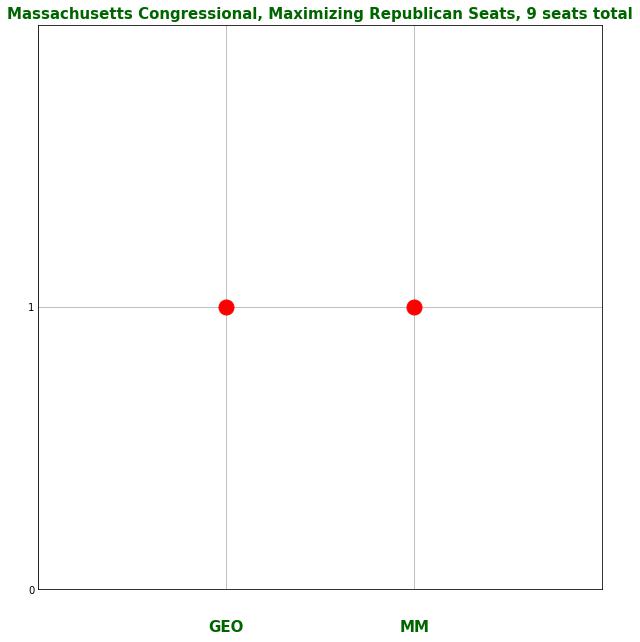}
    \newline
    \includegraphics[width=0.29\linewidth]{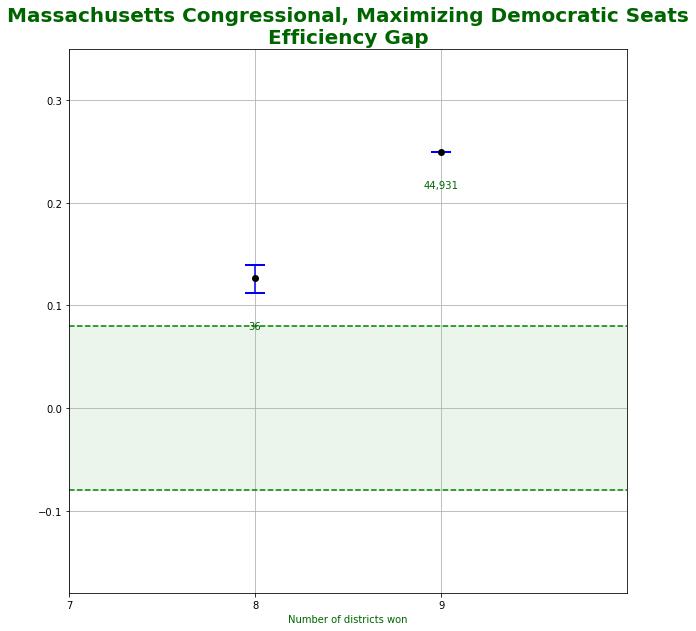}
    \includegraphics[width=0.29\linewidth]{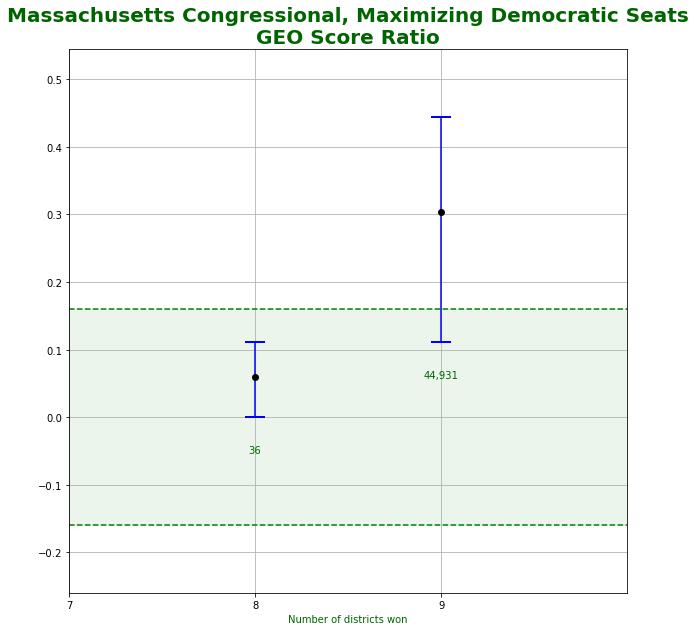}
    \includegraphics[width=0.29\linewidth]{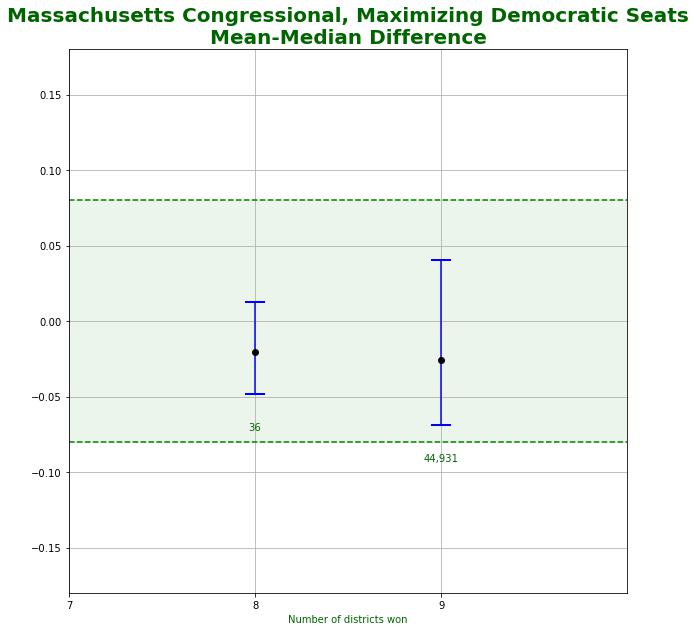}
    \includegraphics[width=0.29\linewidth]{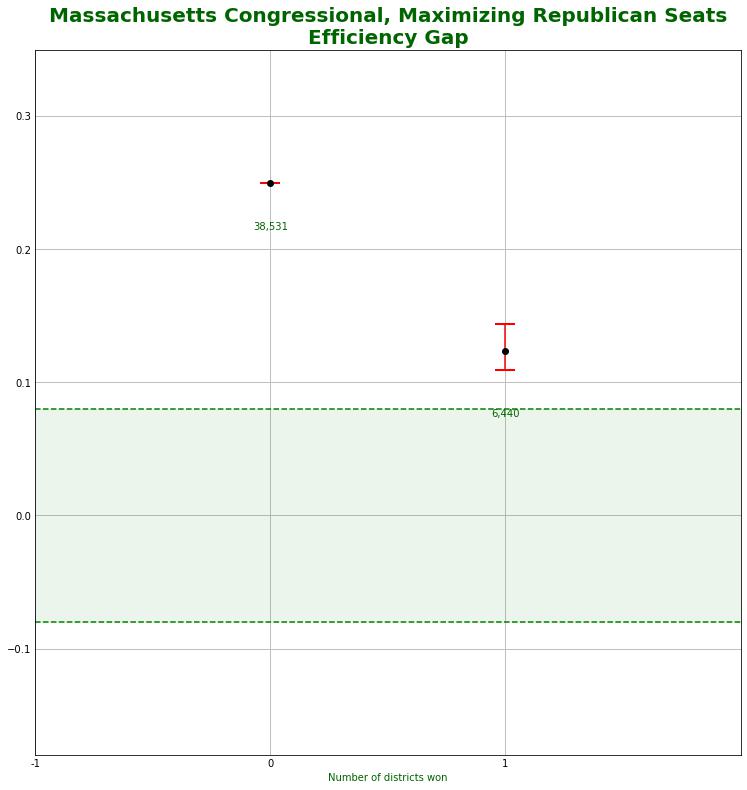}
    \includegraphics[width=0.29\linewidth]{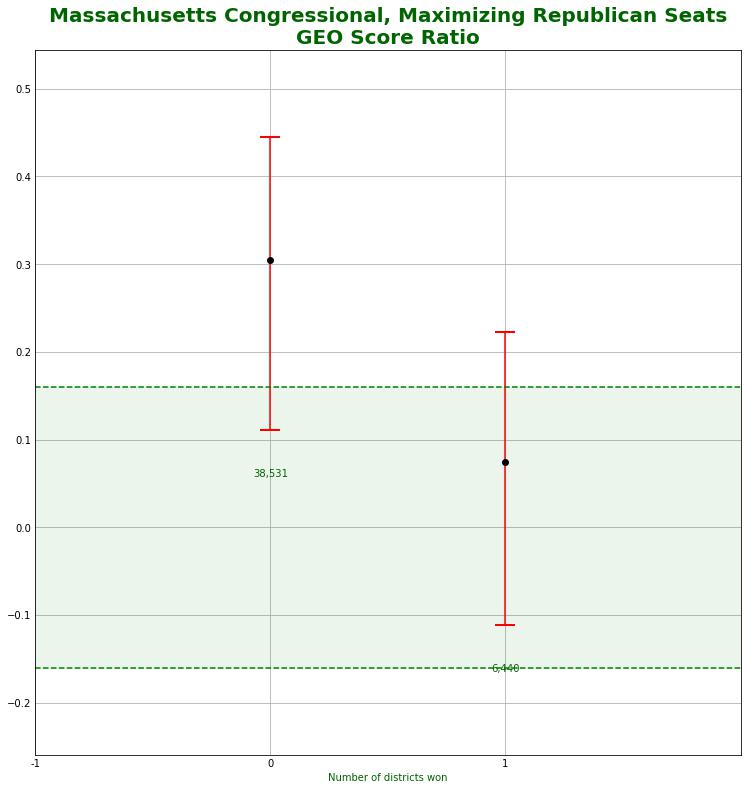}
    \includegraphics[width=0.29\linewidth]{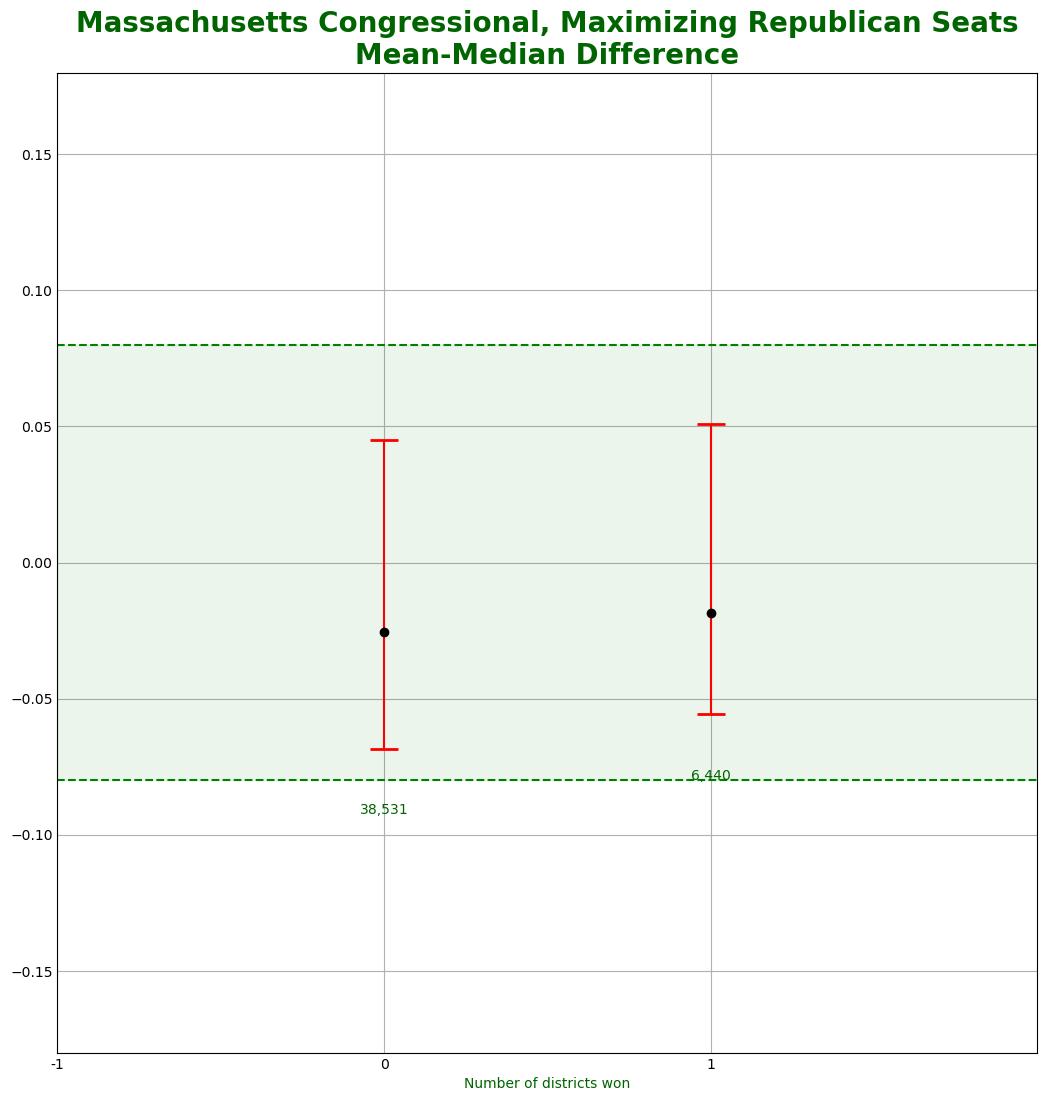}
    \caption{Results of Short Burst runs for Massachusetts Congressional districts.}

    \label{fig:results_MAcong_short_bursts_all}
\end{figure}

\begin{figure}[h]
    \centering
    \includegraphics[width=0.29\linewidth]{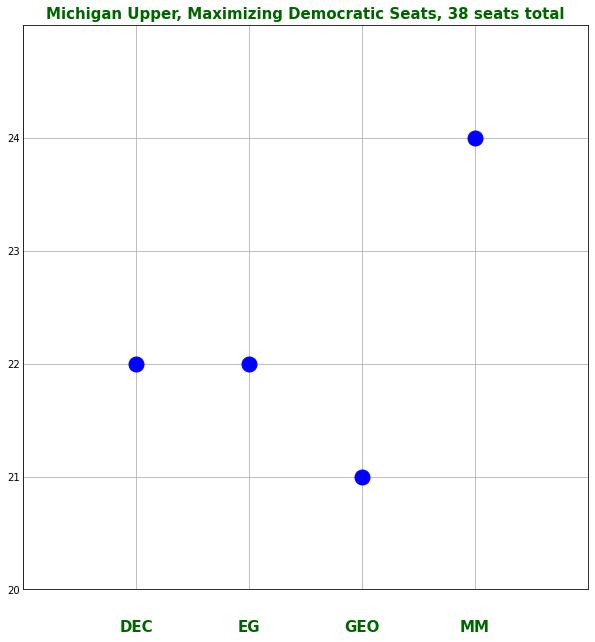}
    \includegraphics[width=0.29\linewidth]{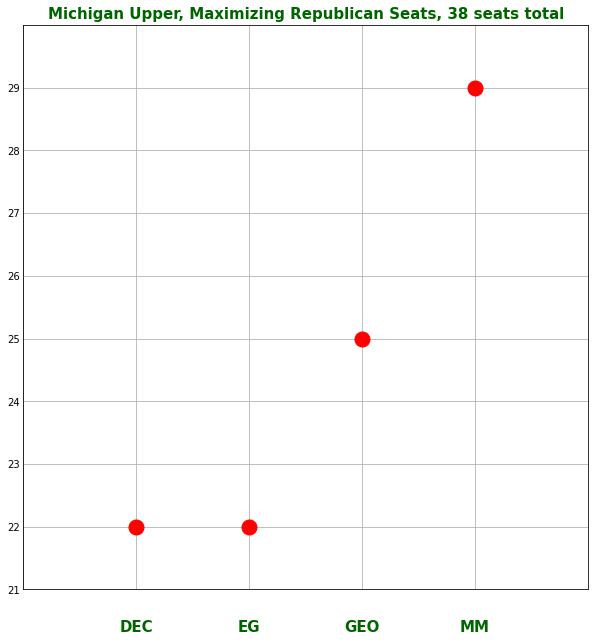}
    \newline
    \includegraphics[width=0.29\linewidth]{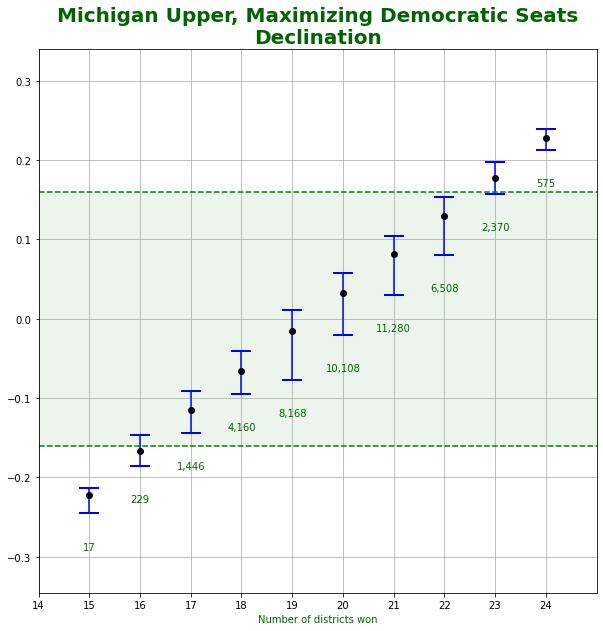}
    \includegraphics[width=0.29\linewidth]{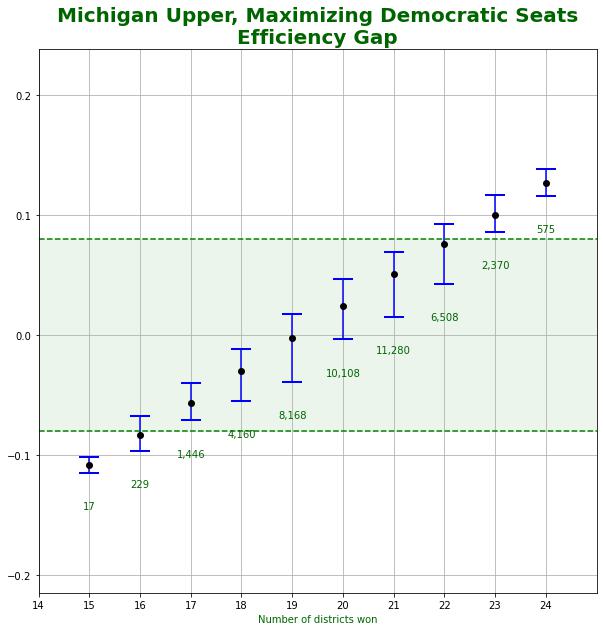}
    \includegraphics[width=0.29\linewidth]{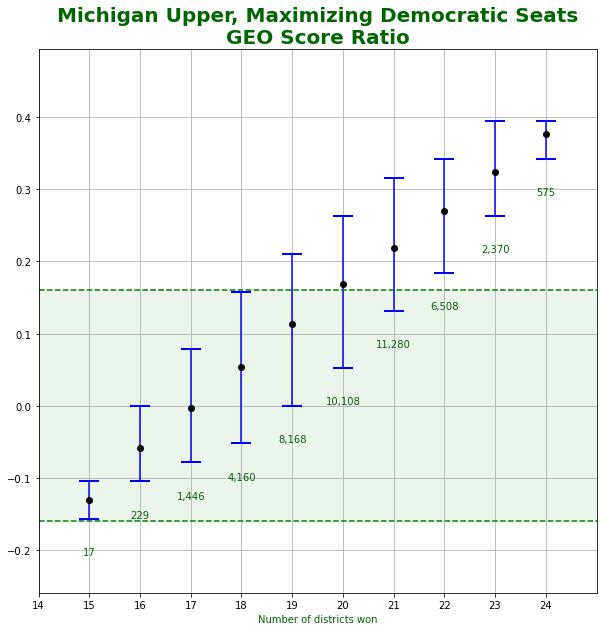}
    \includegraphics[width=0.29\linewidth]{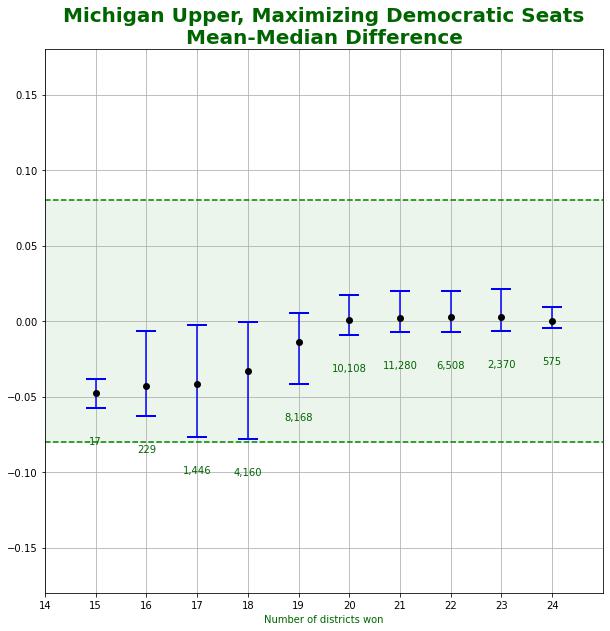}
    \includegraphics[width=0.29\linewidth]{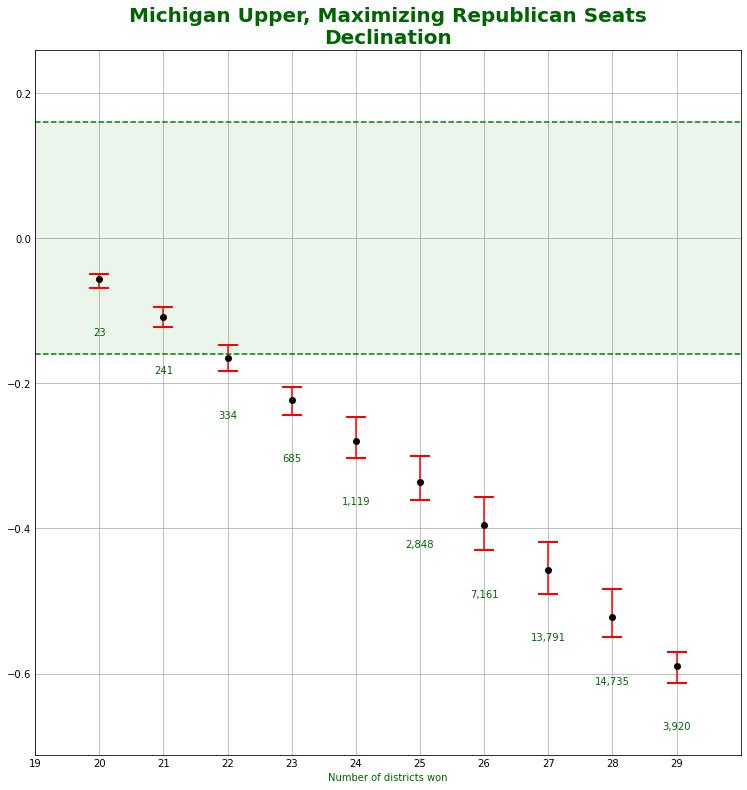}
    \includegraphics[width=0.3\linewidth]{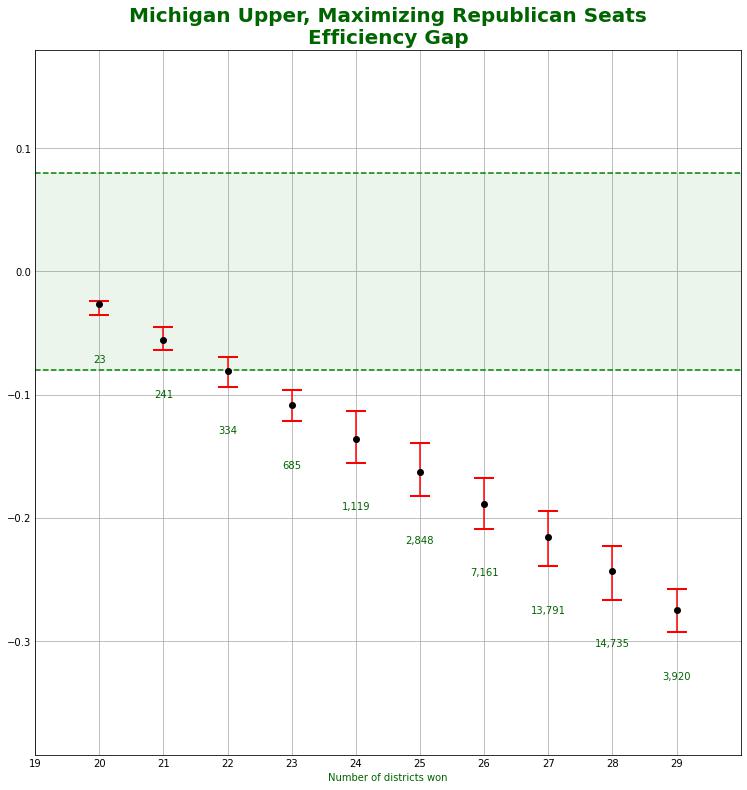}
    \includegraphics[width=0.29\linewidth]{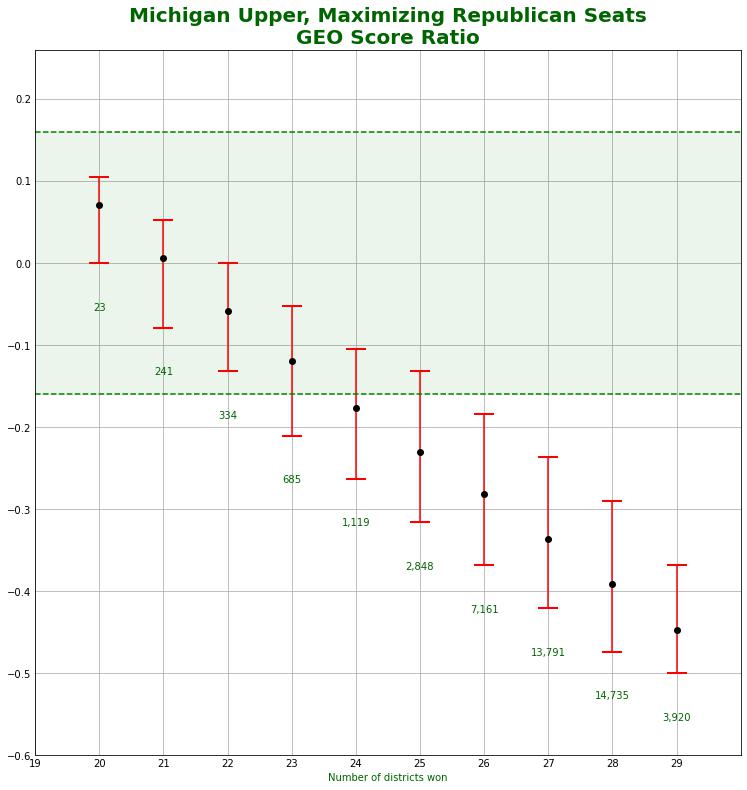}
    \includegraphics[width=0.29\linewidth]
    {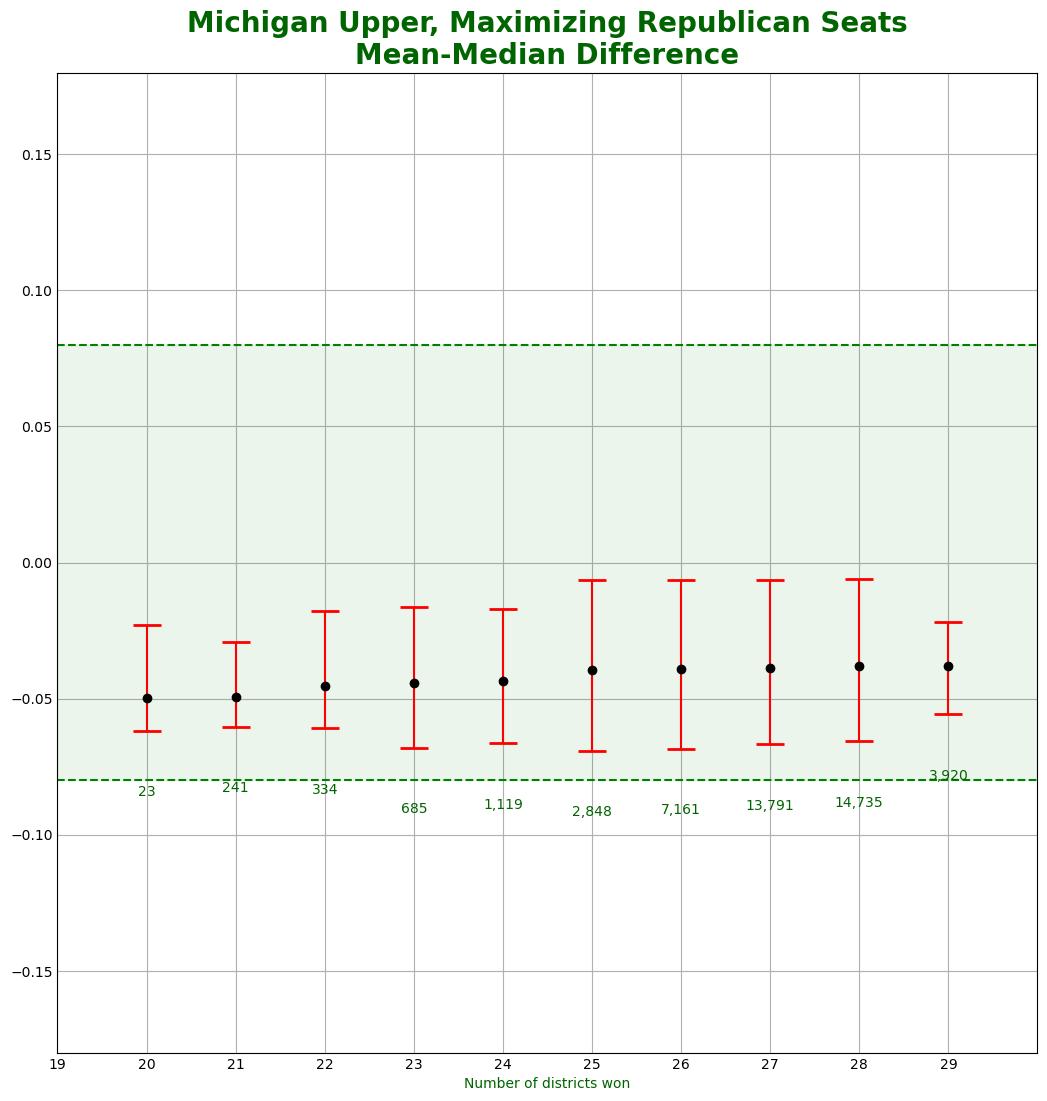}
    \caption{Results of Short Burst runs for Michigan upper house districts.}

\end{figure}

\begin{figure}[h]
    \centering
    \includegraphics[width=0.29\linewidth]{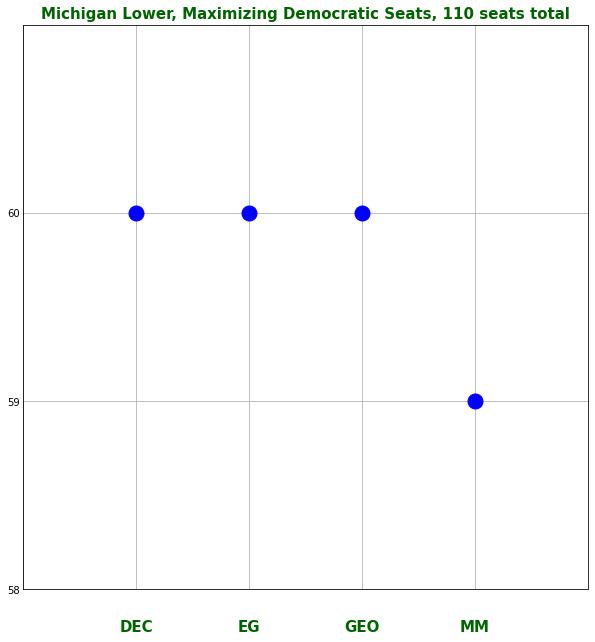}
    \includegraphics[width=0.29\linewidth]{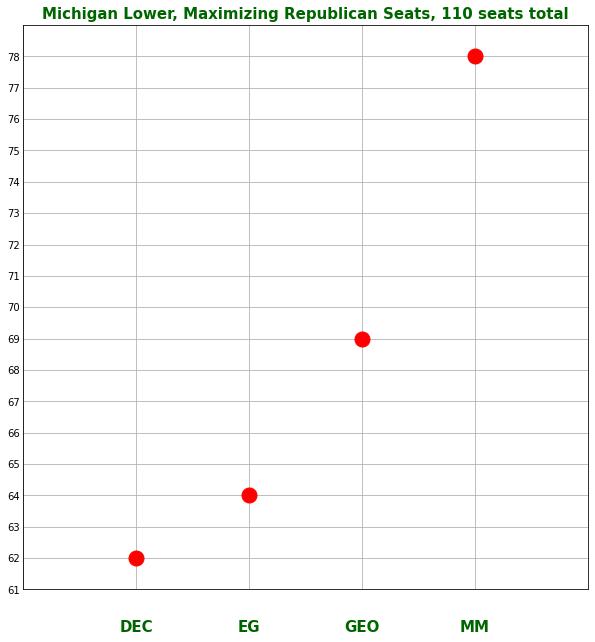}
    \newline
    \includegraphics[width=0.29\linewidth]{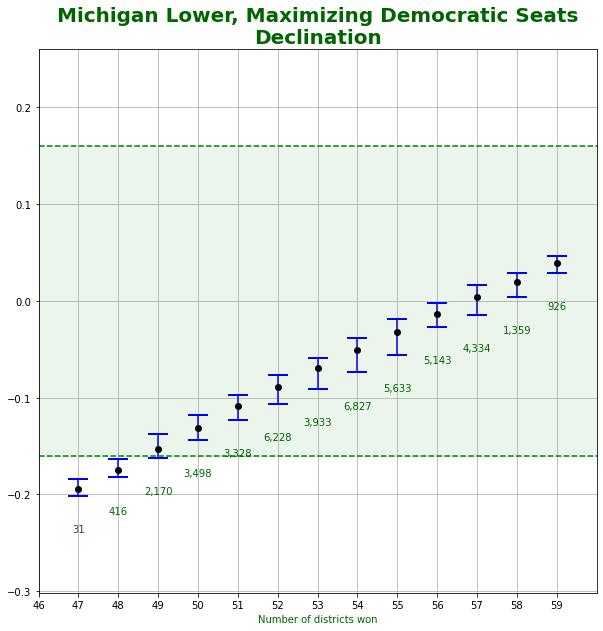}
    \includegraphics[width=0.29\linewidth]{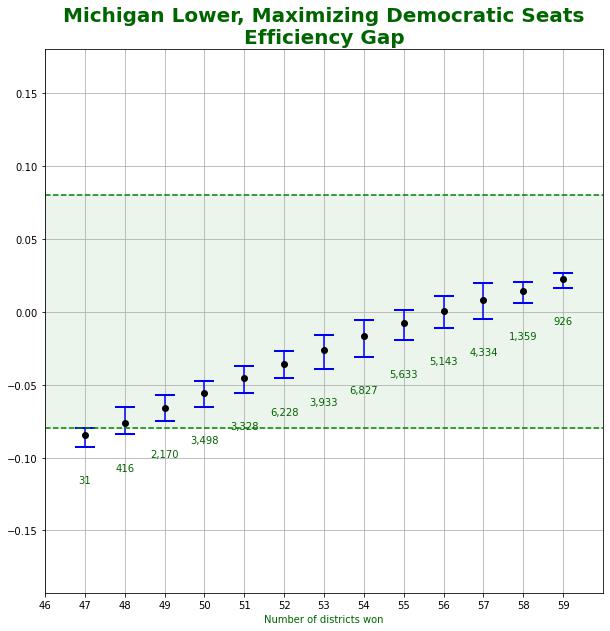}
    \includegraphics[width=0.29\linewidth]{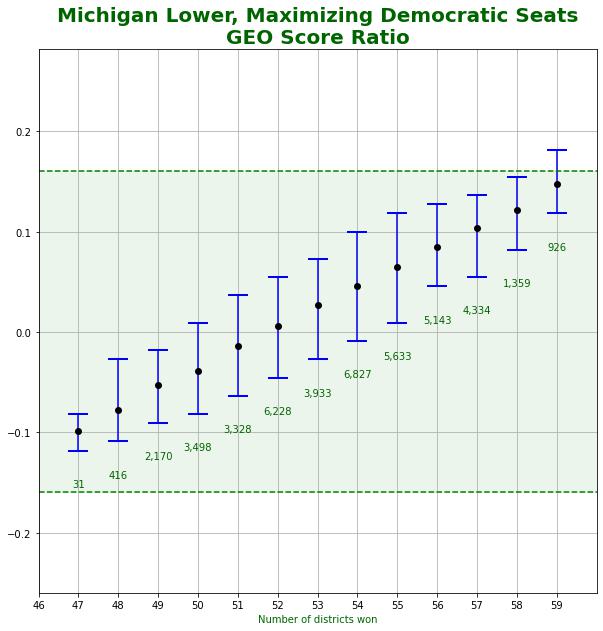}
    \includegraphics[width=0.29\linewidth]{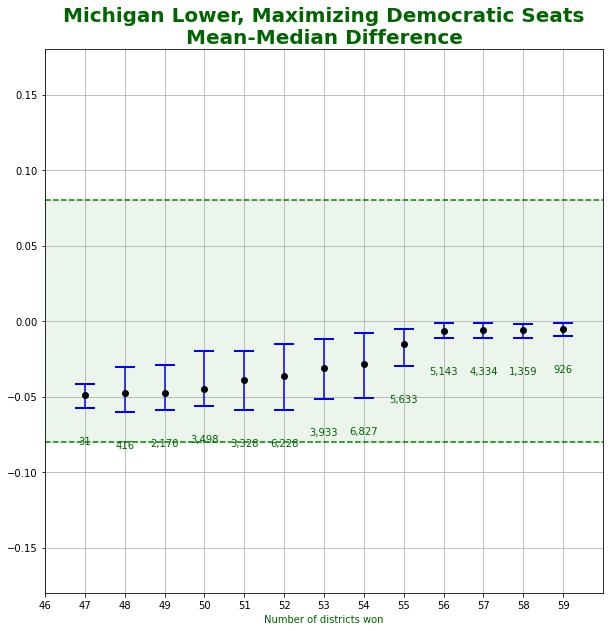}
    \includegraphics[width=0.29\linewidth]{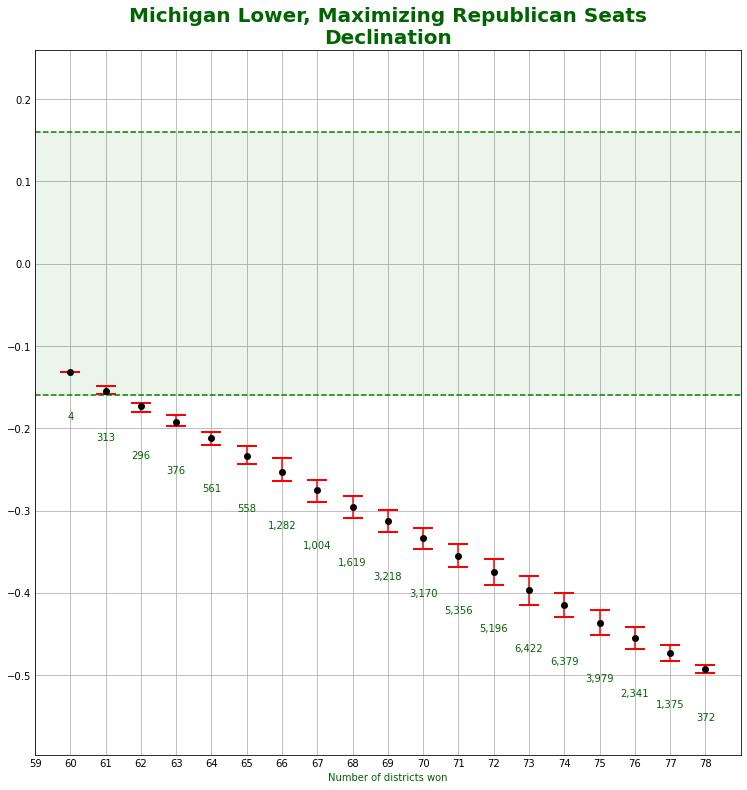}
    \includegraphics[width=0.29\linewidth]{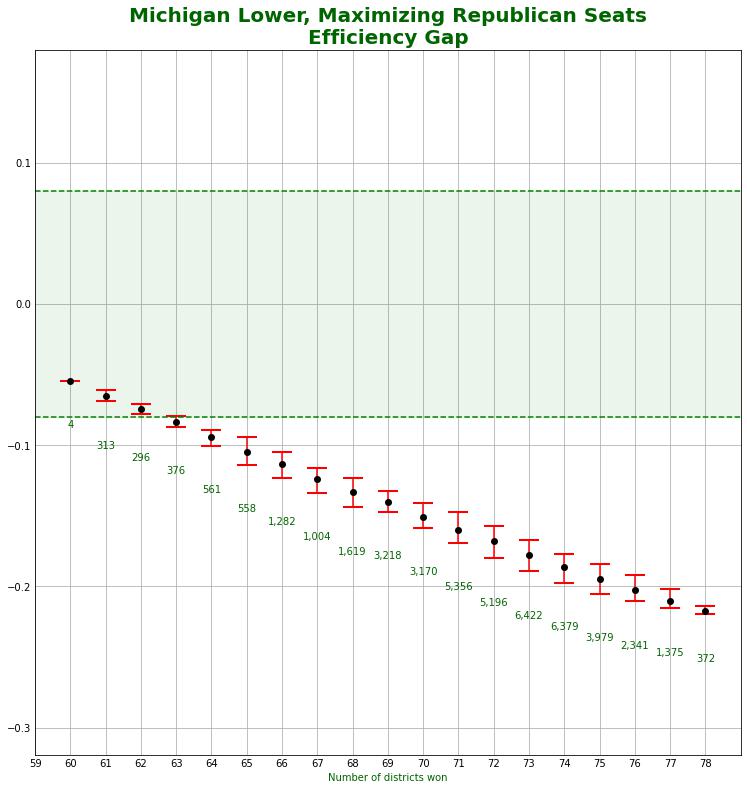}
    \includegraphics[width=0.29\linewidth]{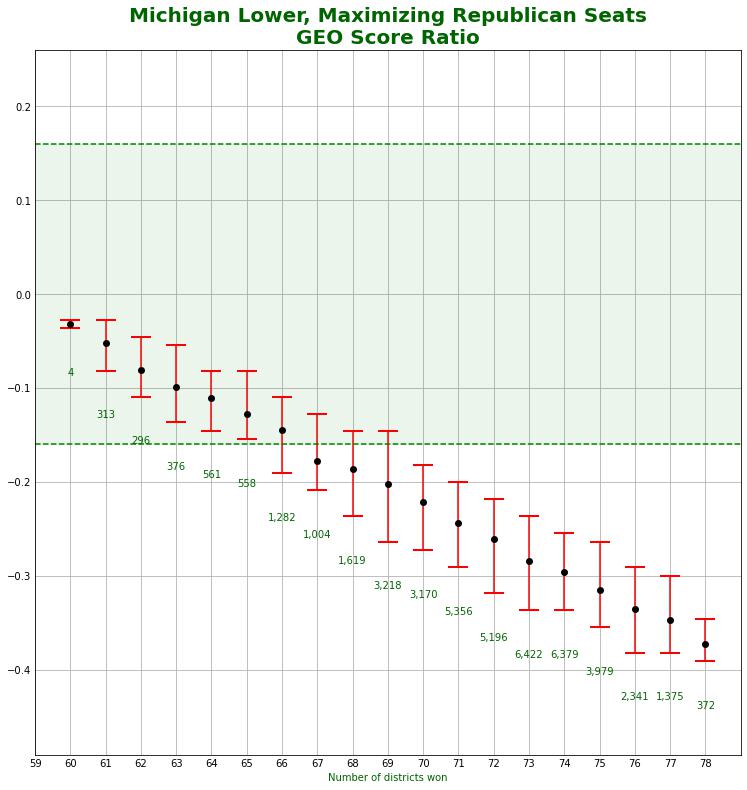}
    \includegraphics[width=0.29\linewidth]{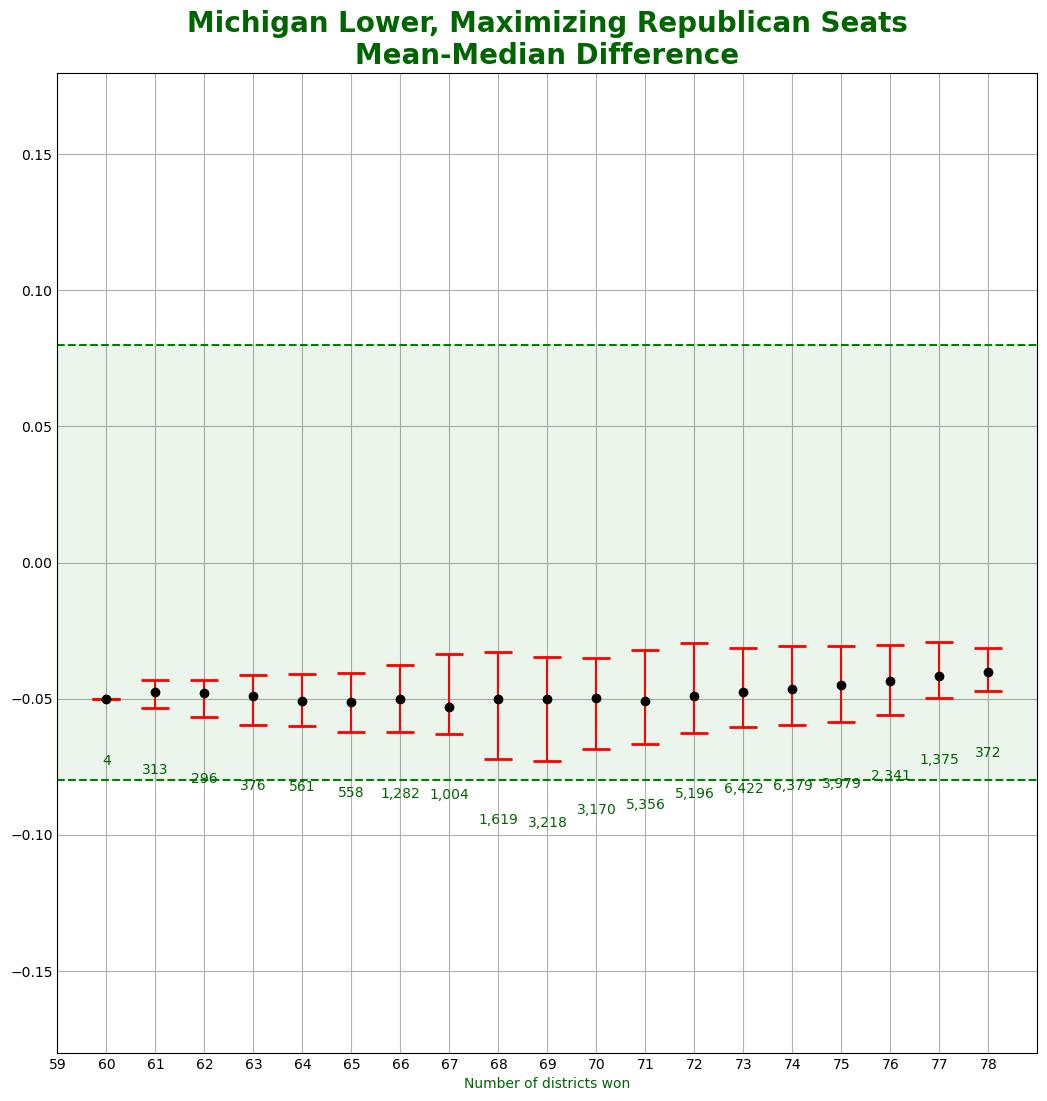}
    \caption{Results of Short Burst runs for Michigan lower house districts.}

    \label{fig:results_MIlower_short_bursts_all}
\end{figure}

\begin{figure}[h]
    \centering
    \includegraphics[width=0.29\linewidth]{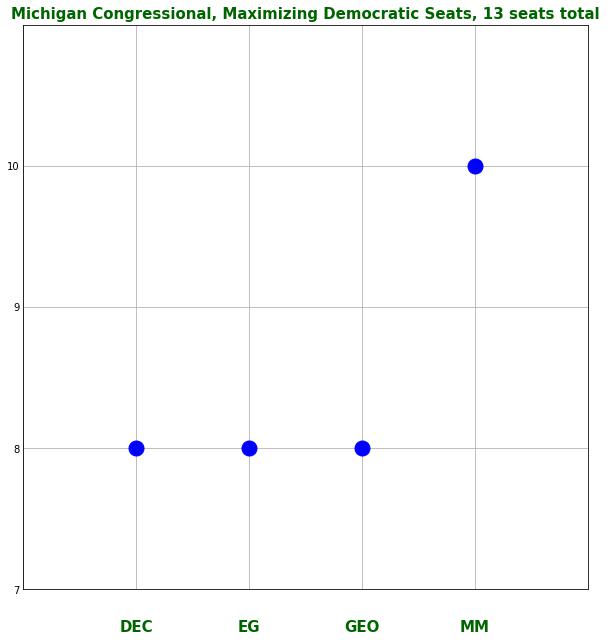}
    \includegraphics[width=0.29\linewidth]{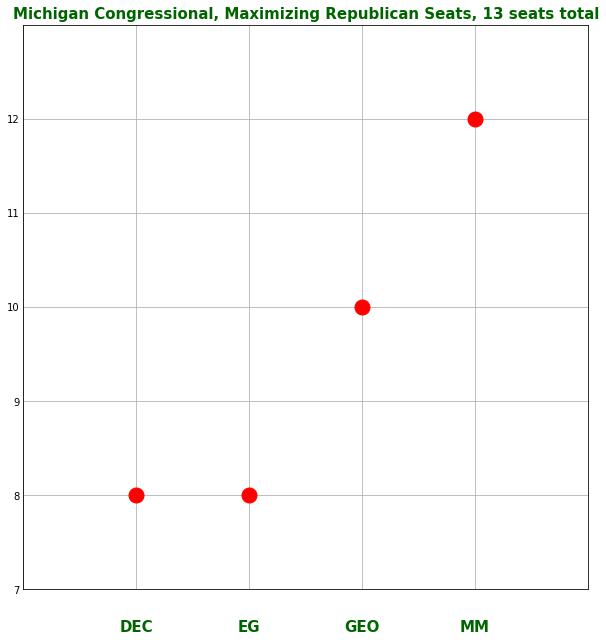}
    \newline
    \includegraphics[width=0.29\linewidth]{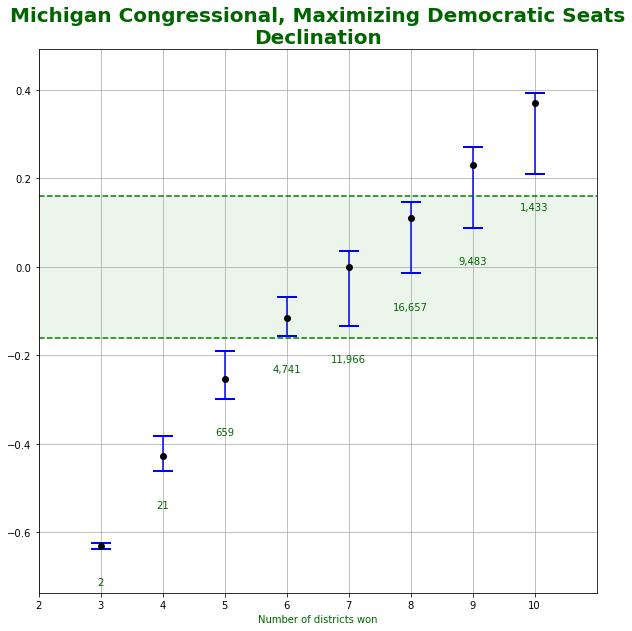}
    \includegraphics[width=0.29\linewidth]{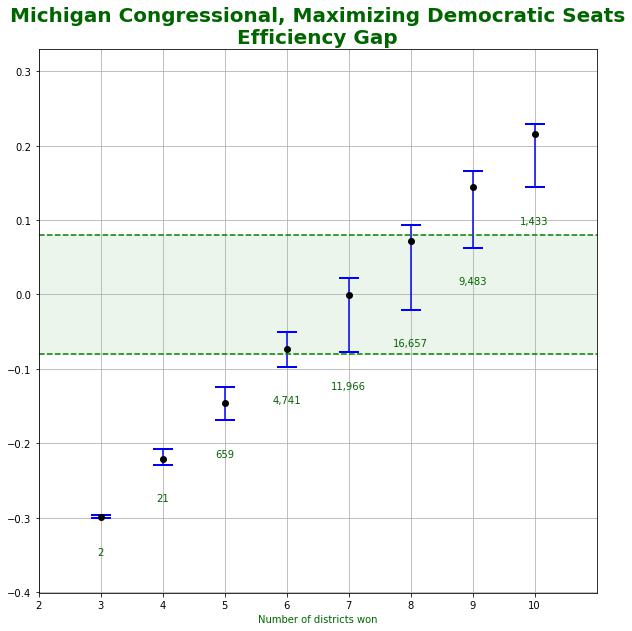}
    \includegraphics[width=0.29\linewidth]{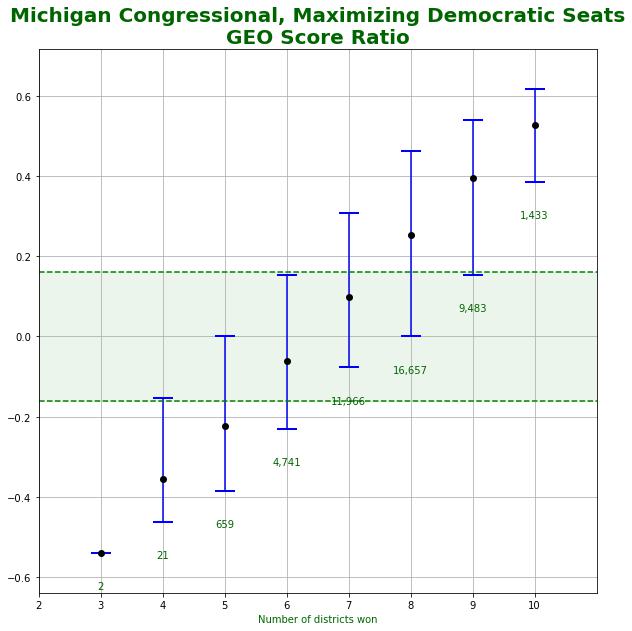}
    \includegraphics[width=0.29\linewidth]{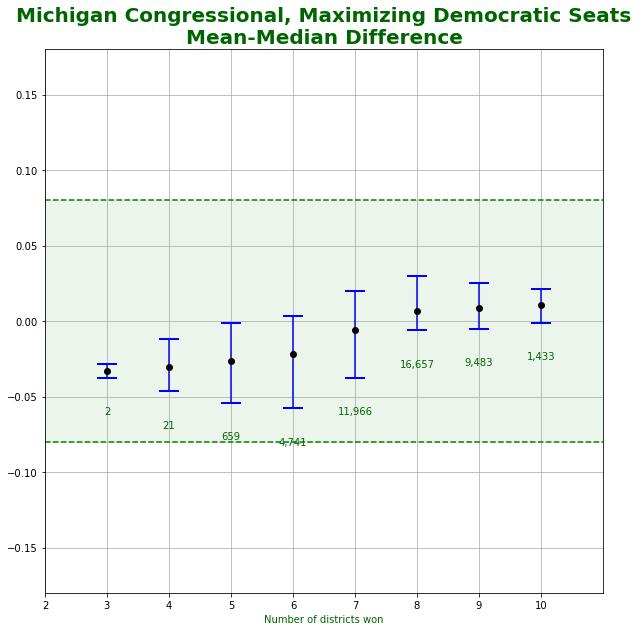}
    \includegraphics[width=0.29\linewidth]{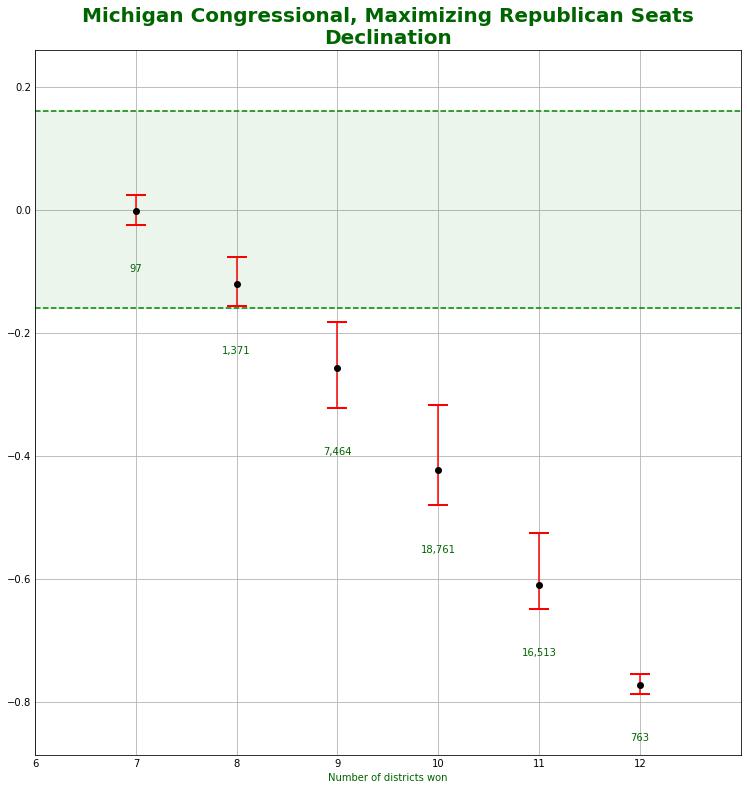}
    \includegraphics[width=0.29\linewidth]{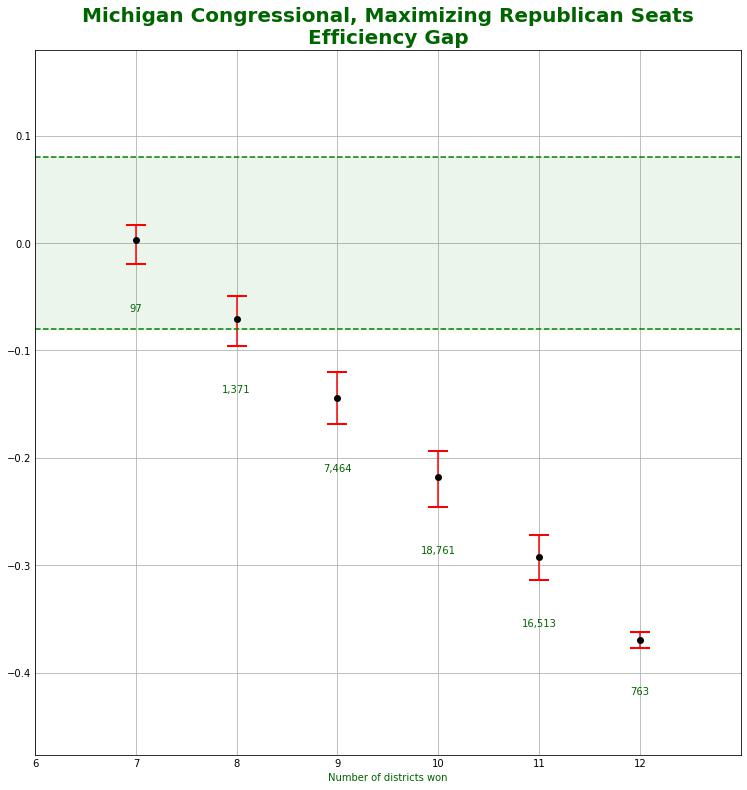}
    \includegraphics[width=0.29\linewidth]{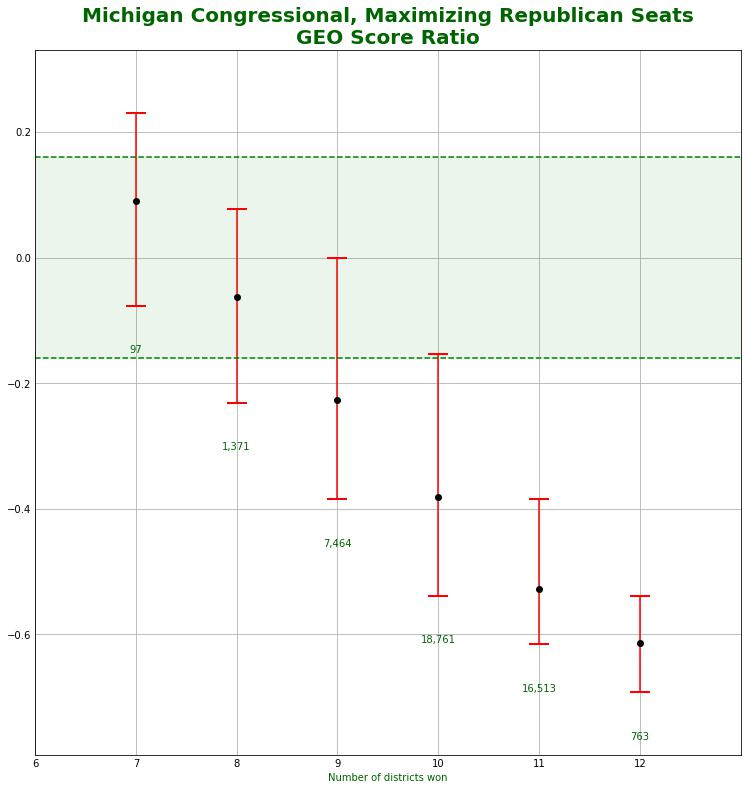}
    \includegraphics[width=0.29\linewidth]{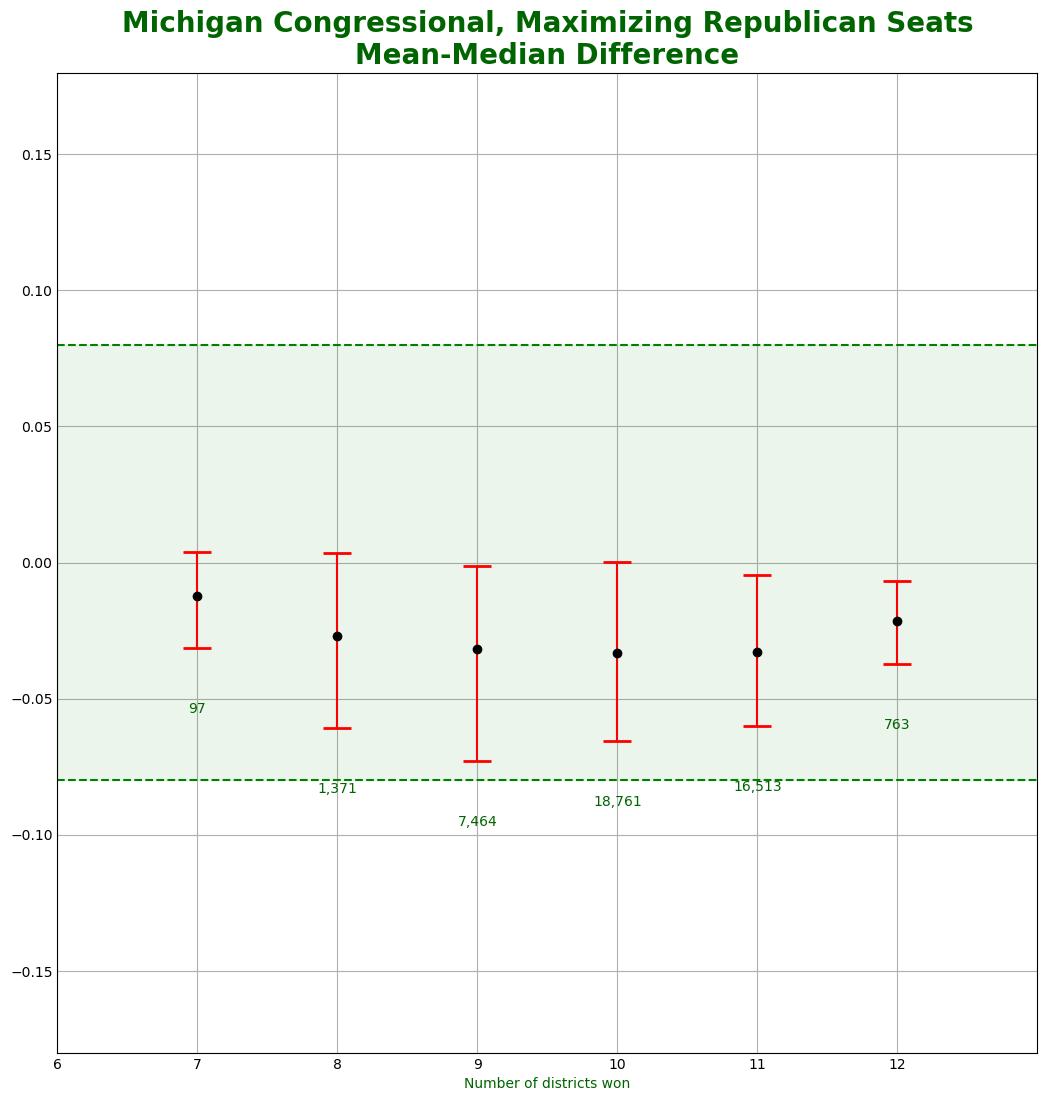}
    \caption{Results of Short Burst runs for Michigan Congressional districts.}

    \label{fig:results_MIcong_short_bursts_all}
\end{figure}

\begin{figure}[h]
    \centering
    \includegraphics[width=0.29\linewidth]{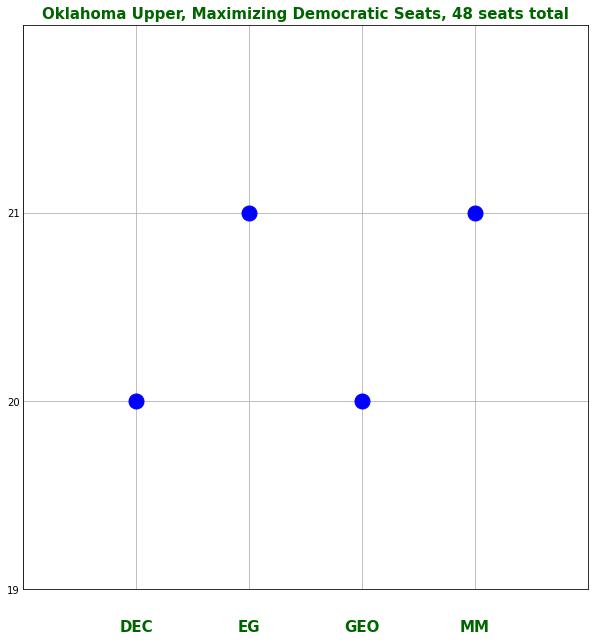}
    \includegraphics[width=0.29\linewidth]{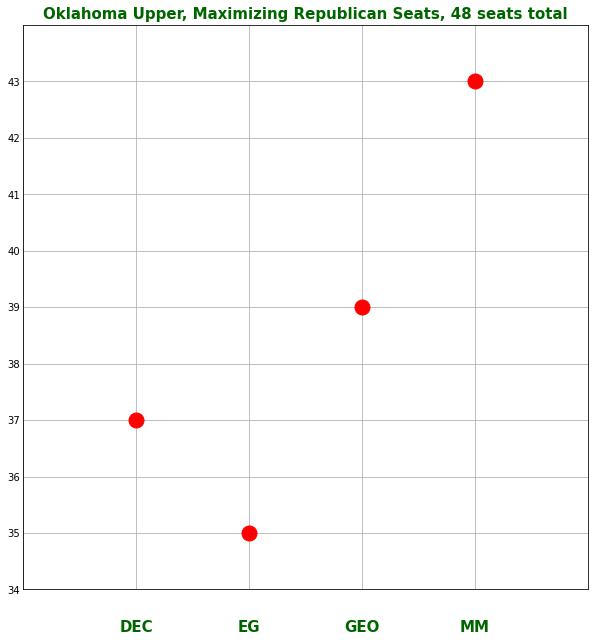}
    \newline
    \includegraphics[width=0.29\linewidth]{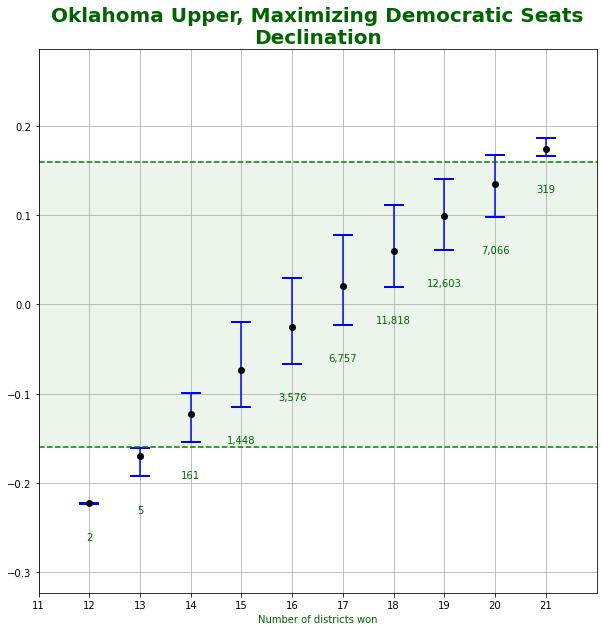}
    \includegraphics[width=0.29\linewidth]{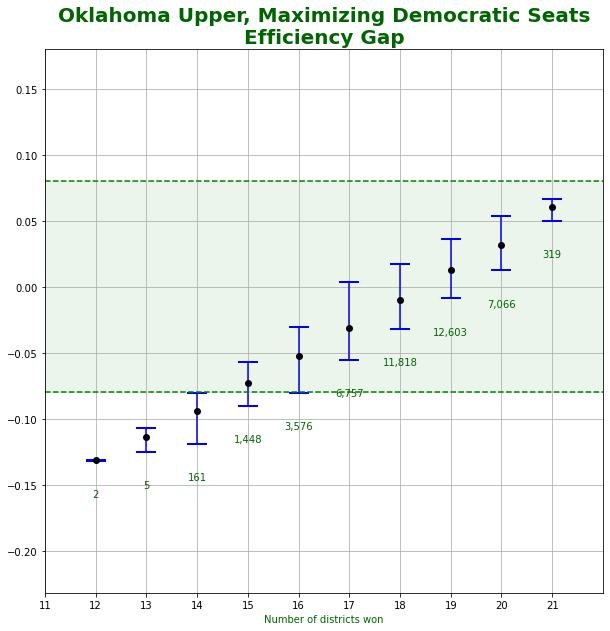}
    \includegraphics[width=0.29\linewidth]{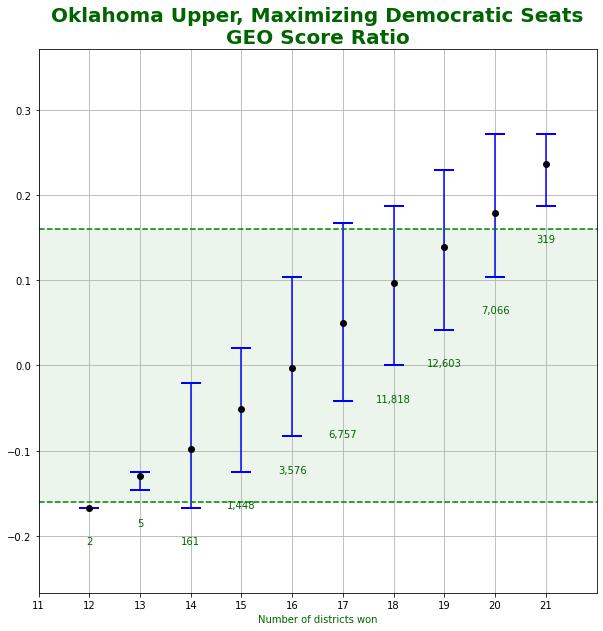}
    \includegraphics[width=0.29\linewidth]{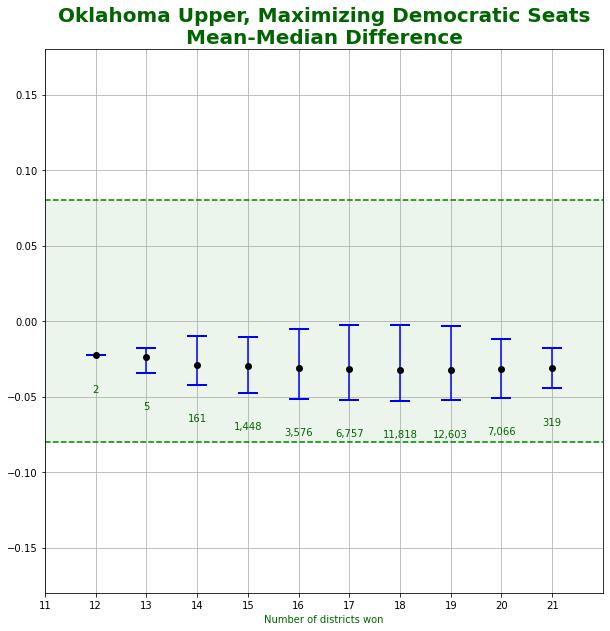}
    \includegraphics[width=0.29\linewidth]{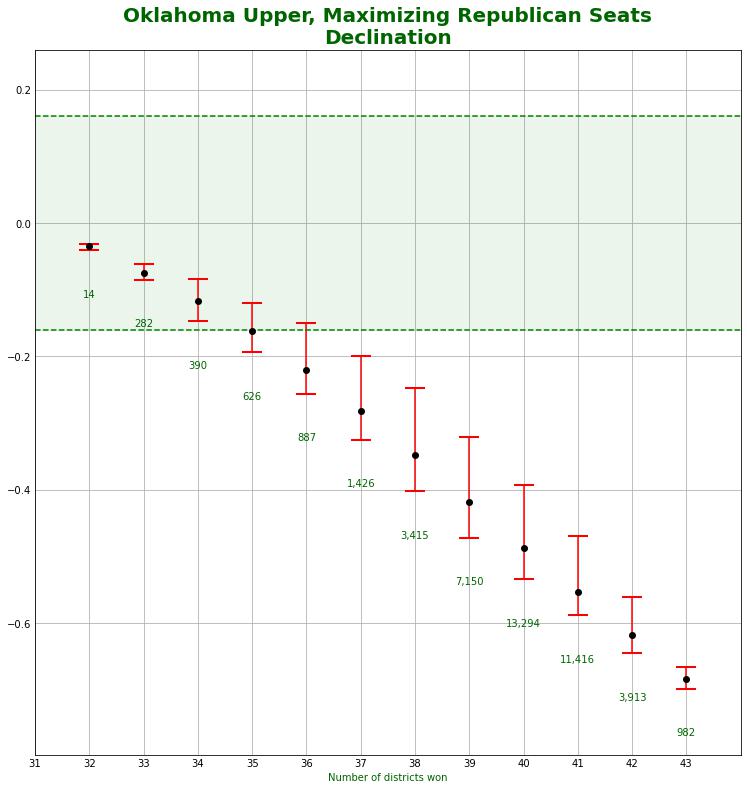}
    \includegraphics[width=0.29\linewidth]{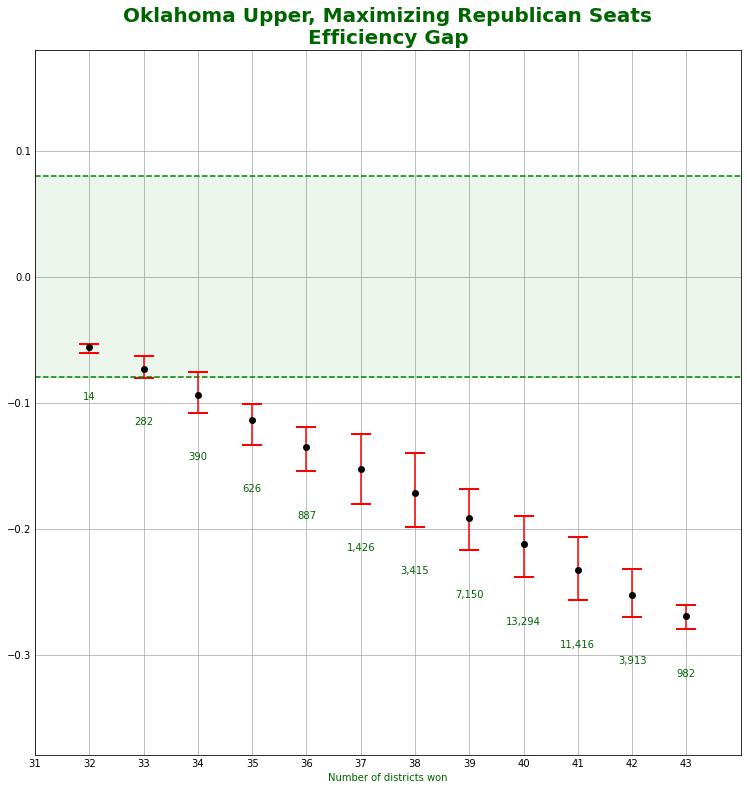}
    \includegraphics[width=0.29\linewidth]{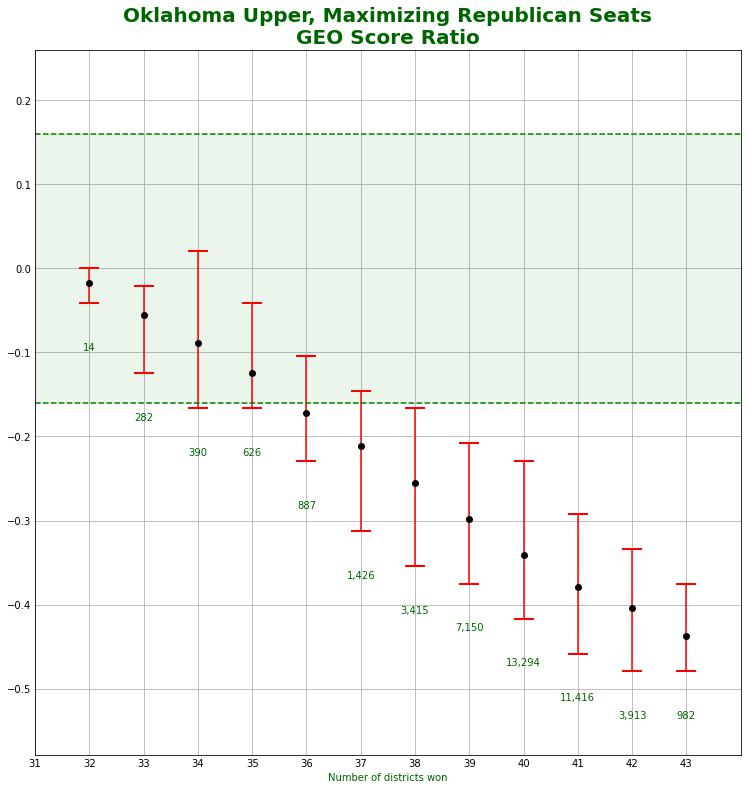}
    \includegraphics[width=0.29\linewidth]{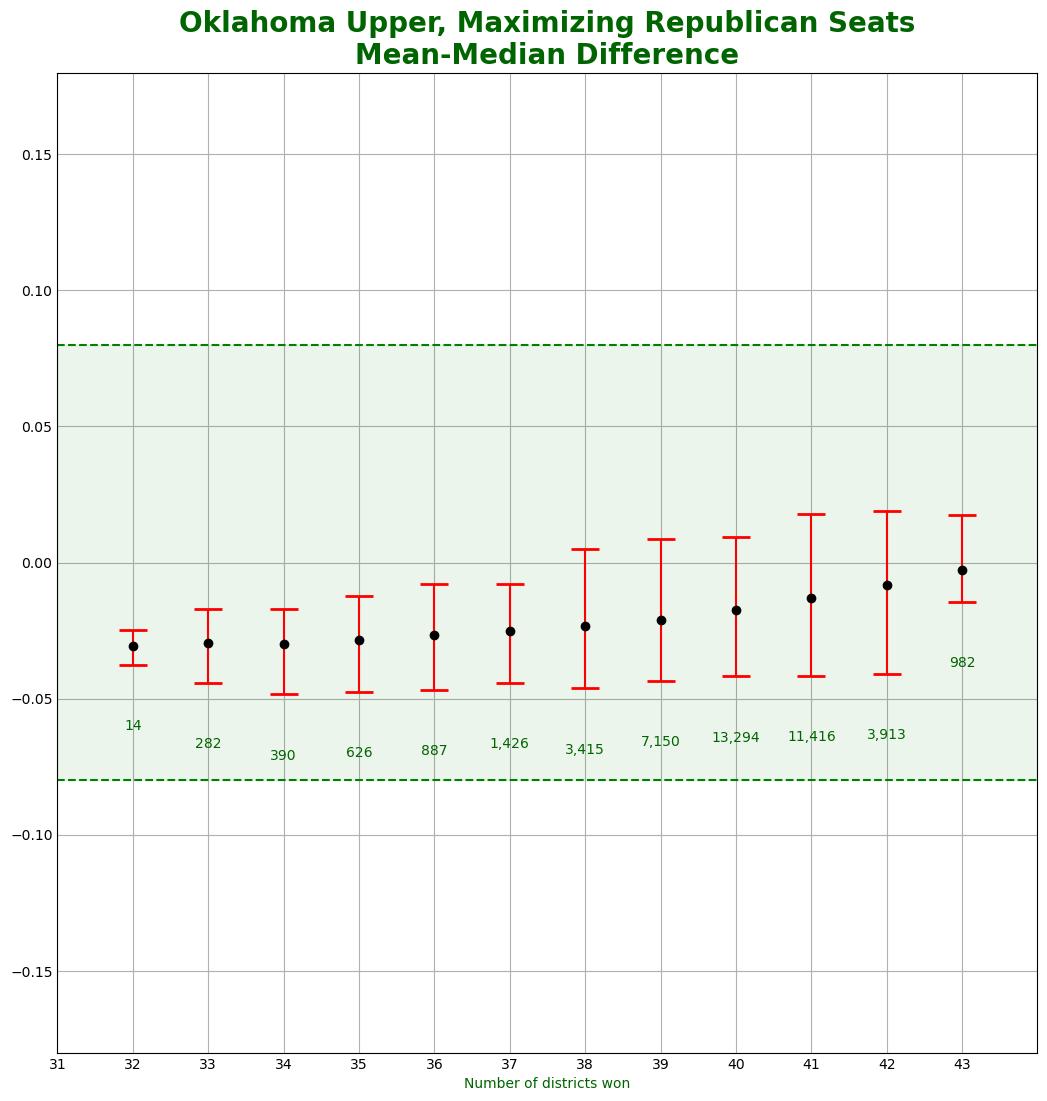}
    \caption{Results of Short Burst runs for Oklahoma upper house districts.}

\end{figure}

\begin{figure}[h]
    \centering
    \includegraphics[width=0.29\linewidth]{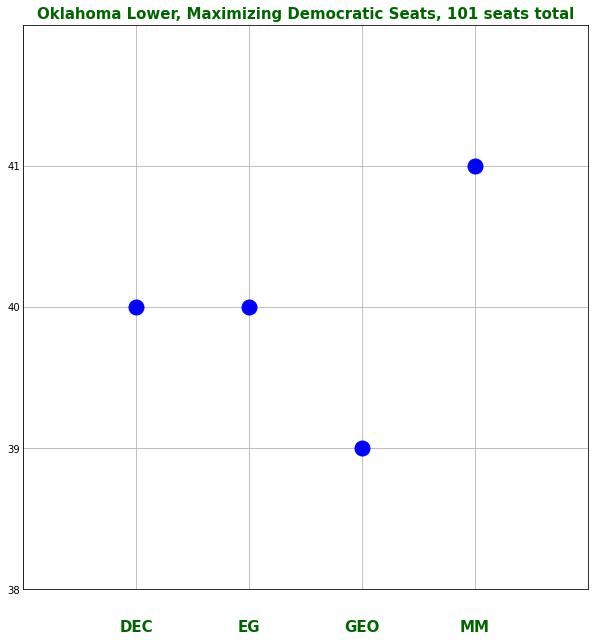}
    \includegraphics[width=0.29\linewidth]{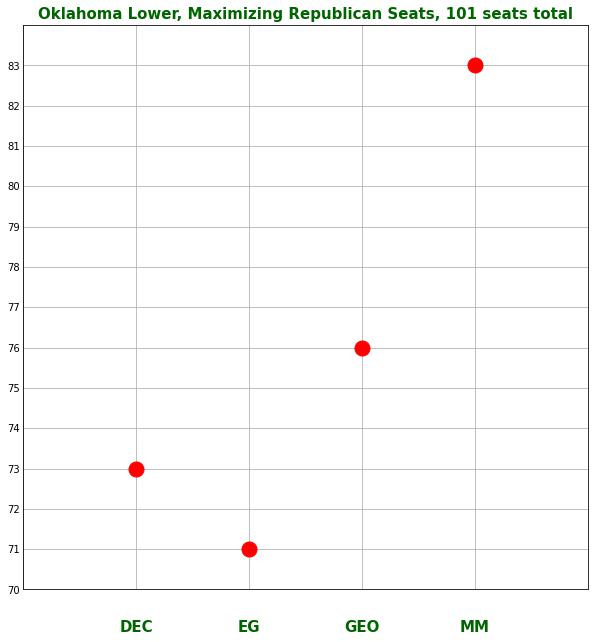}
    \newline
    \includegraphics[width=0.29\linewidth]{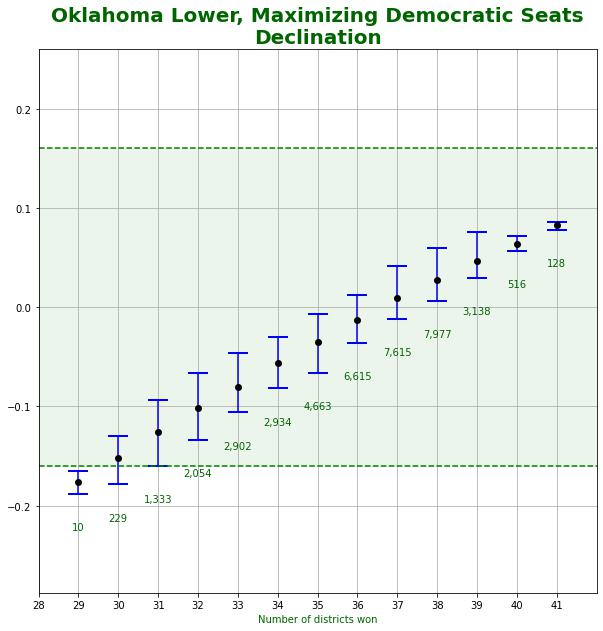}
    \includegraphics[width=0.29\linewidth]{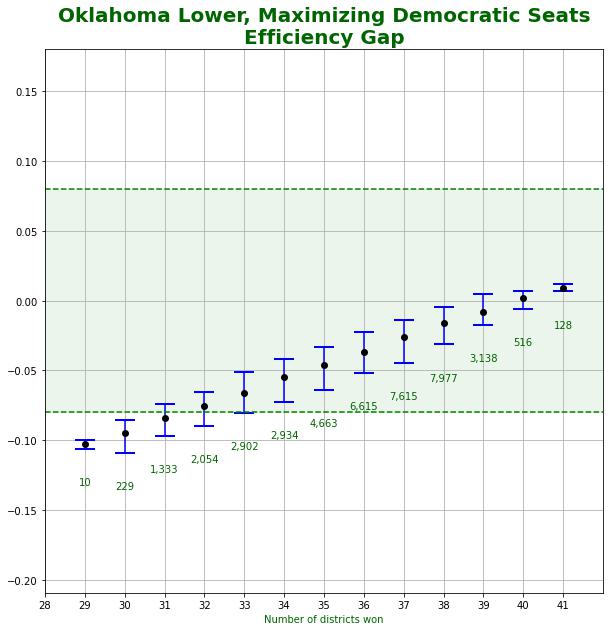}
    \includegraphics[width=0.29\linewidth]{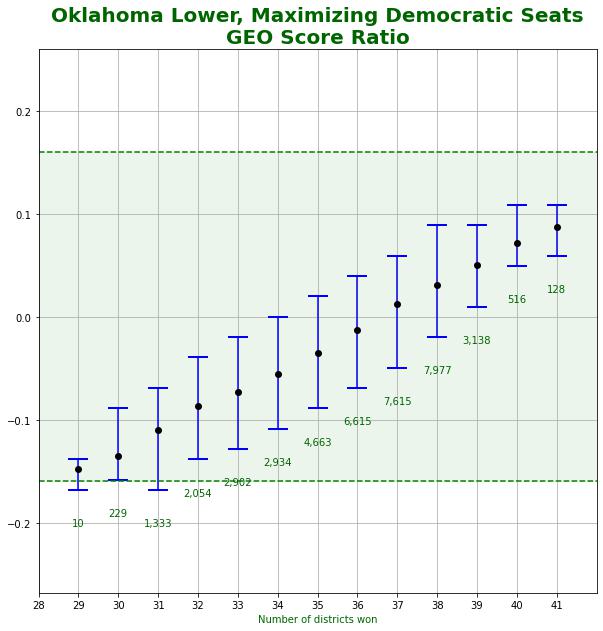}
    \includegraphics[width=0.29\linewidth]{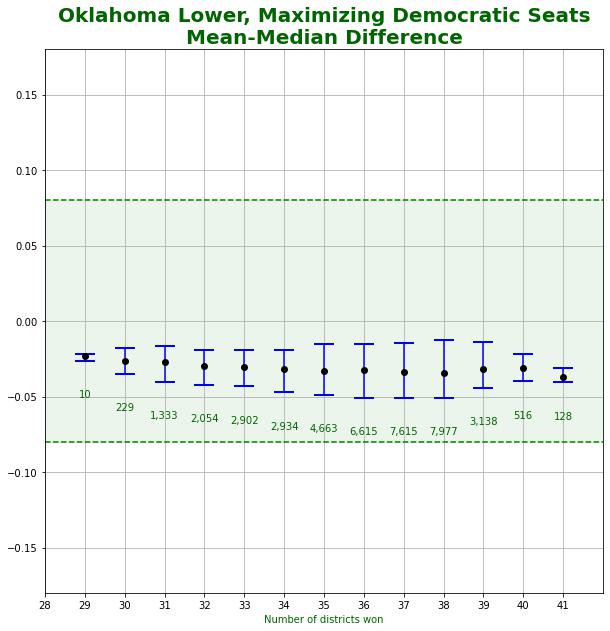}
    \includegraphics[width=0.29\linewidth]{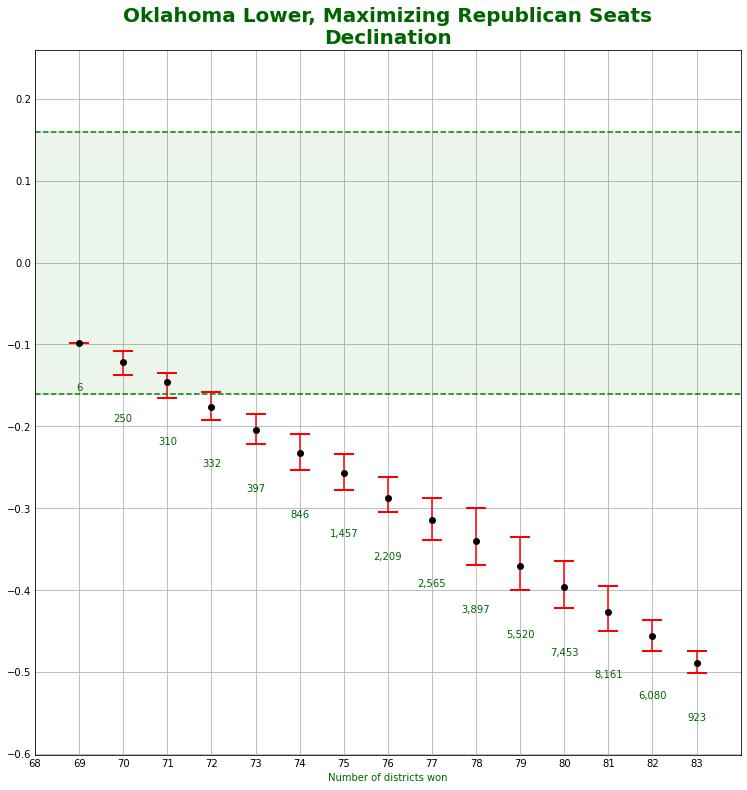}
    \includegraphics[width=0.29\linewidth]{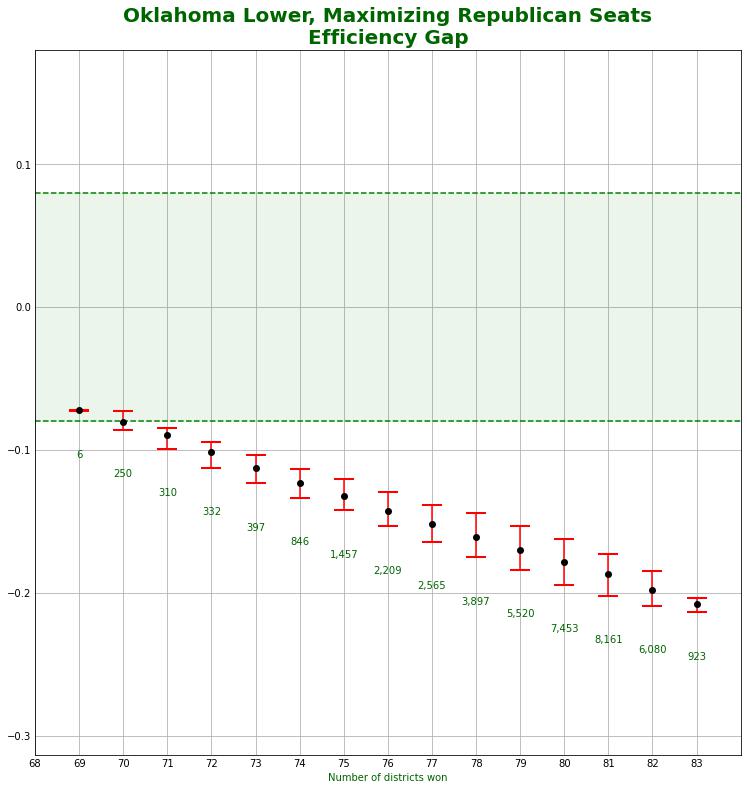}
    \includegraphics[width=0.29\linewidth]{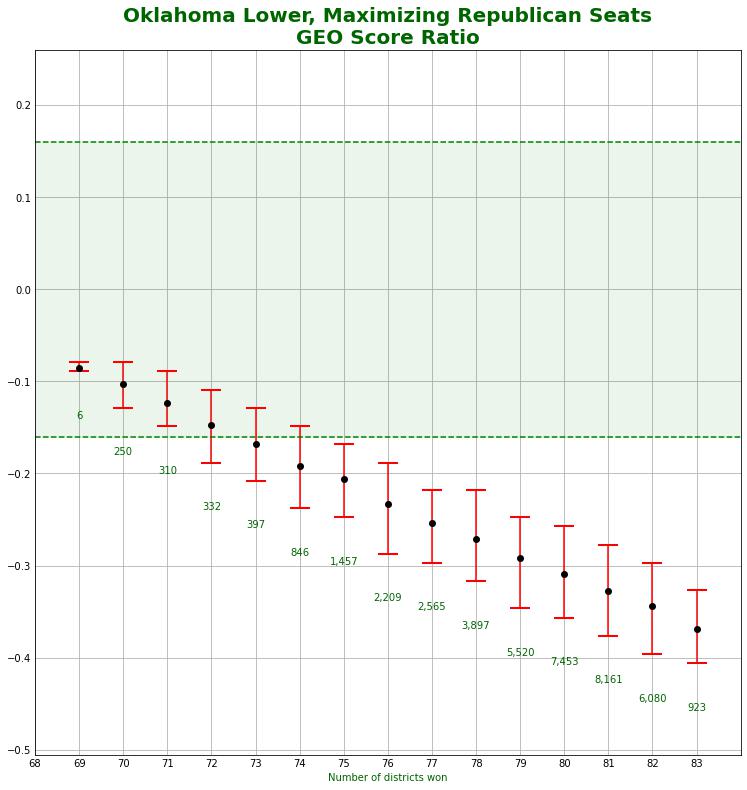}
    \includegraphics[width=0.29\linewidth]{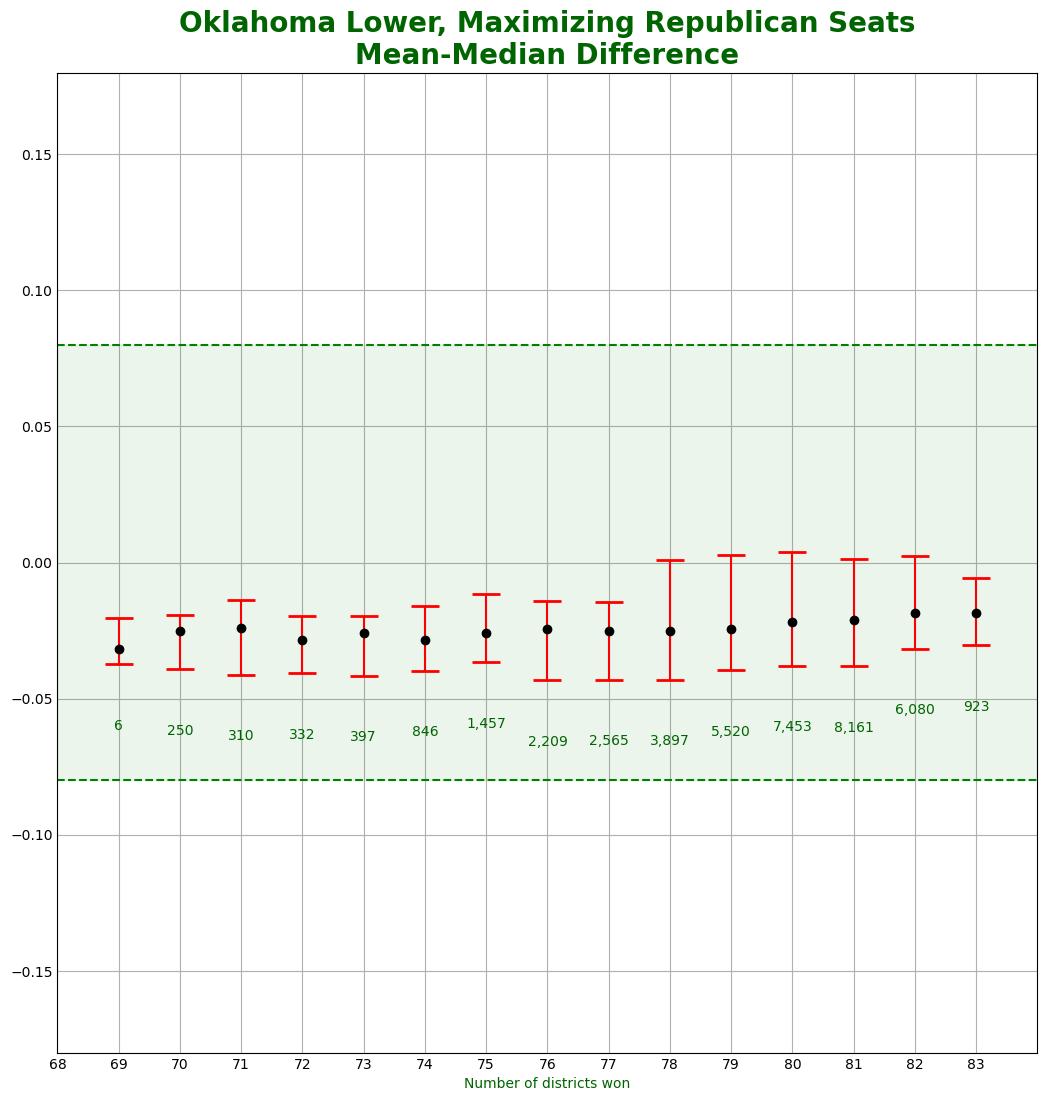}

        \caption{Results of Short Burst runs for Oklahoma lower house districts.}

    \label{fig:results_OKlower_short_bursts_all}
\end{figure}

\begin{figure}[h]
    \centering
    \includegraphics[width=0.29\linewidth]{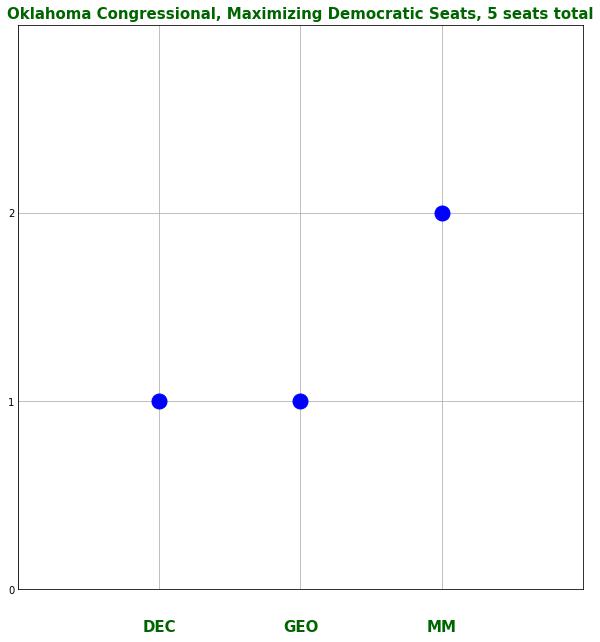}
    \includegraphics[width=0.29\linewidth]{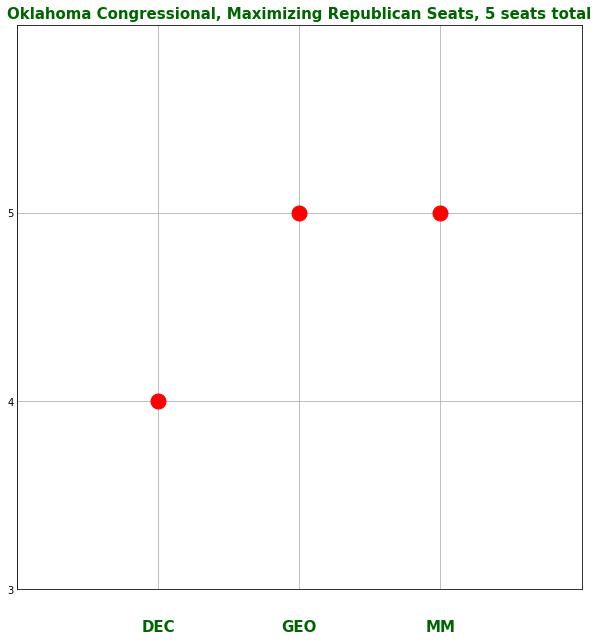}
    \newline
    \includegraphics[width=0.29\linewidth]{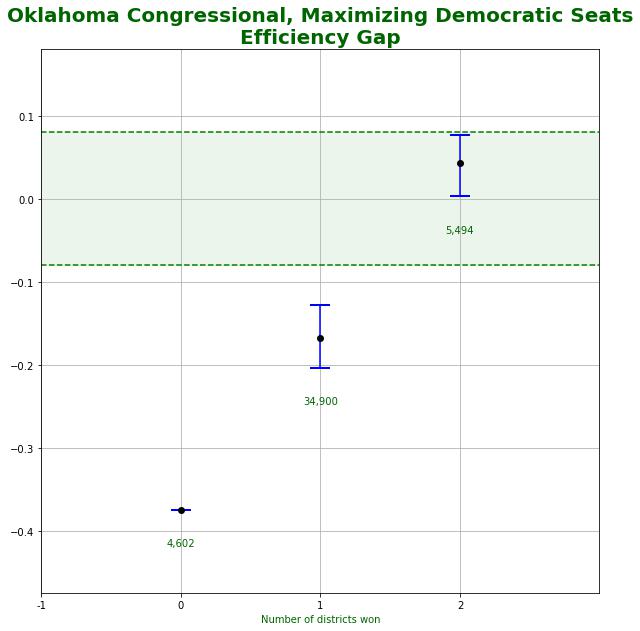}
    \includegraphics[width=0.29\linewidth]{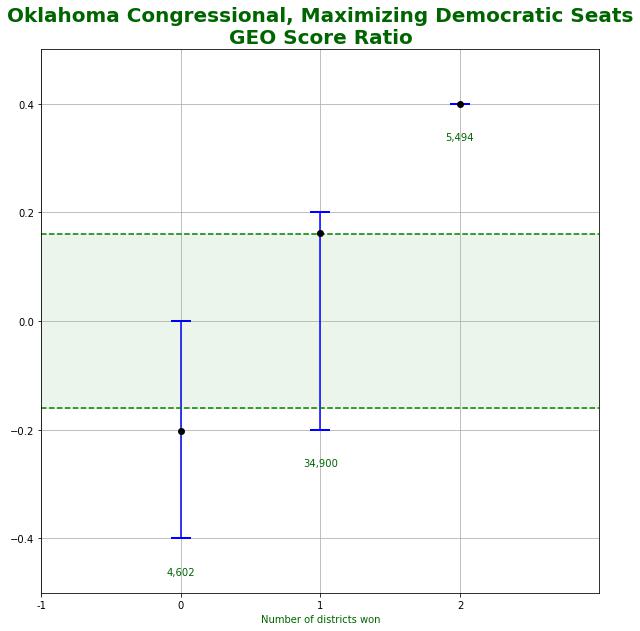}
    \includegraphics[width=0.29\linewidth]{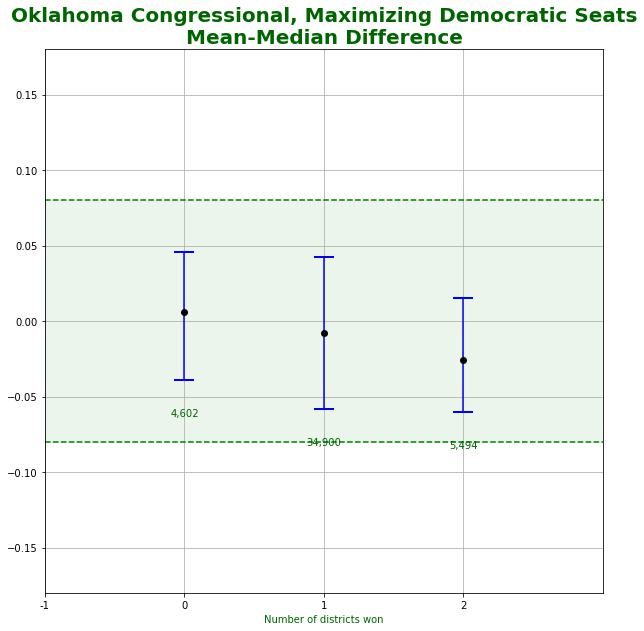}
    \includegraphics[width=0.29\linewidth]{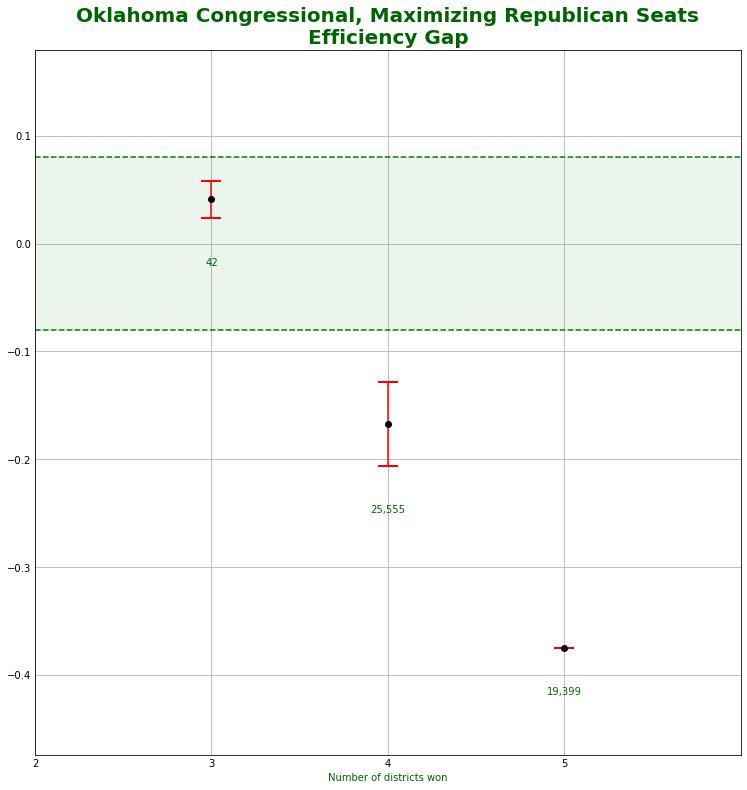}
    \includegraphics[width=0.29\linewidth]{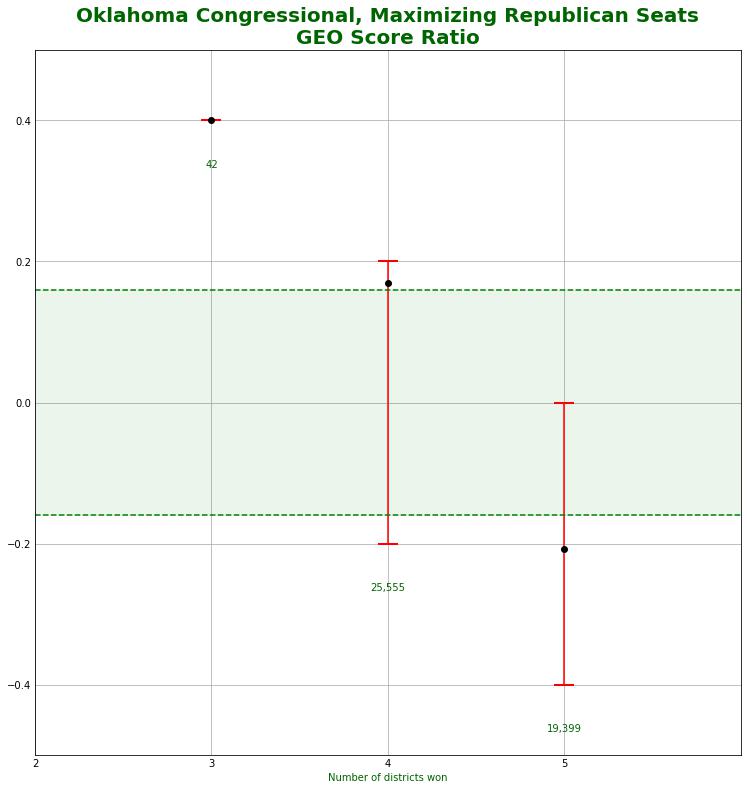}
    \includegraphics[width=0.29\linewidth]{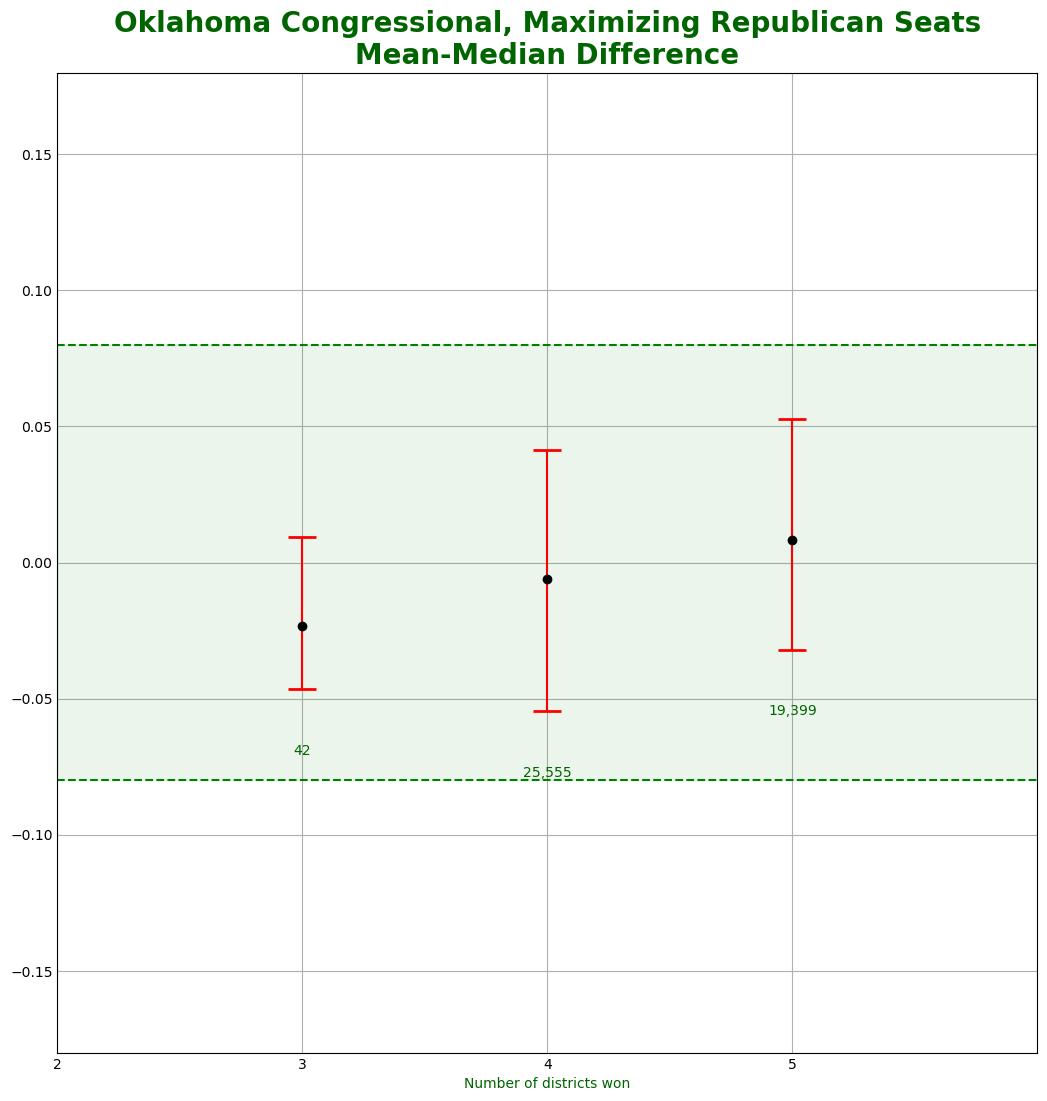}
    \caption{Results of Short Burst runs for Oklahoma Congressional districts.}

    \label{fig:results_OKcong_short_bursts_all}
\end{figure}

\begin{figure}[h]
    \centering
    \includegraphics[width=0.29\linewidth]{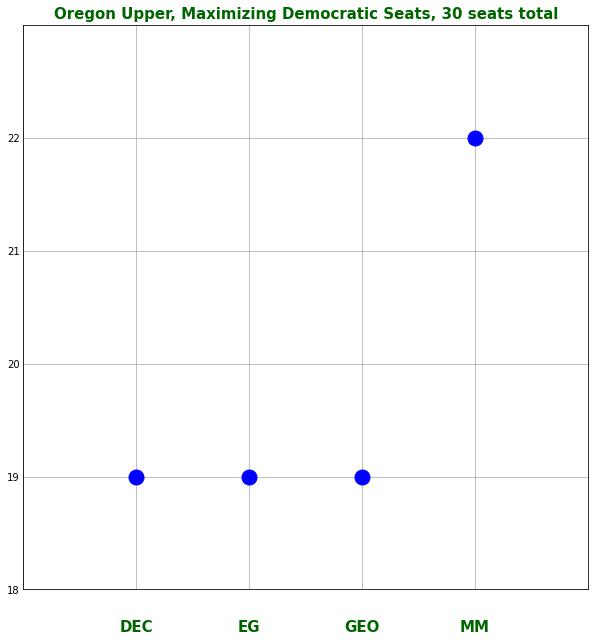}
    \includegraphics[width=0.29\linewidth]{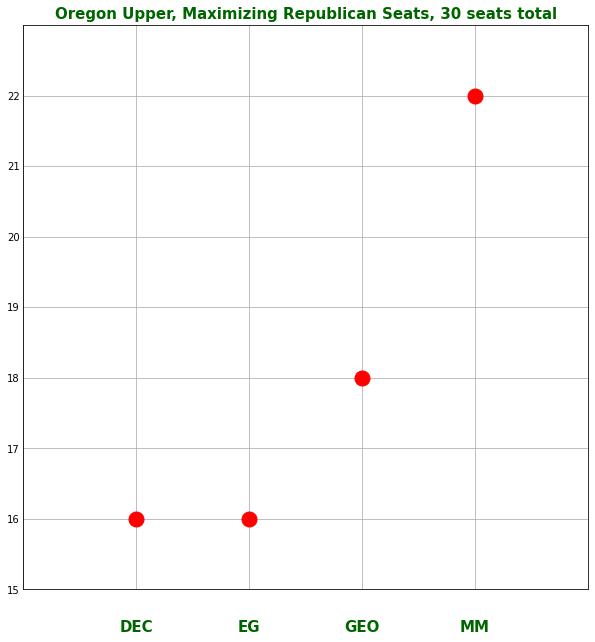}
    \newline
    \includegraphics[width=0.29\linewidth]{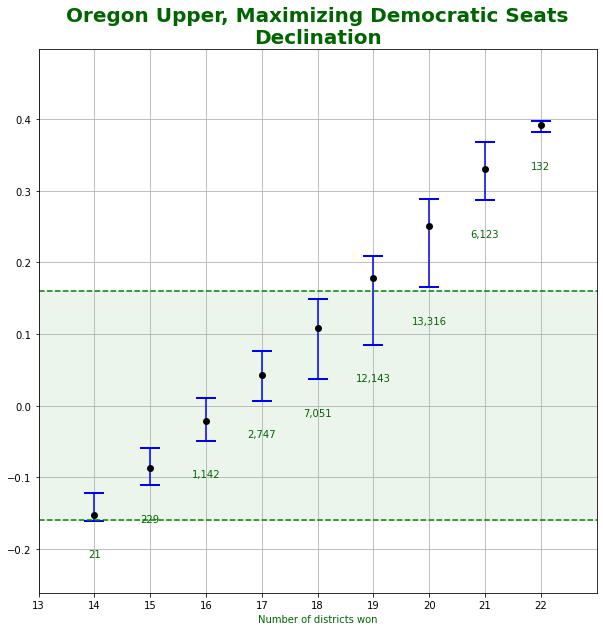}
    \includegraphics[width=0.29\linewidth]{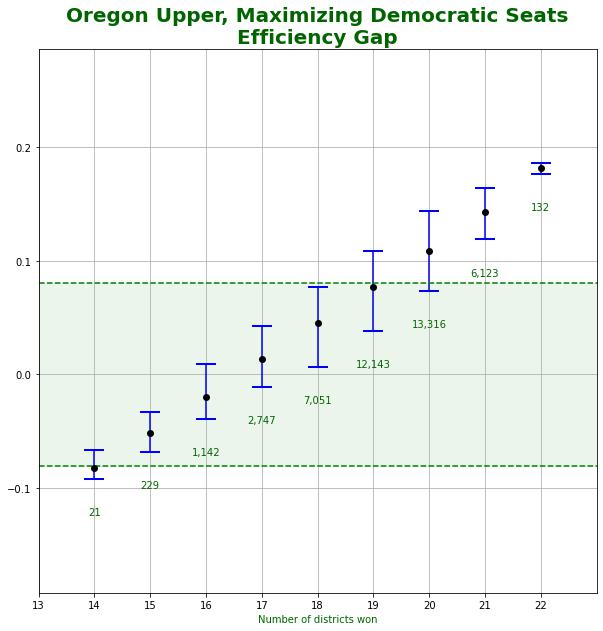}
    \includegraphics[width=0.29\linewidth]{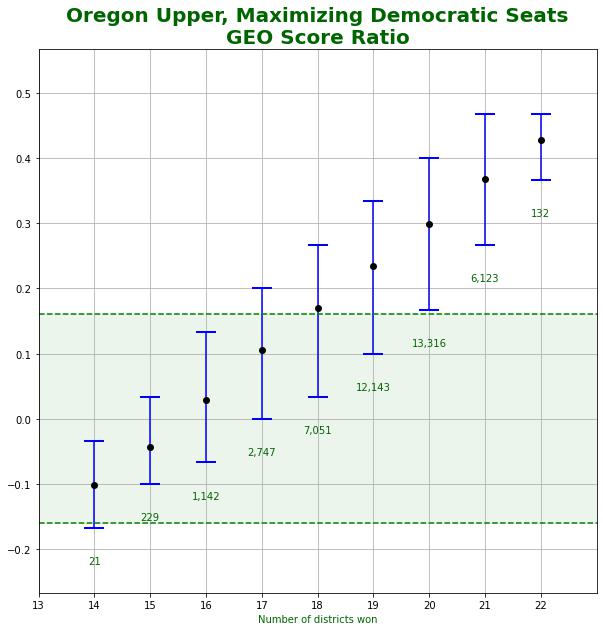}
    \includegraphics[width=0.29\linewidth]{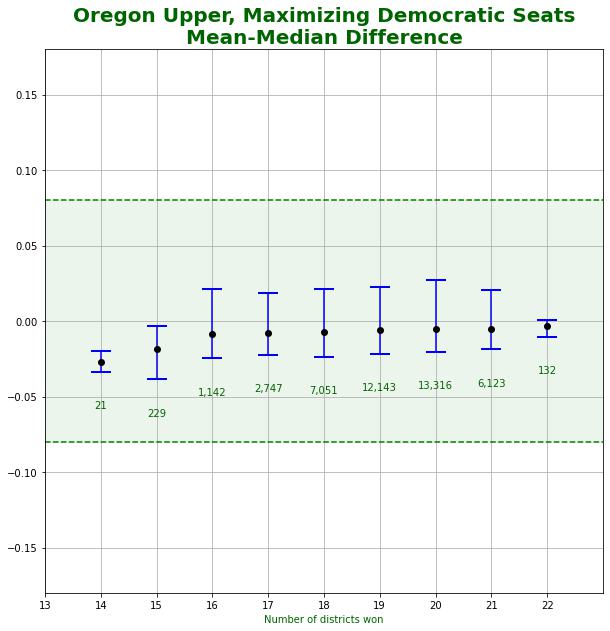}
    \includegraphics[width=0.29\linewidth]{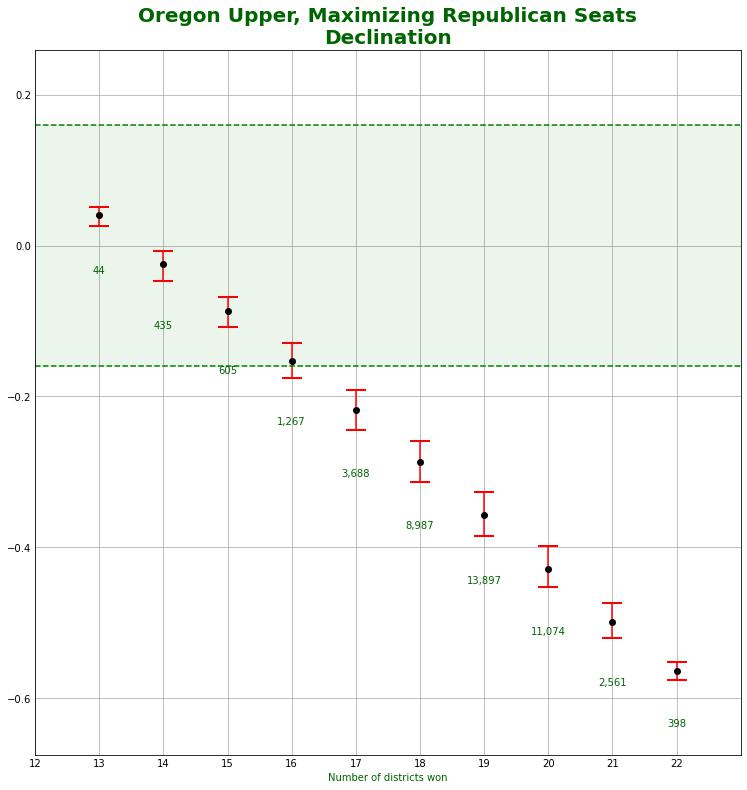}
    \includegraphics[width=0.29\linewidth]{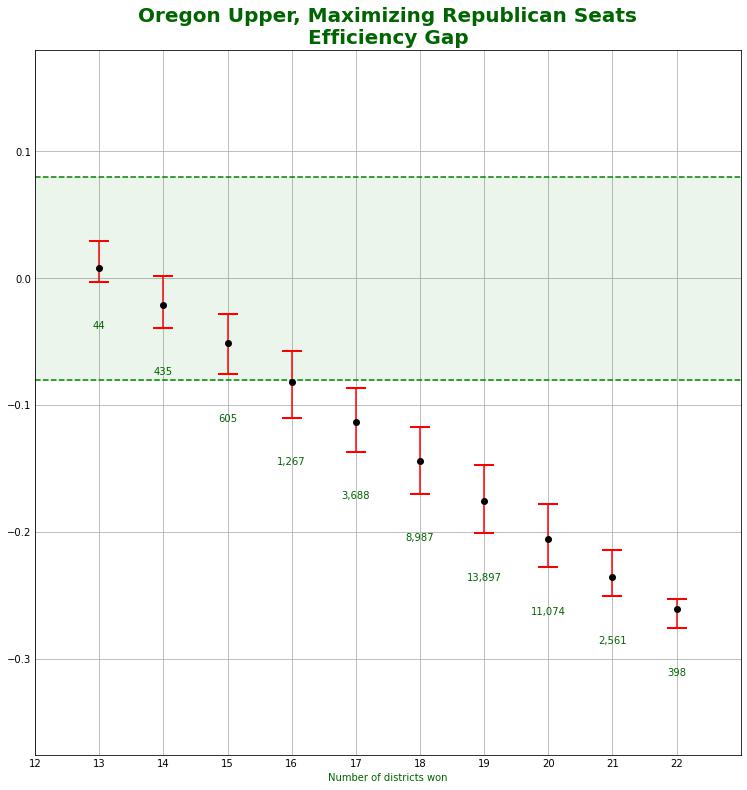}
    \includegraphics[width=0.29\linewidth]{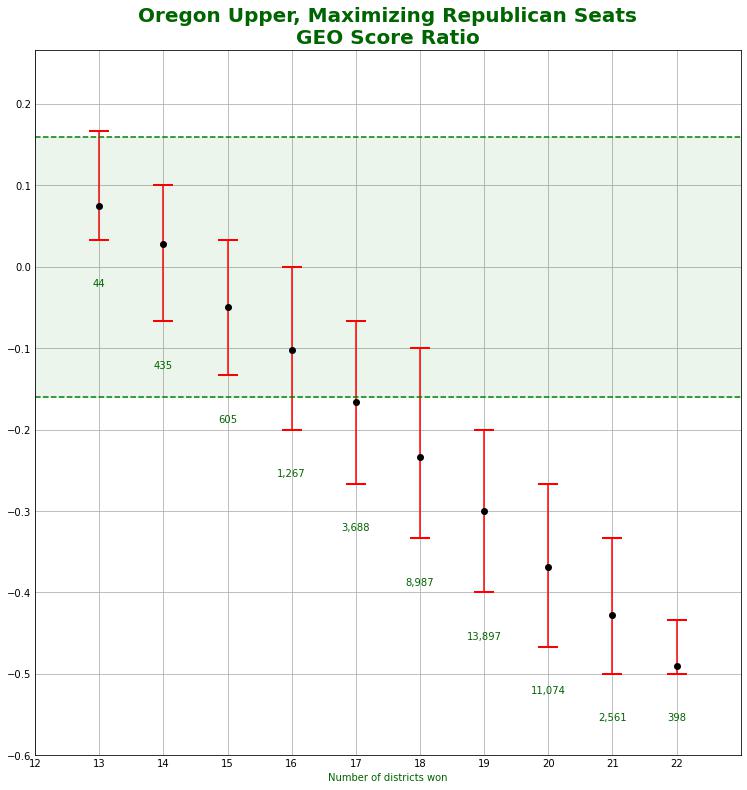}
    \includegraphics[width=0.29\linewidth]{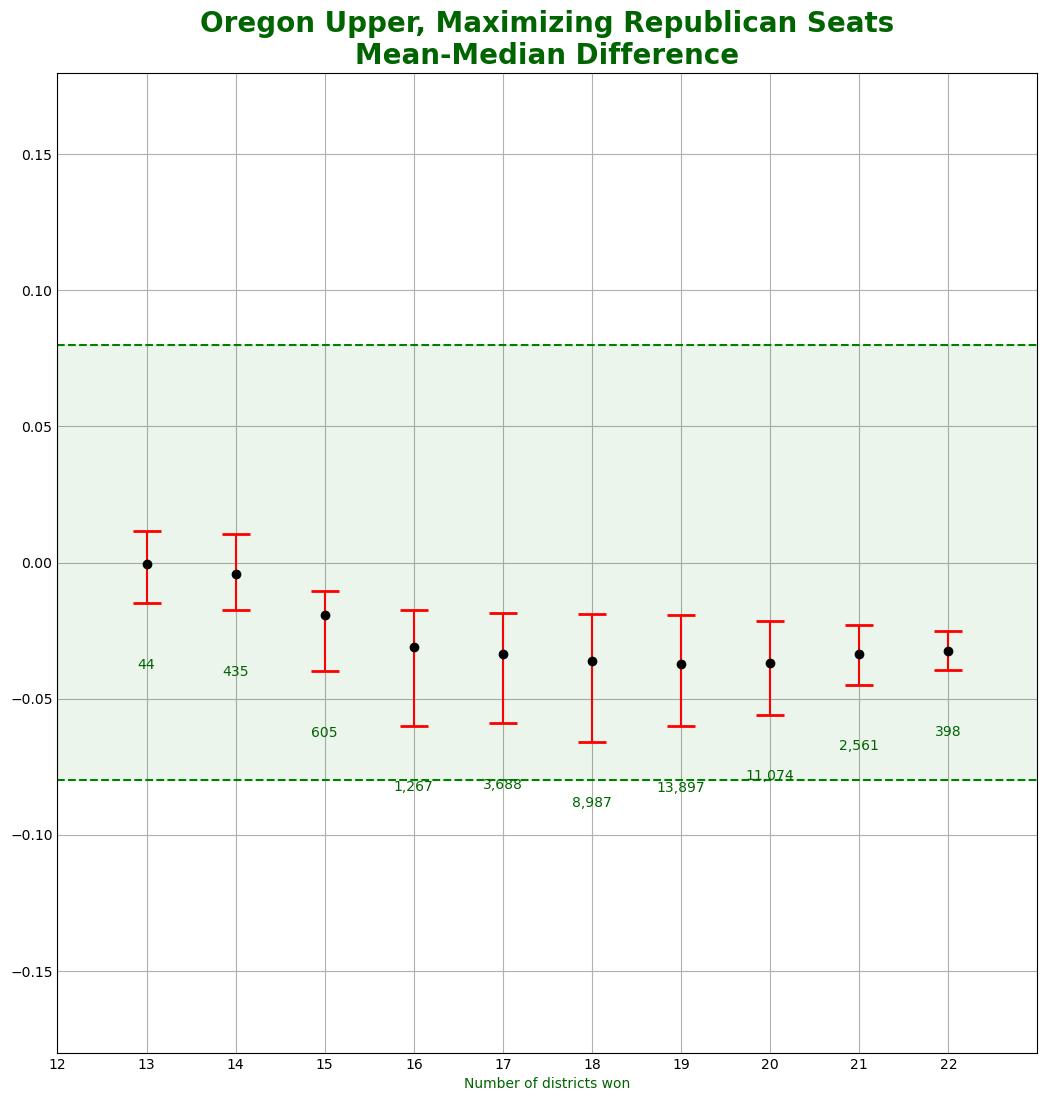}
    \caption{Results of Short Burst runs for Oregon upper house districts.}

\end{figure}

\begin{figure}[h]
    \centering
    \includegraphics[width=0.29\linewidth]{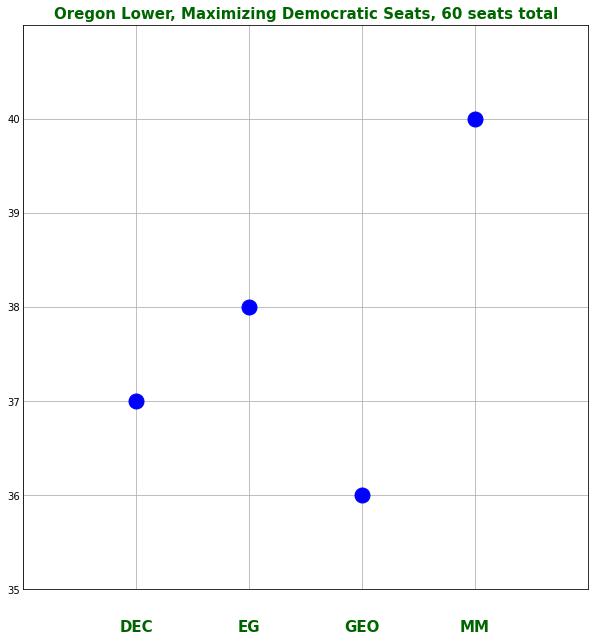}
    \includegraphics[width=0.29\linewidth]{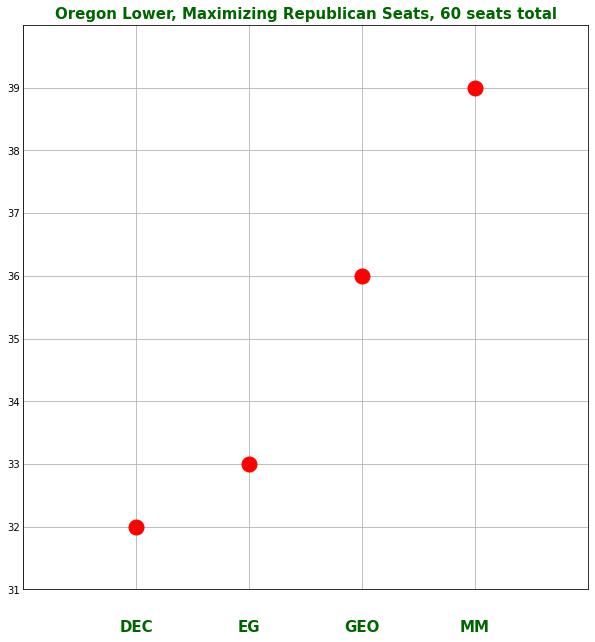}
    \newline
    \includegraphics[width=0.29\linewidth]{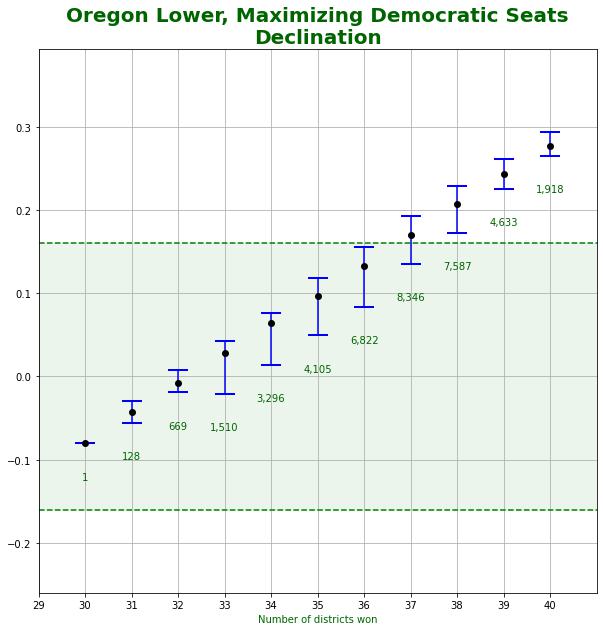}
    \includegraphics[width=0.29\linewidth]{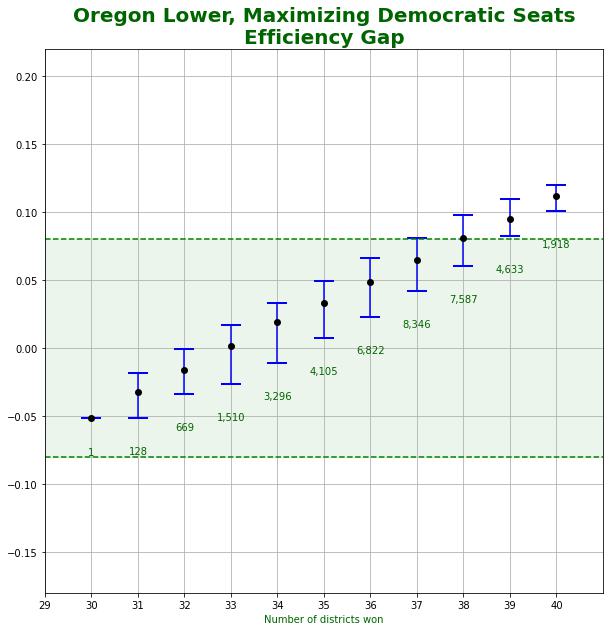}
    \includegraphics[width=0.29\linewidth]{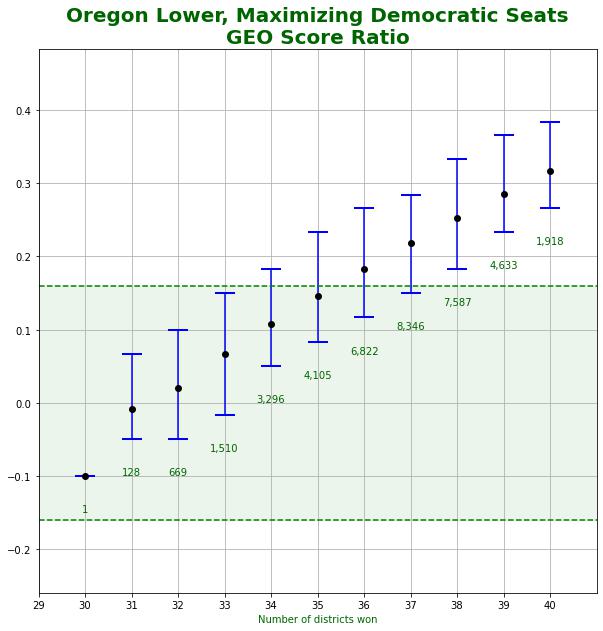}
    \includegraphics[width=0.29\linewidth]{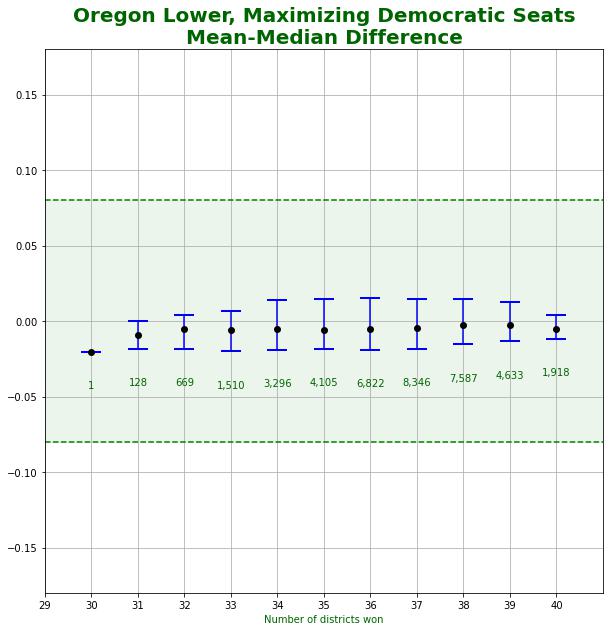}
    \includegraphics[width=0.29\linewidth]{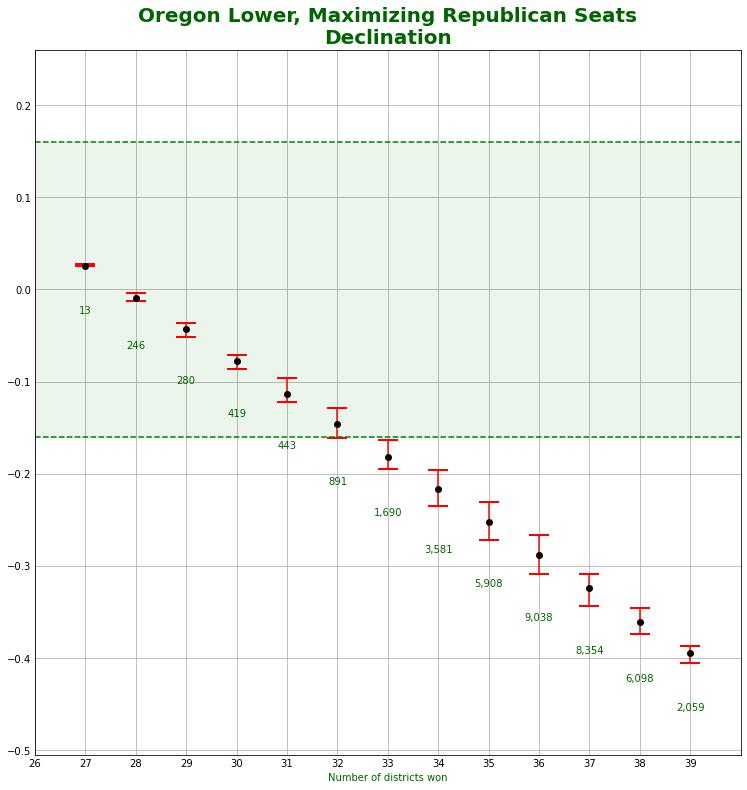}
    \includegraphics[width=0.29\linewidth]{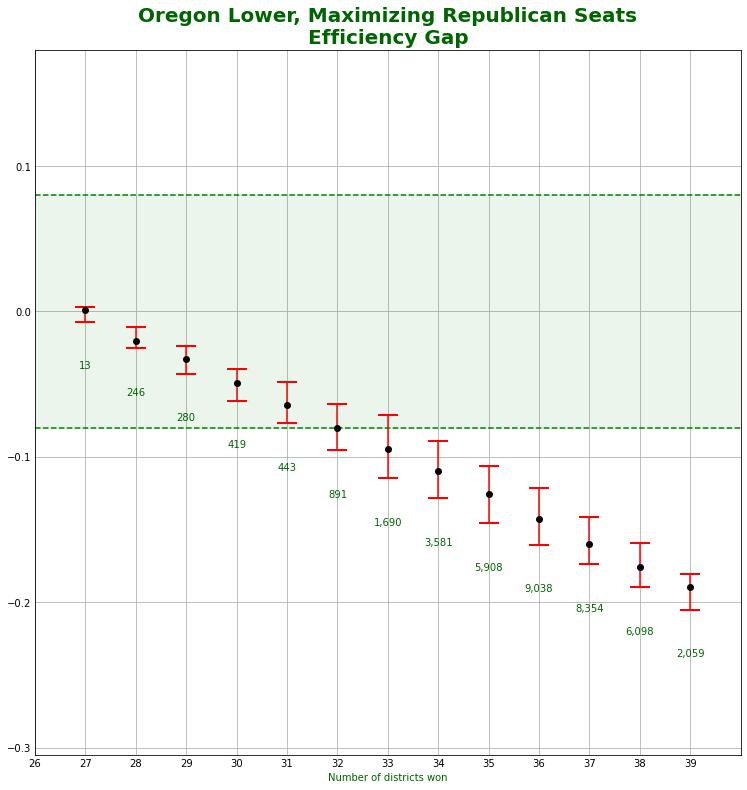}
    \includegraphics[width=0.29\linewidth]{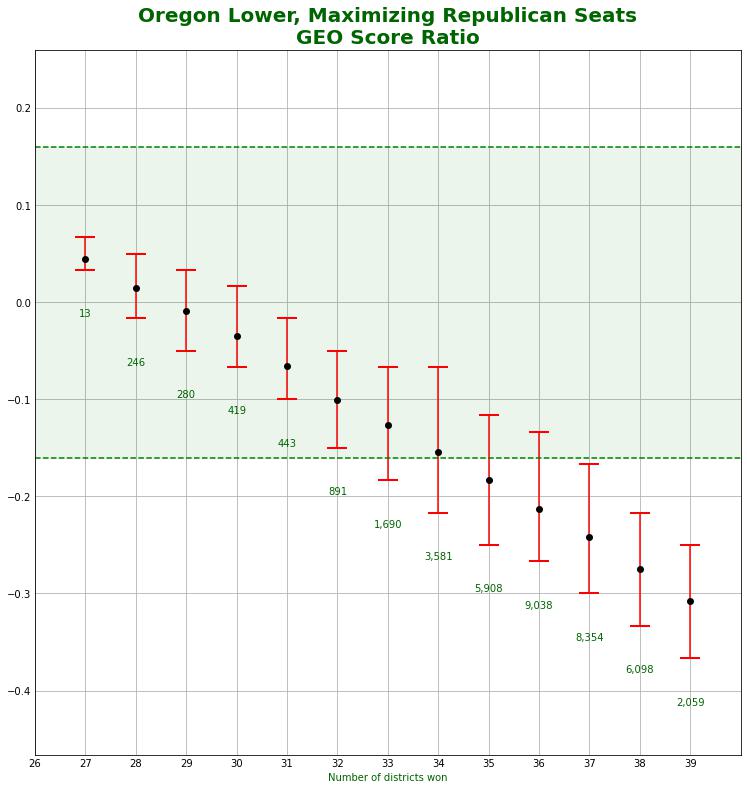}
    \includegraphics[width=0.29\linewidth]{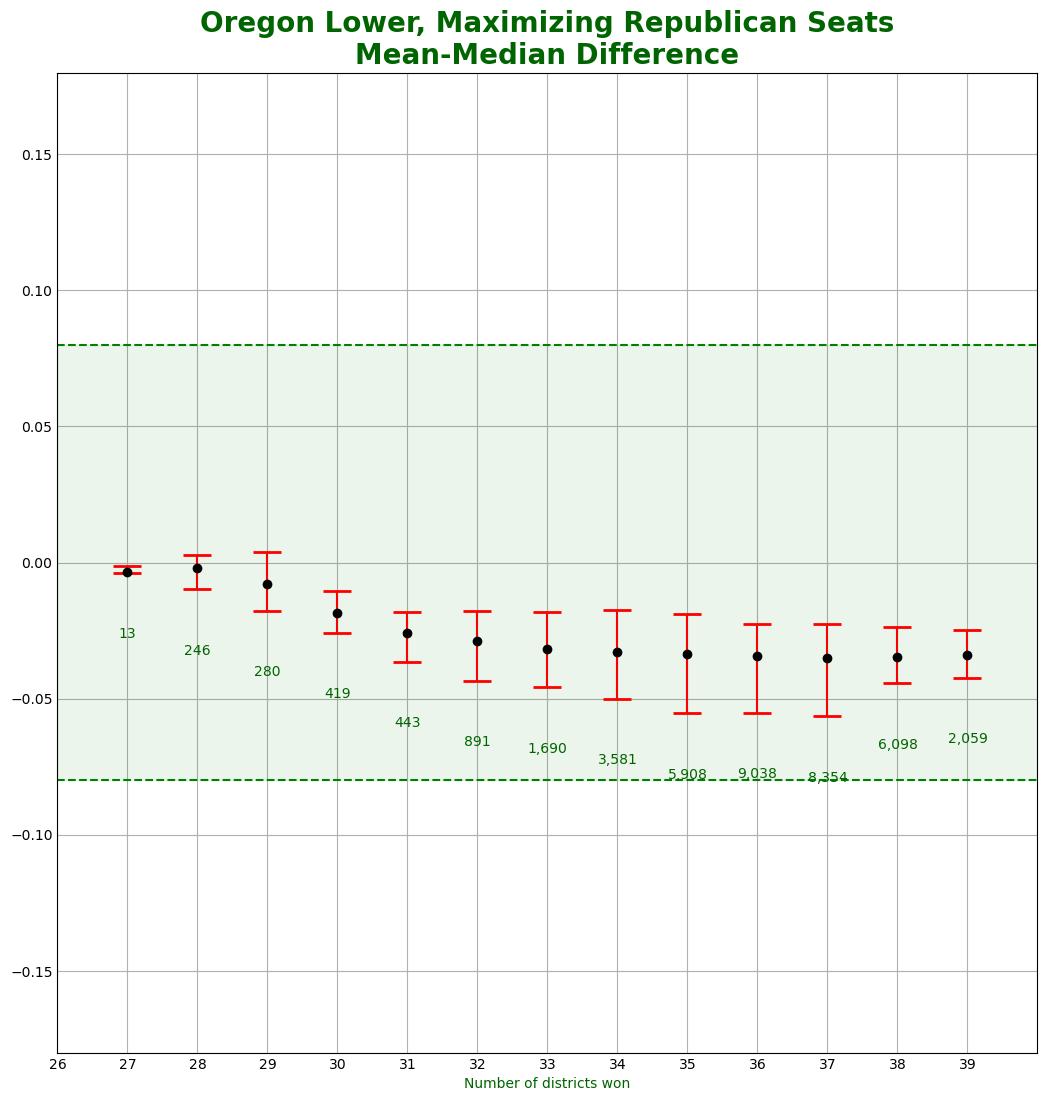}
    \caption{Results of Short Burst runs for Oregon lower house districts.}

    \label{fig:results_ORlower_short_bursts_all}
\end{figure}

\begin{figure}[h]
    \centering
    \includegraphics[width=0.29\linewidth]{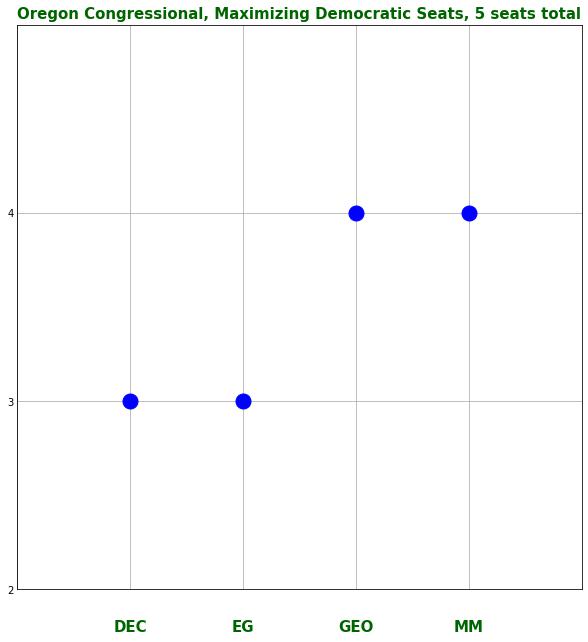}
    \includegraphics[width=0.29\linewidth]{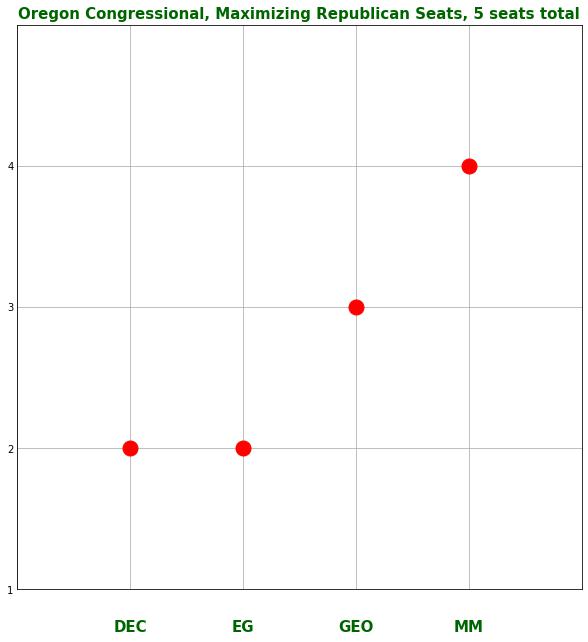}
    \newline
    \includegraphics[width=0.29\linewidth]{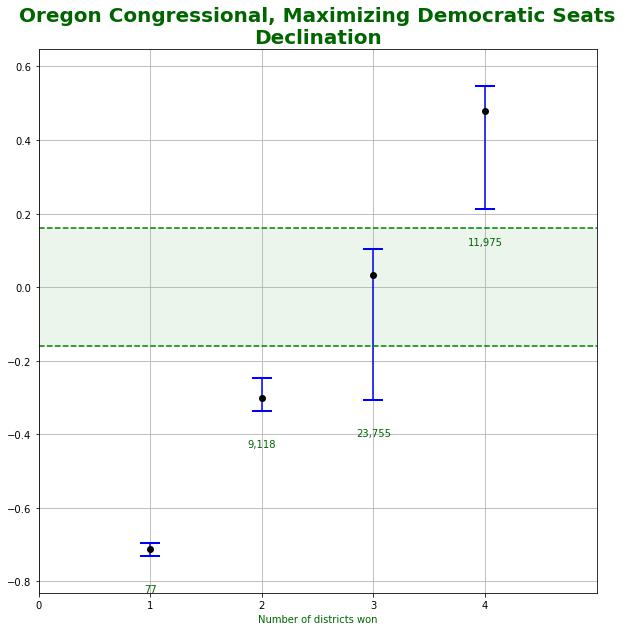}
    \includegraphics[width=0.29\linewidth]{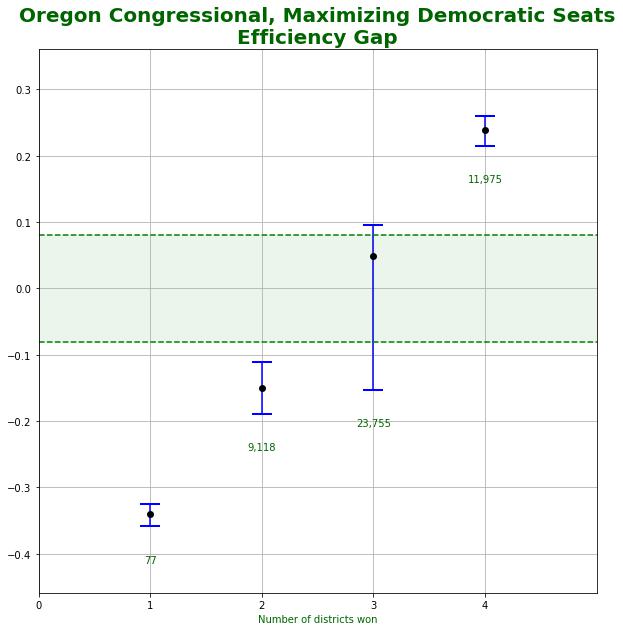}
    \includegraphics[width=0.29\linewidth]{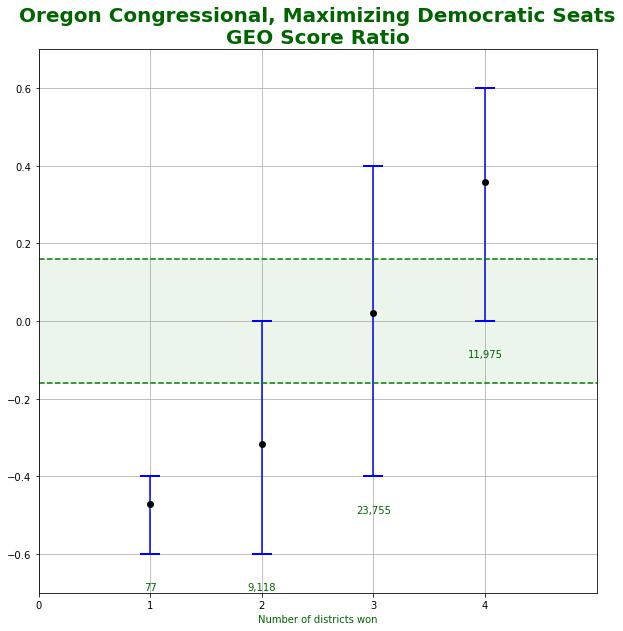}
    \includegraphics[width=0.29\linewidth]{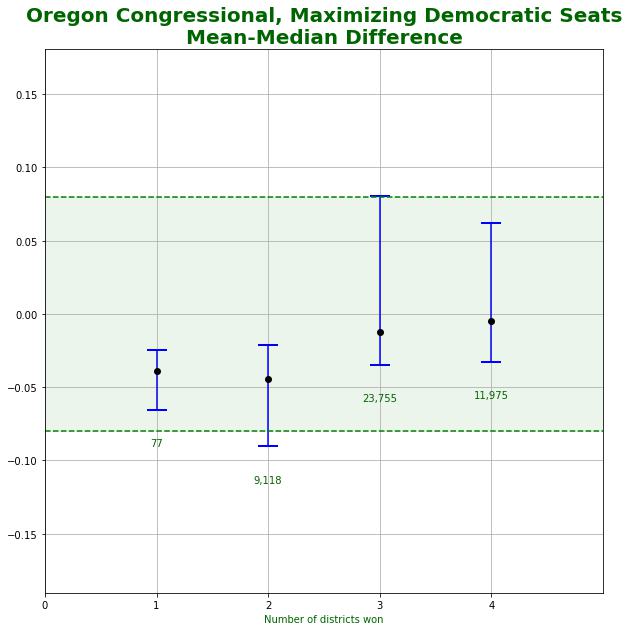}
    \includegraphics[width=0.29\linewidth]{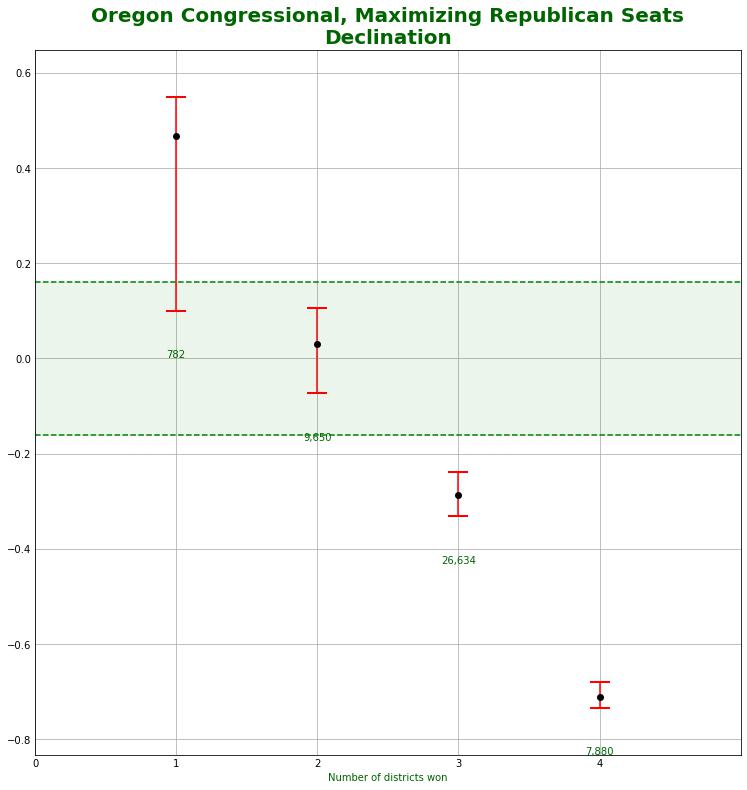}
    \includegraphics[width=0.29\linewidth]{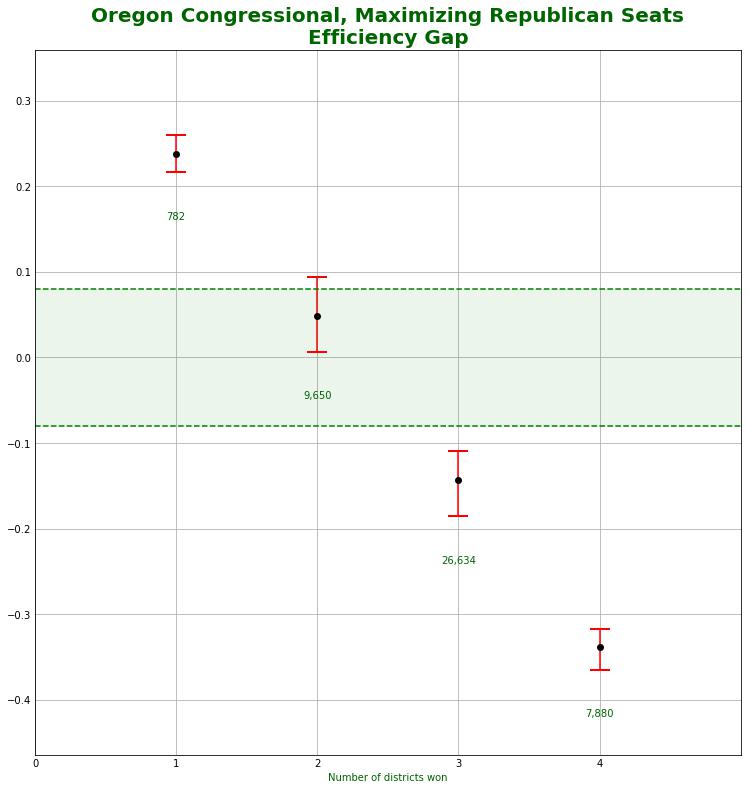}
    \includegraphics[width=0.29\linewidth]{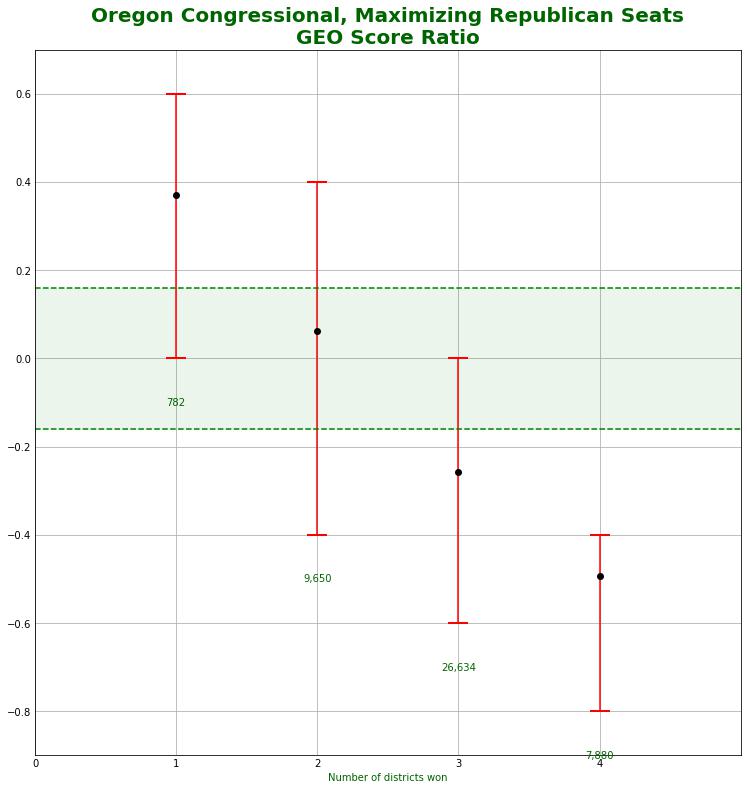}
    \includegraphics[width=0.29\linewidth]{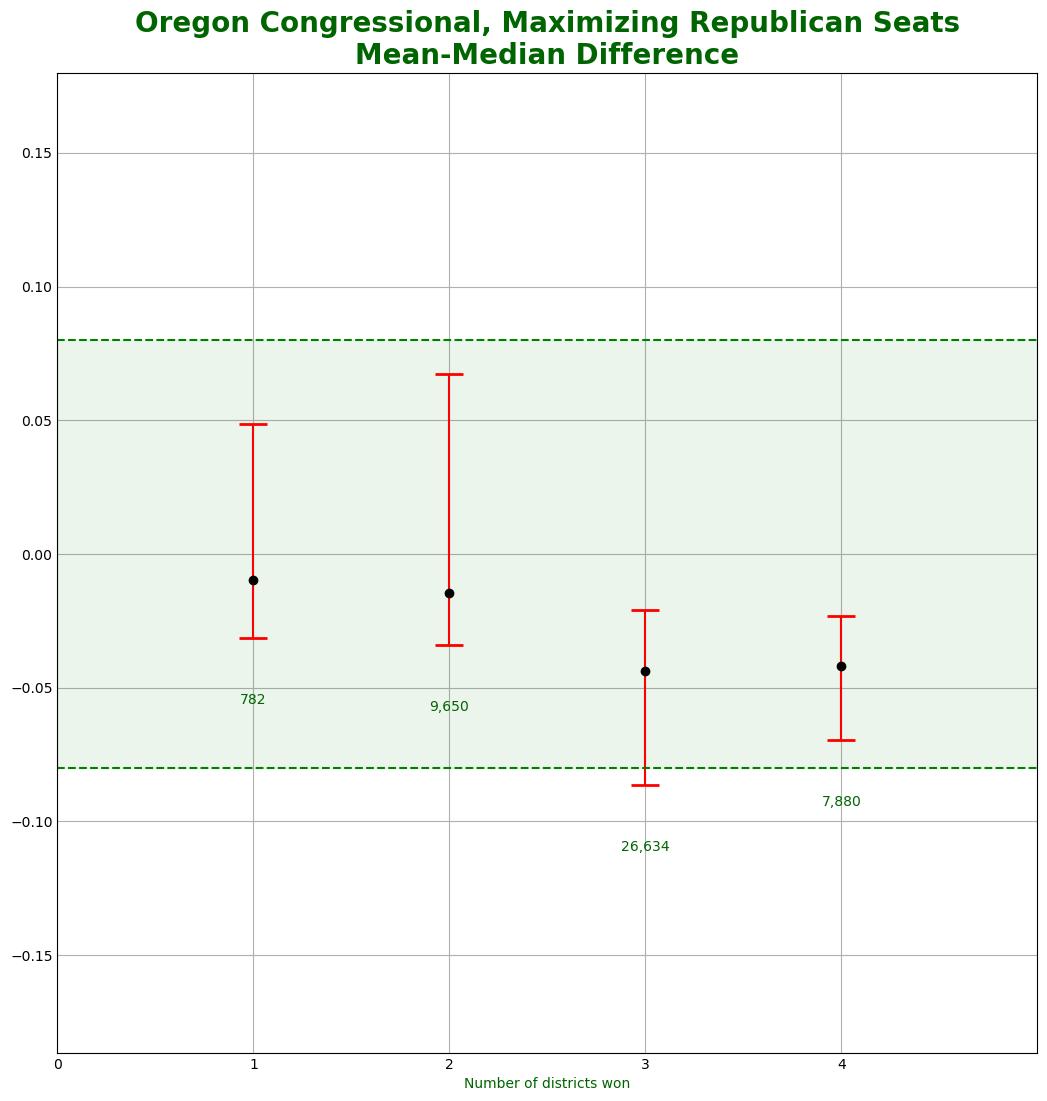}
    \caption{Results of Short Burst runs for Oregon Congressional districts.}

    \label{fig:results_ORcong_short_bursts_all}
\end{figure}

\begin{figure}[h]
    \centering
    \includegraphics[width=0.29\linewidth]{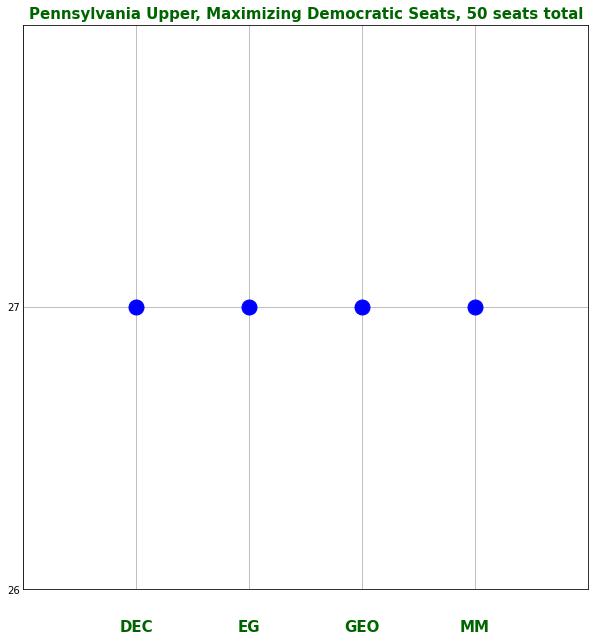}
    \includegraphics[width=0.29\linewidth]{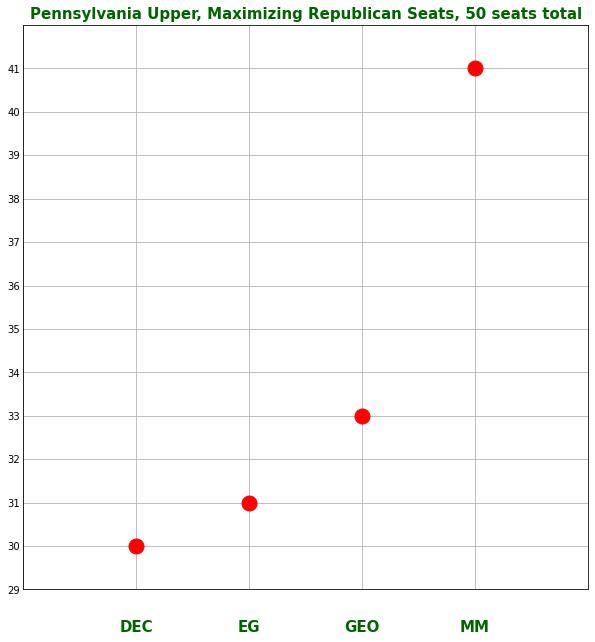}
    \newline
    \includegraphics[width=0.29\linewidth]{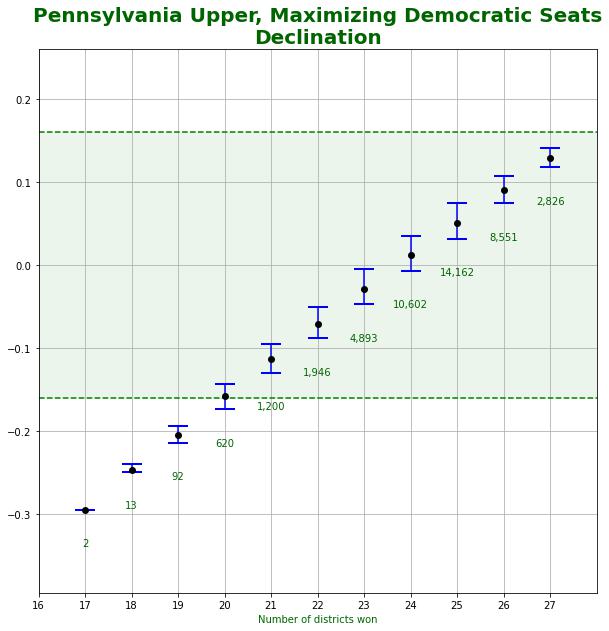}
    \includegraphics[width=0.29\linewidth]{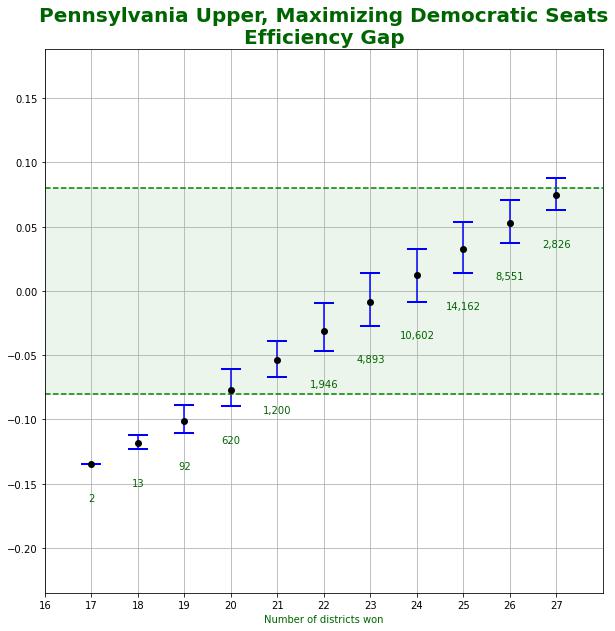}
    \includegraphics[width=0.29\linewidth]{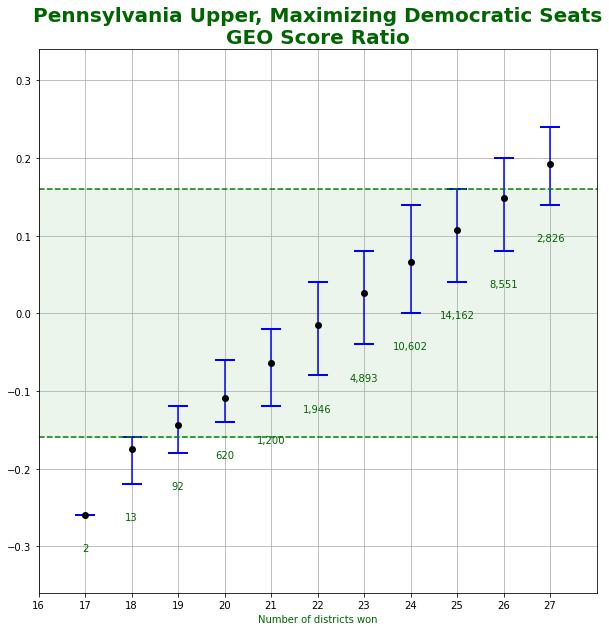}
    \includegraphics[width=0.29\linewidth]{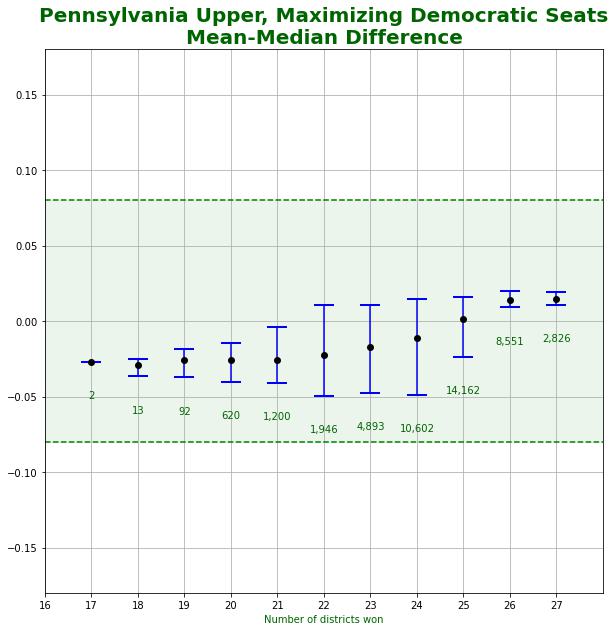}
    \includegraphics[width=0.29\linewidth]{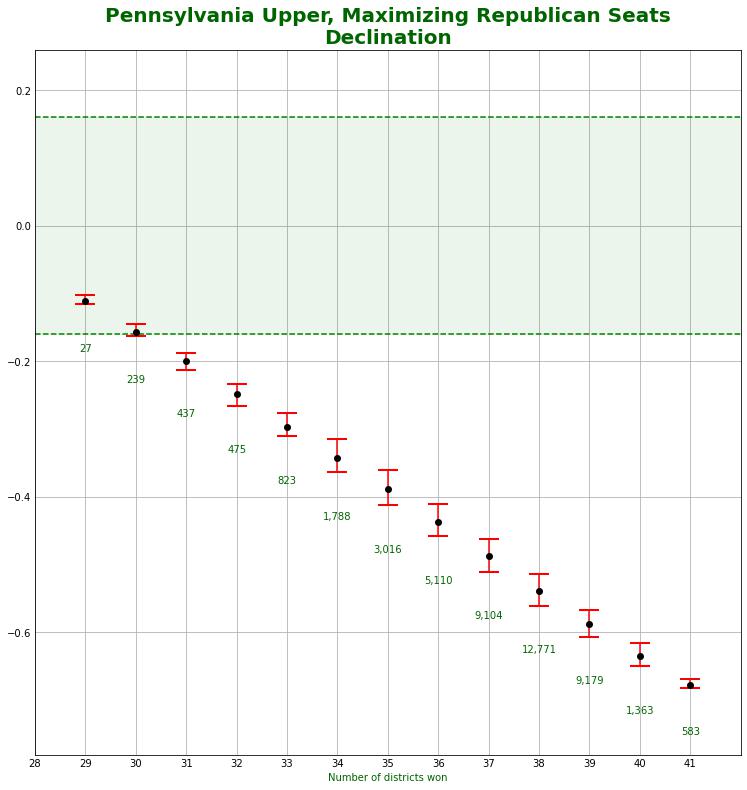}
    \includegraphics[width=0.29\linewidth]{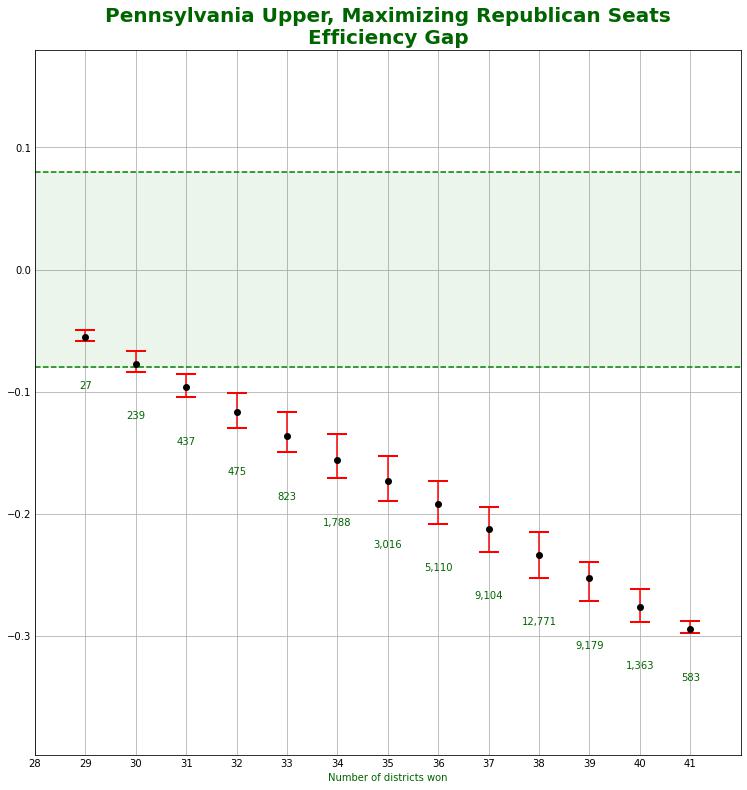}
    \includegraphics[width=0.29\linewidth]{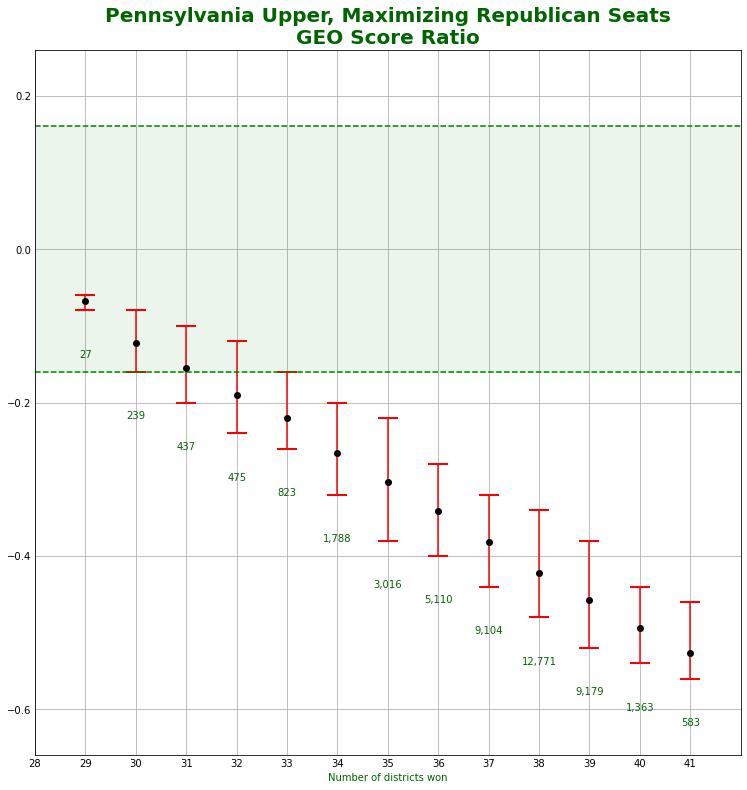}
    \includegraphics[width=0.29\linewidth]{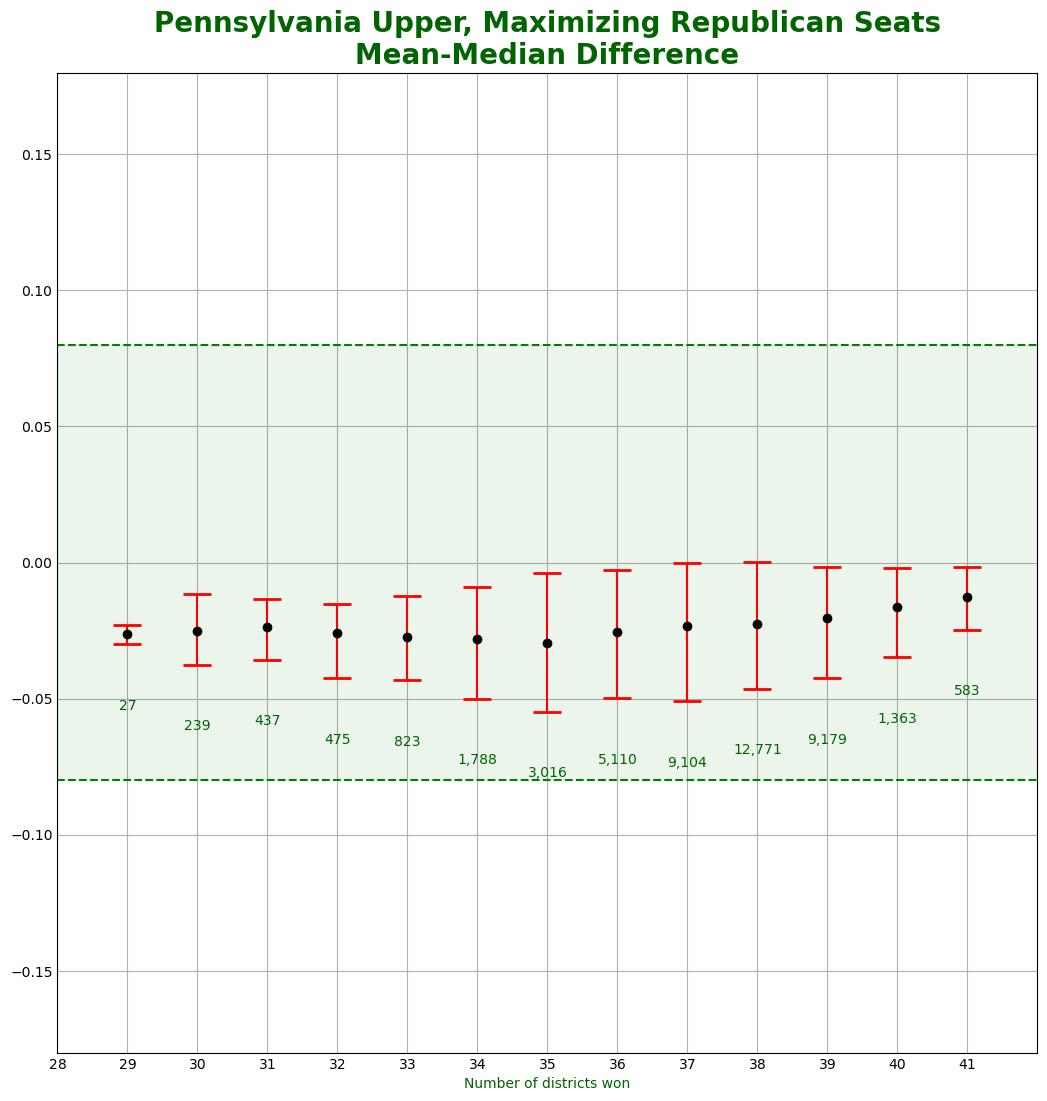}
    \caption{Results of Short Burst runs for Pennsylvania upper house districts.}

\end{figure}

\begin{figure}[h]
    \centering
    \includegraphics[width=0.29\linewidth]{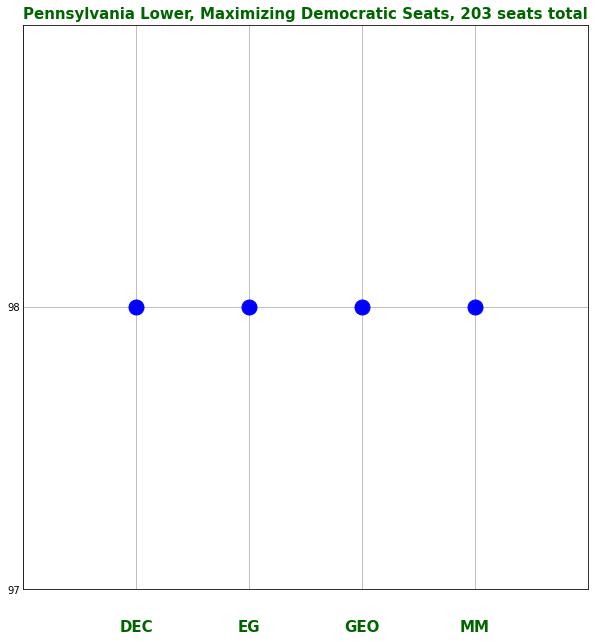}
    \includegraphics[width=0.29\linewidth]{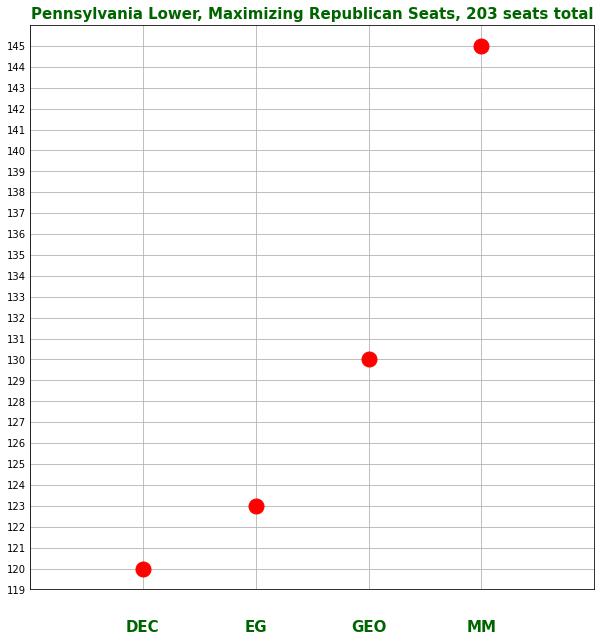}
    \newline
    \includegraphics[width=0.29\linewidth]{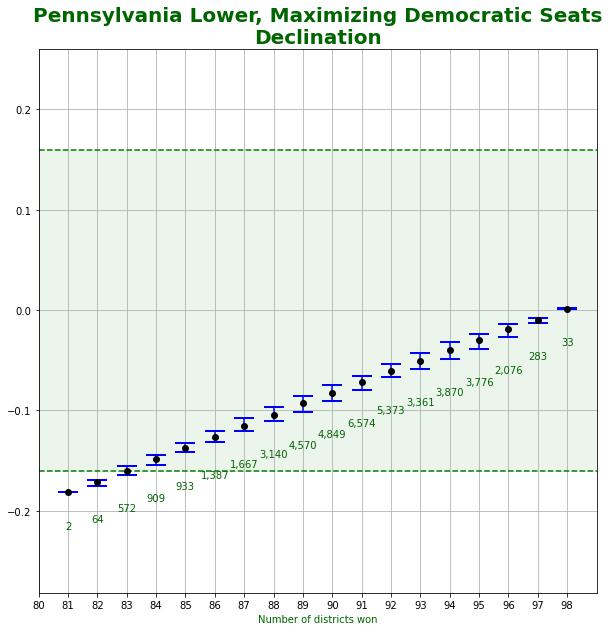}
    \includegraphics[width=0.29\linewidth]{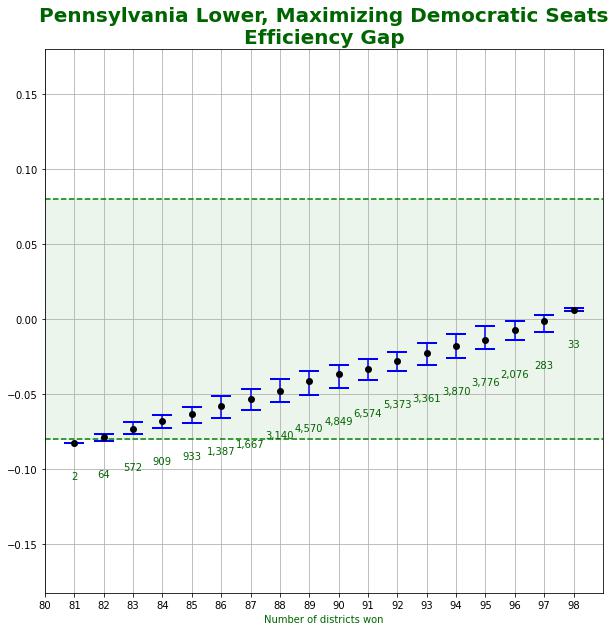}
    \includegraphics[width=0.29\linewidth]{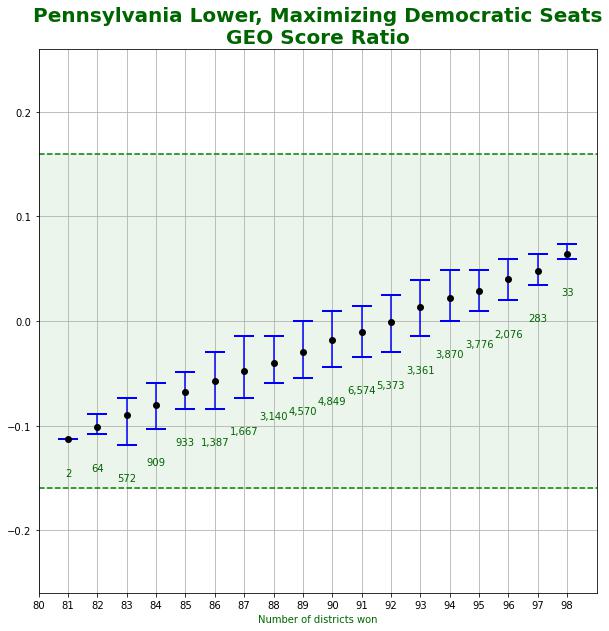}
    \includegraphics[width=0.29\linewidth]{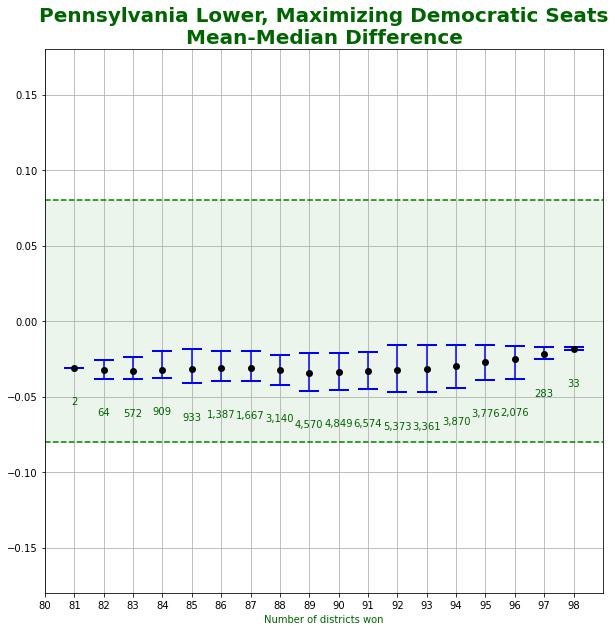}
    \includegraphics[width=0.29\linewidth]{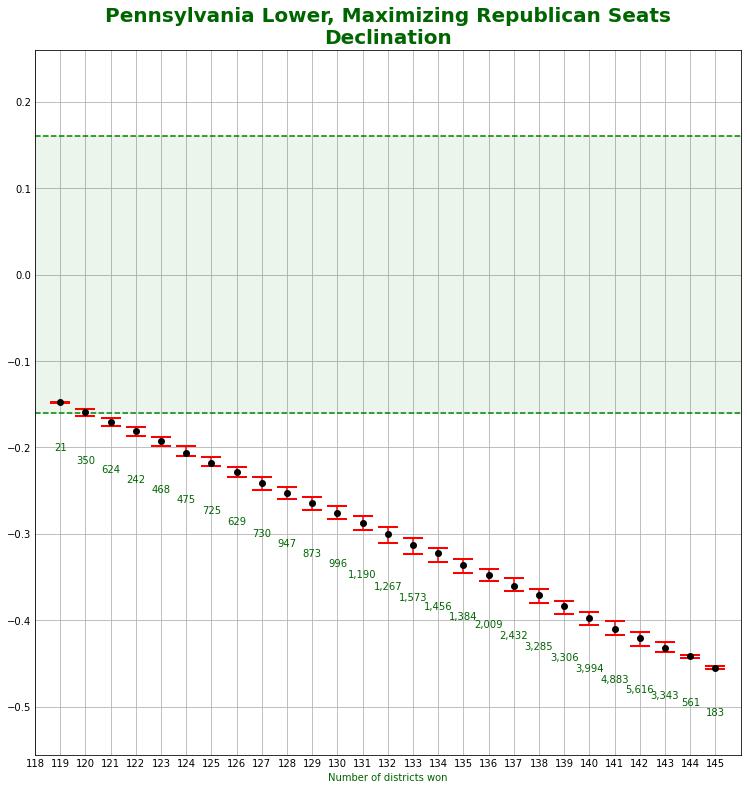}
    \includegraphics[width=0.29\linewidth]{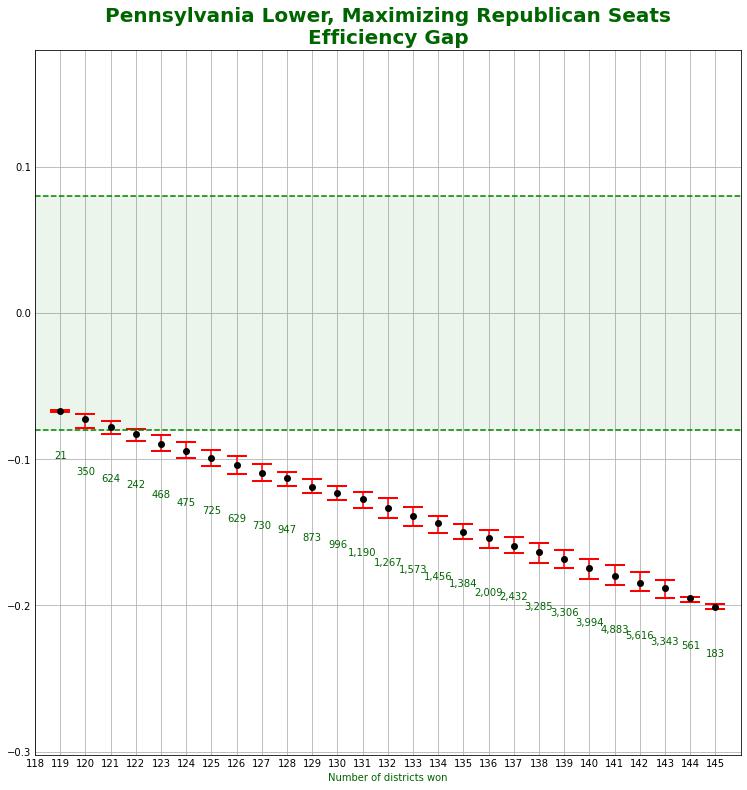}
    \includegraphics[width=0.29\linewidth]{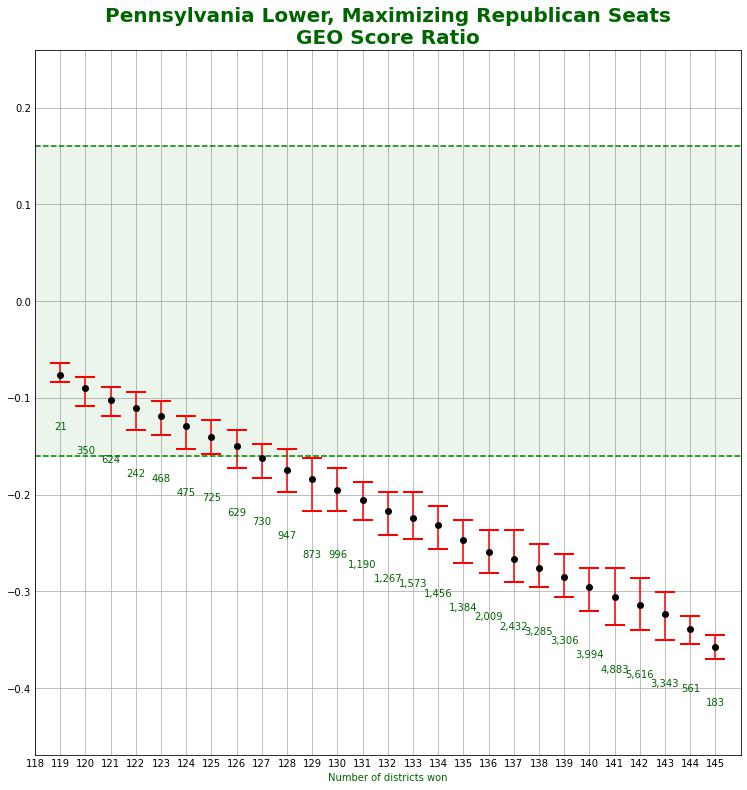}
    \includegraphics[width=0.29\linewidth]{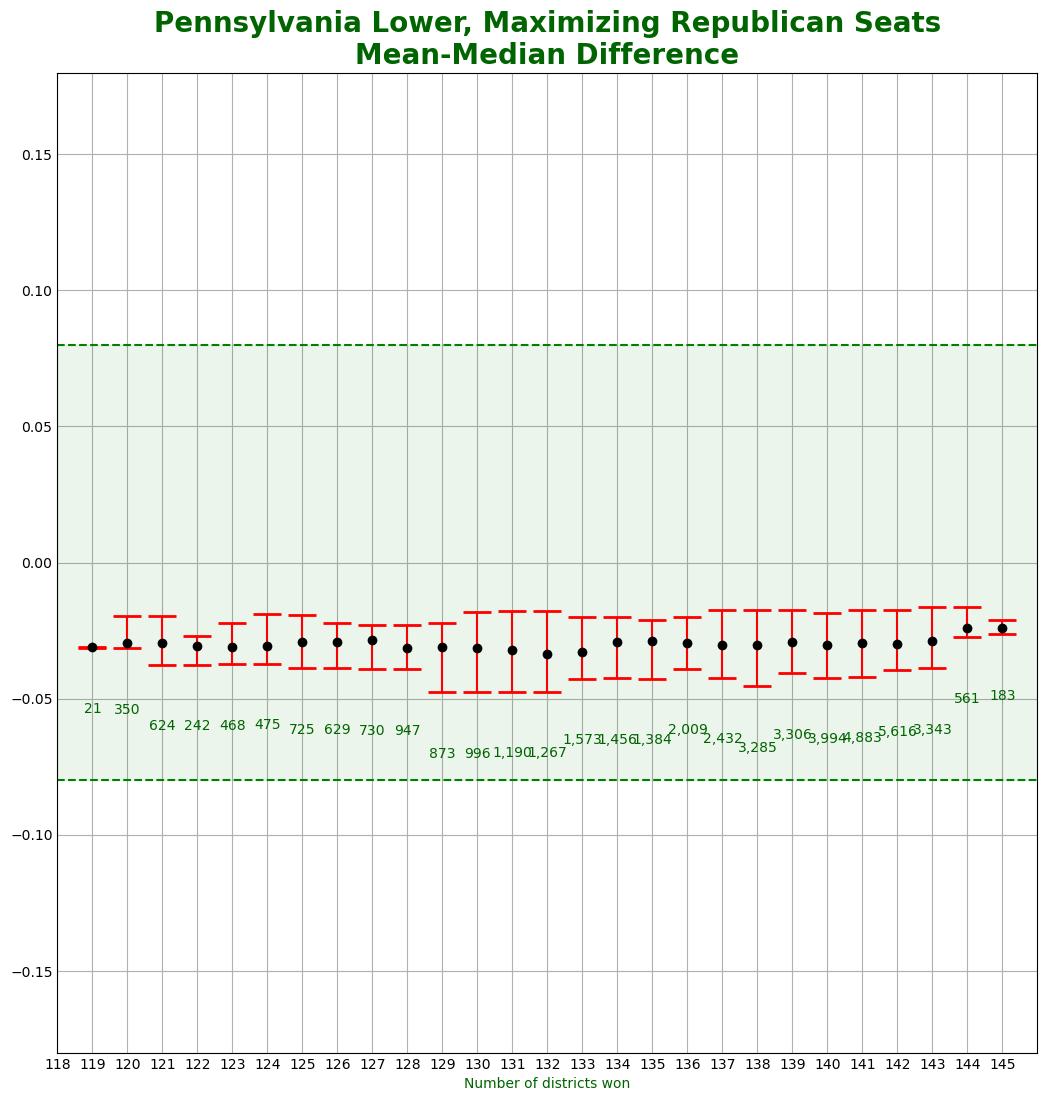}
    \caption{Results of Short Burst runs for Pennsylvania lower house districts.}

    \label{fig:results_PAlower_short_bursts_all}
\end{figure}

\begin{figure}[h]
    \centering
    \includegraphics[width=0.29\linewidth]{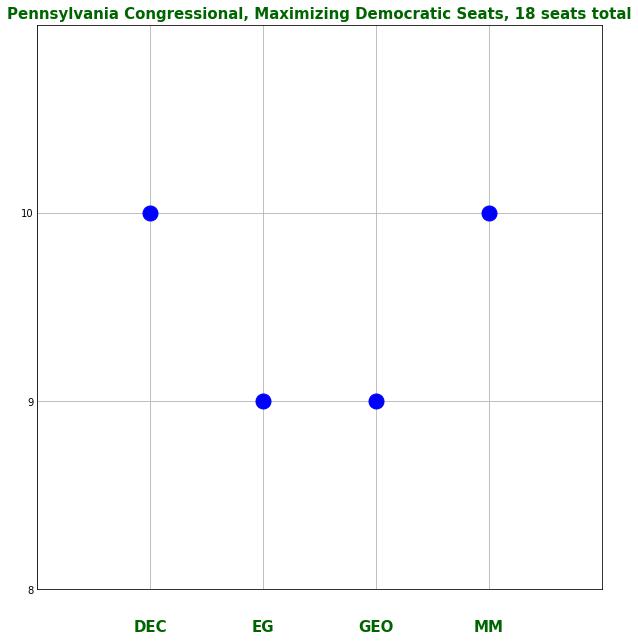}
    \includegraphics[width=0.29\linewidth]{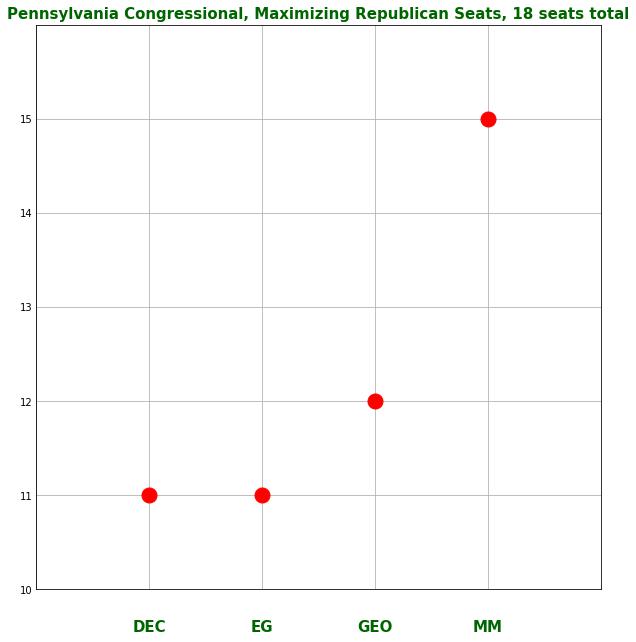}
    \newline
    \includegraphics[width=0.29\linewidth]{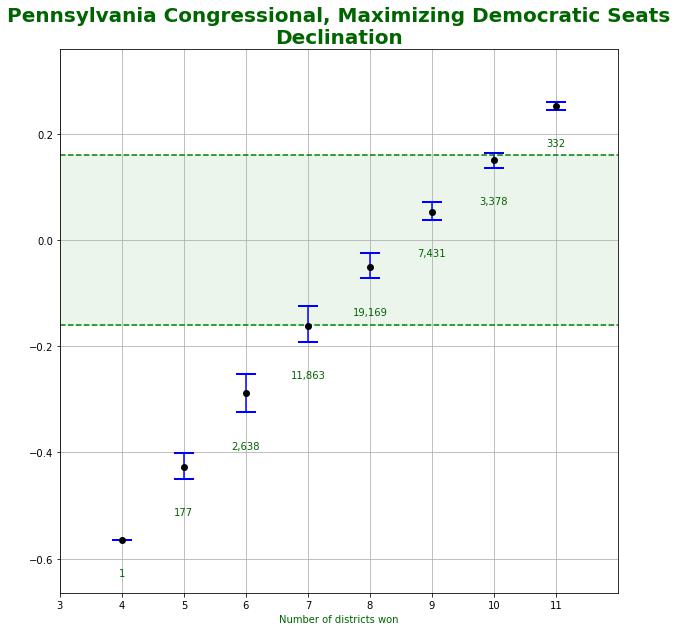}
    \includegraphics[width=0.29\linewidth]{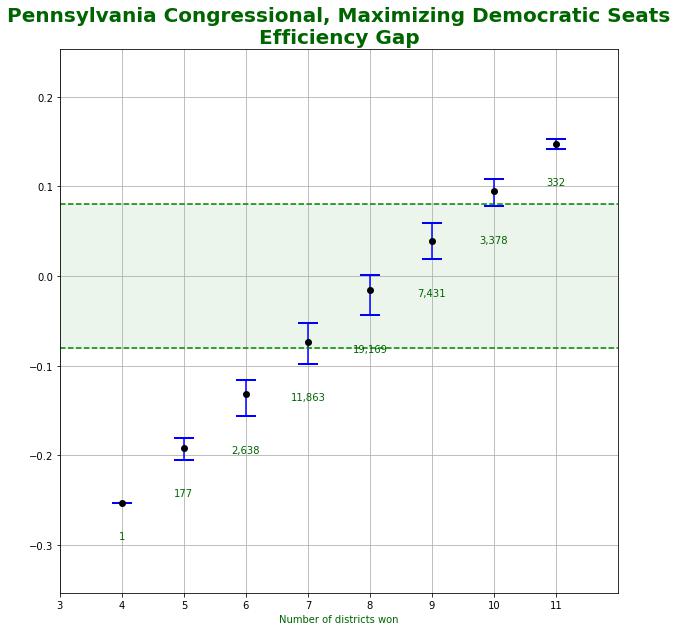}
    \includegraphics[width=0.29\linewidth]{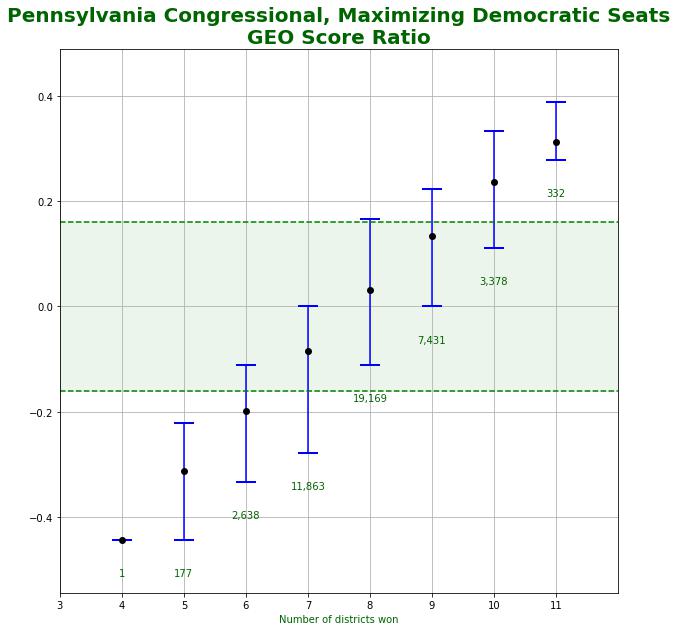}
    \includegraphics[width=0.29\linewidth]{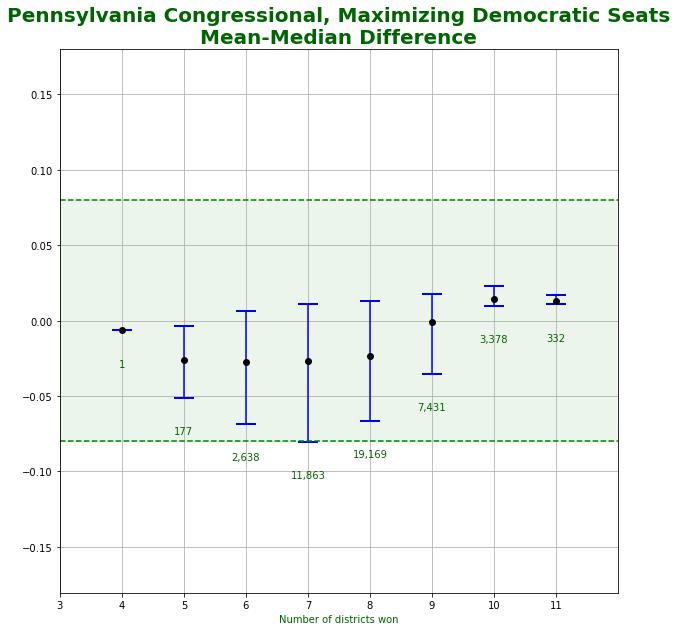}
    \includegraphics[width=0.29\linewidth]{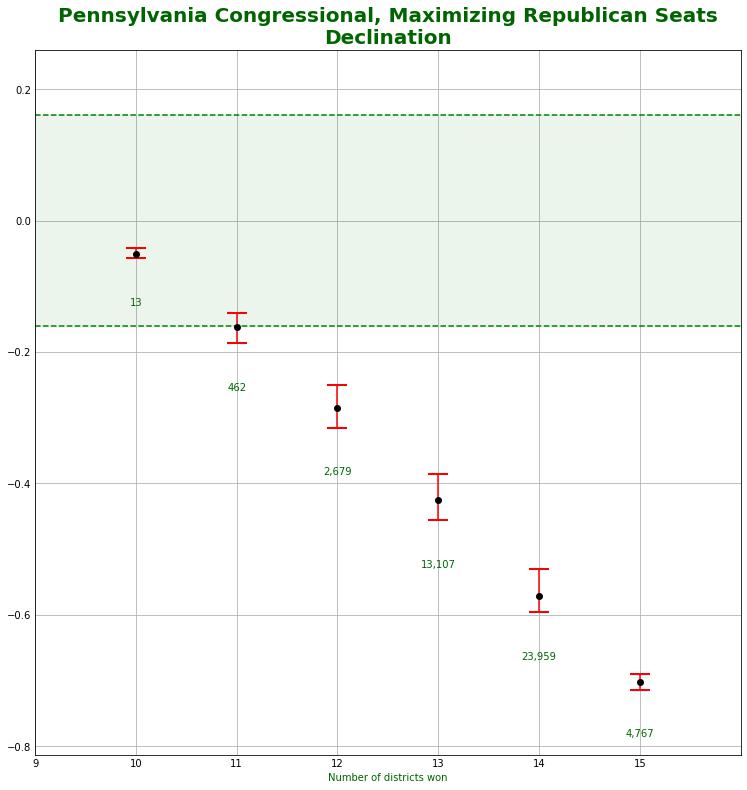}
    \includegraphics[width=0.29\linewidth]{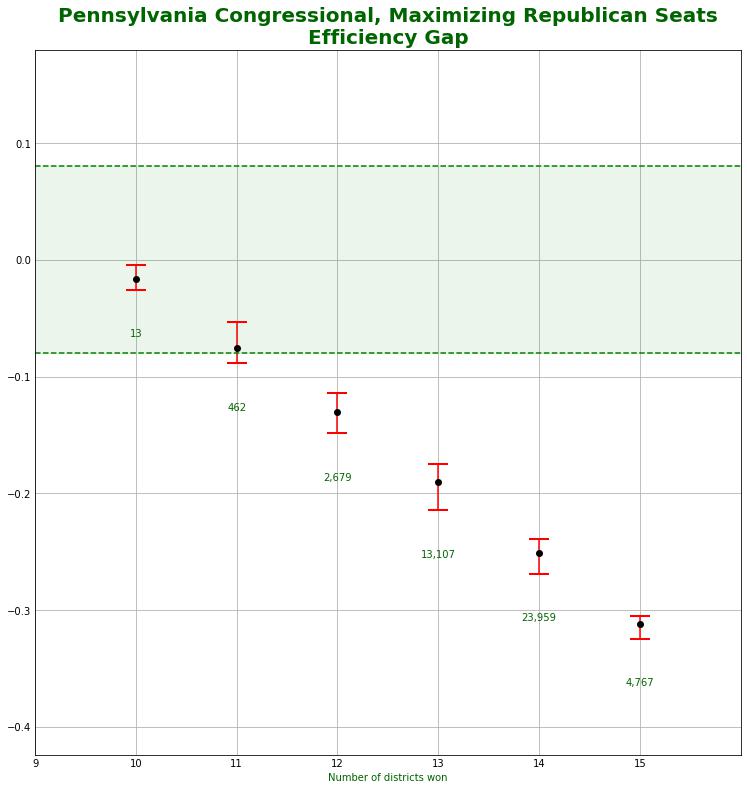}
    \includegraphics[width=0.29\linewidth]{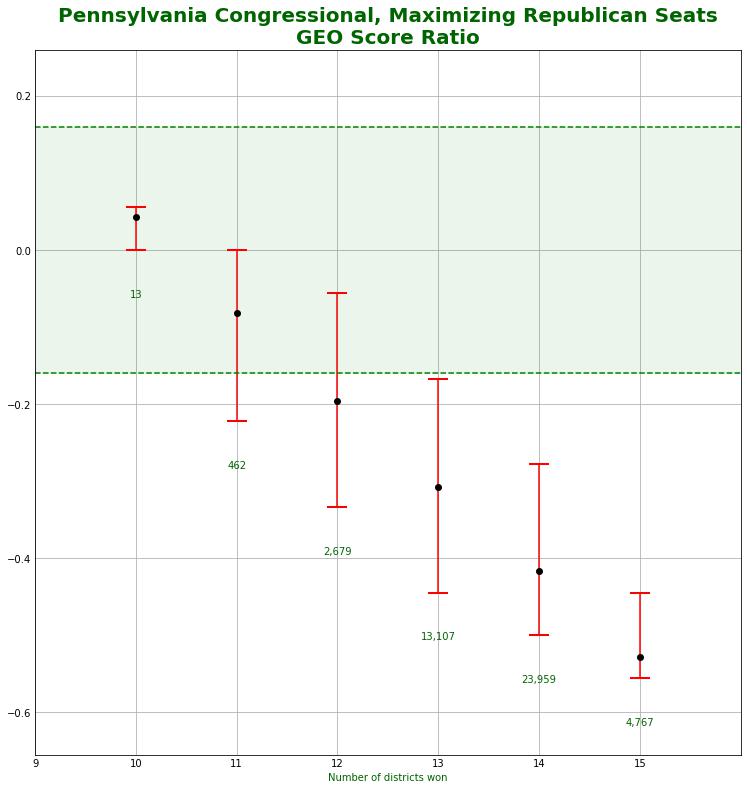}
    \includegraphics[width=0.29\linewidth]{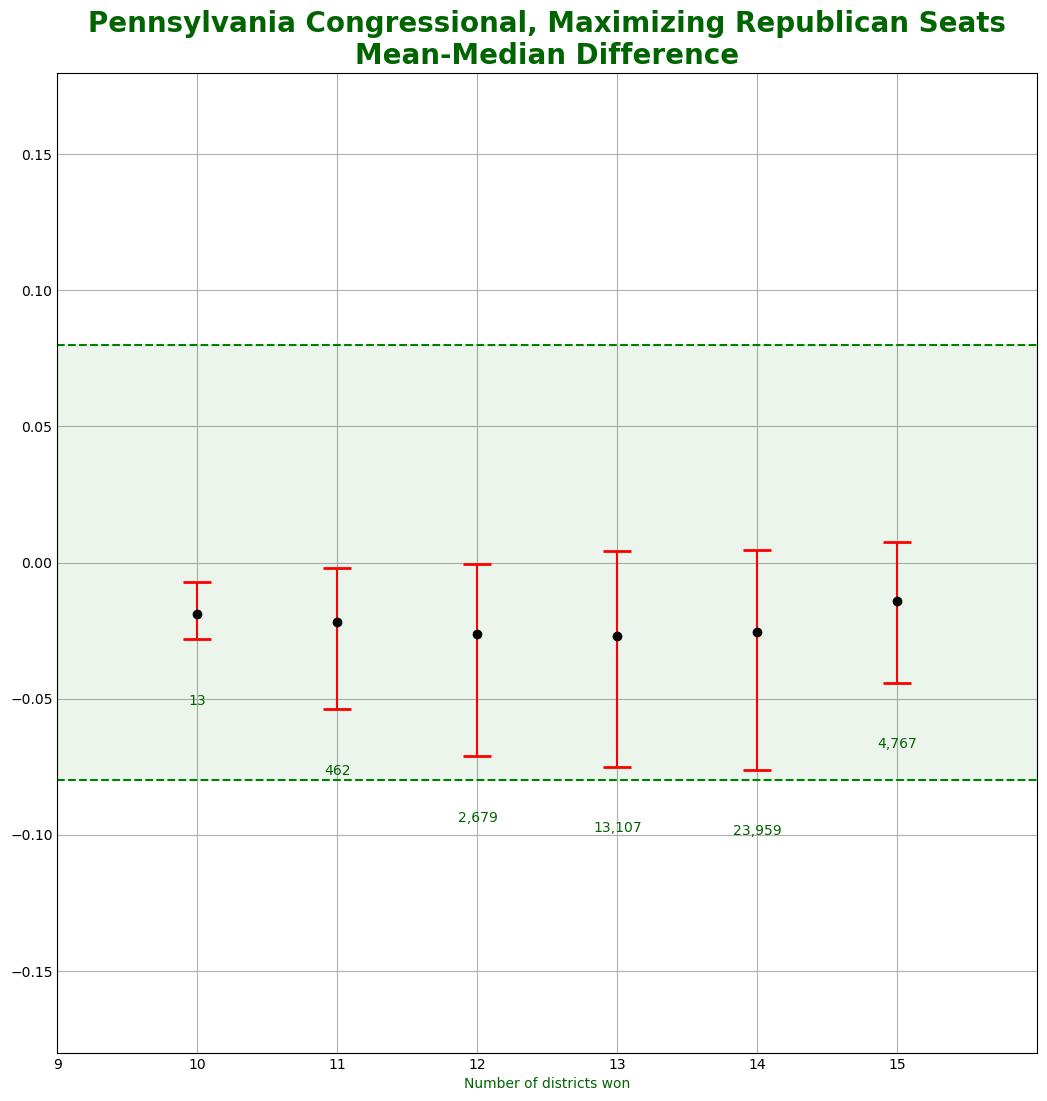}    
    \caption{Results of Short Burst runs for Pennsylvania Congressional districts.}

    \label{fig:results_PAcong_short_bursts_all}
\end{figure}

\end{appendices}

\end{document}